\documentclass[useamsfonts]{pasj01}

\usepackage{longtable}
\usepackage{graphicx,comment}
\usepackage{latexsym}
\usepackage{natbib}
\usepackage{amsmath}
\usepackage{url}
\graphicspath{{figures/}}

\newcommand{\beq}{\begin{equation}}
\newcommand{\eeq}{\end{equation}}

\newcommand\notedo[1]{\todo[color=yellow, inline, size=\small]{To do:
    #1}}
\newcommand\notewrite[1]{\todo[color=orange, inline, size=\small]{To write: #1}}

\newcommand\notecontrib[1]{\todo[color=green, inline, size=\small]{Contributors: #1}}

\newcommand{\revrev}[1]{#1}
\newcommand{\rev}[1]{#1}
\newcommand{\jbrv}[1]{#1}
\newcommand{\al}[1]{#1}
\newcommand{\mtrv}[1]{#1}
\newcommand{\rmrv}[1]{#1}
\newcommand{\hmrv}[1]{#1}
\newcommand{\smrv}[1]{#1}

\newcommand{\simgt}{\lower.5ex\hbox{$\; \buildrel > \over \sim \;$}}
\newcommand{\simlt}{\lower.5ex\hbox{$\; \buildrel < \over \sim \;$}}

\begin{document}
\title{The first-year shear catalog of the Subaru Hyper Suprime-Cam SSP Survey}

\author{Rachel Mandelbaum\altaffilmark{1}}
\author{Hironao Miyatake\altaffilmark{2,3}}
\author{Takashi Hamana\altaffilmark{4}}
\author{Masamune Oguri\altaffilmark{5,6,3}}
\author{Melanie Simet\altaffilmark{7,2}}
\author{Robert Armstrong\altaffilmark{8}}
\author{James Bosch\altaffilmark{8}}
\author{Ryoma Murata\altaffilmark{3,6}}
\author{Fran\c{c}ois Lanusse\altaffilmark{1}}
\author{Alexie Leauthaud\altaffilmark{9}}
\author{Jean Coupon\altaffilmark{10}}
\author{Surhud More\altaffilmark{3}}
\author{Masahiro Takada\altaffilmark{3}}
\author{Satoshi Miyazaki\altaffilmark{4}}
\author{Joshua S.\ Speagle\altaffilmark{11}}
\author{Masato Shirasaki\altaffilmark{4}}
\author{Crist\'obal Sif\'on\altaffilmark{8}}
\author{Song Huang\altaffilmark{3,9}}
\author{Atsushi J.\ Nishizawa\altaffilmark{12}}
\author{Elinor Medezinski\altaffilmark{8}}
\author{Yuki Okura\altaffilmark{13,14}}
\author{Nobuhiro Okabe\altaffilmark{15,16}}
\author{Nicole Czakon\altaffilmark{17}}
\author{Ryuichi Takahashi\altaffilmark{18}}
\author{Will Coulton\altaffilmark{19}}
\author{Chiaki Hikage\altaffilmark{3}}
\author{Yutaka Komiyama\altaffilmark{4,20}}
\author{Robert H.\ Lupton\altaffilmark{8}}
\author{Michael A.\ Strauss\altaffilmark{8}}
\author{Masayuki Tanaka\altaffilmark{4}}
\author{Yousuke Utsumi\altaffilmark{16}}
\altaffiltext{1}{McWilliams Center for Cosmology, Department of Physics, Carnegie Mellon University,
  Pittsburgh, PA 15213, USA}
\altaffiltext{2}{Jet Propulsion Laboratory, California Institute of Technology, Pasadena, CA 91109, USA}
\altaffiltext{3}{Kavli Institute for the Physics and Mathematics of the Universe (Kavli IPMU, WPI),
UTIAS,
  Tokyo Institutes for Advanced Study, The University of Tokyo, Chiba 277-8583, Japan}
\altaffiltext{4}{National Astronomical Observatory of Japan, Mitaka, Tokyo 181-8588, Japan}
\altaffiltext{5}{Research Center for the Early Universe, University of Tokyo, Tokyo 113-0033, Japan}
\altaffiltext{6}{Department of Physics, University of Tokyo, Tokyo 113-0033, Japan}
\altaffiltext{7}{University of California, Riverside, 900 University Avenue, Riverside, CA 92521,
  USA}
\altaffiltext{8}{Department of Astrophysical Sciences, 4 Ivy Lane, Princeton University, Princeton,
  NJ 08544}
\altaffiltext{9}{Department of Astronomy and Astrophysics, University of California, Santa Cruz,
  1156 High Street, Santa Cruz, CA 95064 USA}
\altaffiltext{10}{Department of Astronomy, University of Geneva, ch.\ d'\'Ecogia 16, 1290 Versoix,
  Switzerland}
\altaffiltext{11}{Harvard University, 60 Garden St, Cambridge, MA 02138}
\altaffiltext{12}{Institute for Advanced Research, Nagoya University, Furocho Chikusa-ku, Nagoya,
  464-8602, Japan}
\altaffiltext{13}{RIKEN Nishina Center, 2-1 Hirosawa, Wako, Saitama 351-0198, Japan}
\altaffiltext{14}{RIKEN-BNL Research Center, Department of Physics, Brookhaven National Laboratory,
Bldg.\ 510, Upton, NY, 11792, USA}
\altaffiltext{15}{Department of Physical Science, Hiroshima University, 1-3-1 Kagamiyama,
  Higashi-Hiroshima, Hiroshima 739-8526, Japan}
\altaffiltext{16}{Hiroshima Astrophysical Science Center, Hiroshima University, Higashi-Hiroshima,
  Kagamiyama 1-3-1, 739-8526, Japan}
\altaffiltext{17}{Academia Sinica Institute of Astronomy and Astrophysics, P.O.\ Box 23-141, Taipei
  10617, Taiwan}
\altaffiltext{18}{Faculty of Science and Technology, Hirosaki University, 3 Bunkyo-cho, Hirosaki,
  Aomori 036-8561, Japan}
\altaffiltext{19}{Department of Physics, Jadwin Hall, Washington Road, Princeton University, Princeton,
  NJ 08544}
\altaffiltext{20}{Department of Astronomy, School of Science, Graduate University for Advanced
  Studies (SOKENDAI), 2-21-1 Osawa, Mitaka, Tokyo 181-8588, Japan}
\email{rmandelb@andrew.cmu.edu}
\maketitle

\begin{abstract}
  We present and characterize the catalog of galaxy shape measurements that will be used for
  cosmological weak lensing measurements in the Wide layer of the \rmrv{first year of the} Hyper Suprime-Cam (HSC) survey.
  The catalog covers an area of 136.9~deg$^2$ split into six fields, with a mean $i$-band seeing of
  $0.58$\arcsec\ \rmrv{and $5\sigma$ point-source depth of $i\sim 26$}.  Given conservative galaxy
  selection criteria for first year science, the depth  and
  excellent image quality results in unweighted and weighted source number densities of 24.6 and
  21.8~arcmin$^{-2}$, \mtrv{respectively.}  
We define the requirements for cosmological weak lensing science with this catalog,
  \revrev{then focus on characterizing potential systematics in the catalog using a series of internal null tests for
  problems with point-spread function (PSF) modeling, shear estimation, and other aspects of the image processing.  
We find that the PSF models narrowly meet requirements for weak lensing science with this
  catalog, with fractional PSF model size residuals of approximately $0.003$ (requirement: 0.004) and the PSF model shape
  correlation function $\rho_1<3\times 10^{-7}$ (requirement: $4\times 10^{-7}$) at 0.5$^\circ$
  scales.  A variety of galaxy shape-related null tests are statistically consistent with zero,
  but star-galaxy shape correlations reveal additive systematics on on $>1^\circ$ scales that are sufficiently large as to require mitigation
  in cosmic shear measurements.}
Finally, we discuss the
  dominant systematics and the planned algorithmic changes to reduce them in future data reductions.
\end{abstract}

\section{Introduction}

The currently accepted cosmological model that is broadly consistent with multiple observations,
$\Lambda$CDM, is dominated by dark ingredients: dark matter, which we observe through its
gravitational effects, and dark energy, the presence of which was inferred due to the accelerated
expansion of the universe as detected using supernovae
\citep{1998AJ....116.1009R,1999ApJ...517..565P}.  Weak gravitational lensing provides us with a way
of observing the total matter density (including dark matter), via the deflections of light due to
intervening matter along the line-of-sight, which both magnifies and distorts galaxy shapes
\citep[for recent reviews, see][]{2013PhR...530...87W,2015RPPh...78h6901K}.
The lensing measurement that is commonly used to constrain the amplitude and growth of matter
fluctuations is `cosmic shear', the auto-correlation of galaxy shape distortions.
\rmrv{When measured in redshift bins (`tomography'), cosmic shear is particularly powerful at
  tracing structure growth as a function of time.}
Since the initial detections of cosmic shear a decade ago
\citep{2000MNRAS.318..625B, 2000A&A...358...30V,2001ApJ...552L..85R, 2002ApJ...572...55H},
\mtrv{ever larger}
datasets and
\mtrv{increasingly}
sophisticated measurement techniques have led to steadily
decreasing errors, both statistical and systematic
\citep[e.g., most
recently,][]{2013MNRAS.432.2433H,2016PhRvD..94b2002B,2016ApJ...824...77J,2017MNRAS.465.1454H}.

What has driven the development of ever-larger lensing surveys is the realization more than a decade
ago that weak lensing measurements of structure growth - particularly as a function of time - can
place powerful constraints on the initial amplitude of matter fluctuations, the matter density, and
the nature of dark energy \citep[e.g.,][]{2002PhRvD..65b3003H, 2002PhRvD..65f3001H,TakadaJain:04,
  2004PhRvD..70l3515B, 2004ApJ...600...17B, 2004PhRvD..69h3514I,2004ApJ...601L...1T}. \mtrv{Moreover,}
the
scale dependence of structure growth can be used to constrain the neutrino mass
\citep[e.g.,][]{2003PhRvL..91d1301A}.  The galaxy-shear cross-correlation function (or galaxy-galaxy
lensing) can be combined with galaxy clustering to provide information about structure growth
and dark energy \citep[e.g.,][]{2013MNRAS.432.1544M,2015ApJ...806....2M,2017MNRAS.467.3024L,2017MNRAS.464.4045K} and,
when \mtrv{combined} with redshift-space distortions, about gravity on cosmological scales
\citep[e.g.,][]{2016MNRAS.456.2806B,2017MNRAS.465.4853A}.
In addition, weak lensing by clusters of galaxies also contains information about dark energy
\citep[e.g.,][]{2011PhRvD..83b3008O}, and provides \rmrv{an} important means
\mtrv{to calibrate}
mass-observable
relations of clusters for using cluster abundances to constrain cosmology
\citep[e.g.,][]{2010ApJ...709...97L,Okabeetal:10,2014ApJ...794..136D,2014MNRAS.443.1973V,2015MNRAS.449..685H,2016MNRAS.461.3794O,2016JCAP...08..013B}.

Currently there are three ongoing \mtrv{wide-area} sky surveys that have weak lensing among their primary science
cases: the Kilo-Degree Survey\footnote{\texttt{http://kids.strw.leidenuniv.nl/}} \citep[KiDS:][]{2013ExA....35...25D}, the Dark Energy Survey\footnote{\texttt{http://www.darkenergysurvey.org/}}
(DES), and the survey that is the subject of this paper: the Hyper Suprime-Cam
survey\footnote{\texttt{http://hsc.mtk.nao.ac.jp/}}
(HSC).  In the context of these other surveys, the unique aspect of the HSC survey is its
combination of depth and high-resolution imaging that gives it a longer redshift baseline.  For
low-redshift cosmological constraints, the primary consideration is area, making DES more powerful;
while for higher-redshift constraints, the depth and resolution of HSC gives it the best
constraining power.  Moreover, the excellent image quality in HSC should enable the reduction of
systematic uncertainties in weak lensing shear, which is important to avoid a systematics-dominated
measurement.
In the coming decade, three larger surveys will begin that will place even
stronger cosmological constraints than is possible with ongoing surveys:
Euclid\footnote{\texttt{http://sci.esa.int/euclid/},
  \texttt{http://www.euclid-ec.org}}
  \citep{2011arXiv1110.3193L},
  LSST\footnote{\texttt{http://www.lsst.org/lsst/}}
  \citep{2009arXiv0912.0201L}, and
  WFIRST\footnote{\texttt{http://wfirst.gsfc.nasa.gov}}
  \citep{2015arXiv150303757S}.  As the deepest of the ongoing weak lensing surveys, the HSC survey may be
  considered a path-finder for LSST in many \rmrv{respects}, as it will encounter many of the issues faced in
  the LSST image processing when it comes to the challenges posed by deep ground-based
  images\rmrv{, albeit with far fewer exposures at any given point within the survey footprint.}

While weak lensing is a powerful cosmological measurement, it is \rmrv{also very technically
  challenging due to the small size of the shear signals, which are dwarfed by the noise introduced
  by the much larger intrinsic shapes of galaxies (shape noise).  When averaging over the large
  galaxy samples needed to make this statistical measurement, it is also} important to ensure that systematic
errors are reduced below the statistical floor so that the cosmological constraints are not biased.
Observationally, there are several sources of bias related to the process of inferring coherent
galaxy shape distortions \citep[e.g.,][]{2015MNRAS.450.2963M,2016MNRAS.460.2245J}, which is
typically done by measuring shapes for each galaxy and then taking appropriate weighted averages or
correlation functions\footnote{But see \citet{2014MNRAS.438.1880B} and \citet{2016MNRAS.459.4467B}
  for examples of methodology that do not work from per-galaxy shapes and \rmrv{rather} infer shear only for the
  ensemble \mtrv{of galaxies}, avoiding certain systematic errors in the process.}.  The redshift distribution of the lensed galaxies must be well-understood so as to
properly interpret the observed shape distortions in terms of mass density and structure growth, so
this is another possible source of systematic \mtrv{error}
\citep[e.g.,][]{2016PhRvD..94d2005B,2017MNRAS.465L..20S}.  Finally, there are
\mtrv{several astrophysical}
uncertainties, such as intrinsic alignments of galaxy shapes
\citep{2015SSRv..193....1J,2015SSRv..193...67K,2015SSRv..193..139K,2015PhR...558....1T} and the
impact of baryonic effects on the matter power spectrum
\citep{2013PhRvD..87d3509Z,2015MNRAS.454.1958M}.  

The goal of this paper is to address the first of
these problems: the difficulty in robustly inferring weak lensing shear from the galaxy images in
\mtrv{the context}
of the convolution by the point-spread function (PSF) and other image processing issues.
Here we \rev{focus primarily on} internal tests (within the catalog \rev{without reference to simulations}) 
to demonstrate that shear-related systematics in the HSC first-year shear catalog, \mtrv{constructed based on the
data taken between 2014 March and 2016 April},
are
reduced to below the level needed for first-year HSC lensing science. \rev{Some systematics cannot
  be assessed using internal tests; we refer to additional papers that characterize those
  systematics and their contributions to the error budget.}

We begin in Section~\ref{sec:software} with a summary of the software used for analysis of the HSC
survey images for shear inference and null testing.  In Section~\ref{sec:requirements}, we define the
requirements on the PSF modeling, shear inference, and other aspect\rmrv{s} of the image analysis to ensure
that the first-year HSC survey weak lensing analysis is not dominated by systematics.  We show tests
of the PSF modeling process in Section~\ref{sec:psf}, and of the shape measurements and shear inference
in Section~\ref{sec:shear}.  Simulations used to characterize the shear catalog are described in
Section~\ref{sec:sims}.  Tests of other aspect\rmrv{s} of the image processing are shown in
Section~\ref{sec:earlystages}.  \rev{\rmrv{While the photometric redshifts for the HSC wide survey are characterized in \citet{2017arXiv170405988T}, and
their performance for weak lensing \rmrv{will be} quantified in other papers, we briefly
comment on issues related to photometric redshifts for the shear catalog in
Section~\ref{sec:photoz}.}   We summarize the key elements of the systematic error budget and areas
for future work in  Section~\ref{sec:summary}.}

\section{Data and analysis software}\label{sec:software}

In this section, we \rmrv{define} the dataset used for first
year science, and the key software used to analyze it and produce the shear catalog described in
this paper.  As there are separate papers describing \mtrv{the survey overview and design of HSC
survey \citep{SurveyOverview},}
the HSC camera \citep{CameraPaper:inprep} and
the HSC analysis pipeline \citep{PipelinePaper:inprep}, our discussions of these will be brief.  We
refer interested readers to those papers for more detail, as well as to
\citet{2017arXiv170208449A} -- hereafter the HSC DR1 paper --
for more general information about the dataset.

\subsection{First year dataset}\label{subsec:firstyear}


\begin{figure*}
\begin{center}
\includegraphics[clip,width=6.3in]{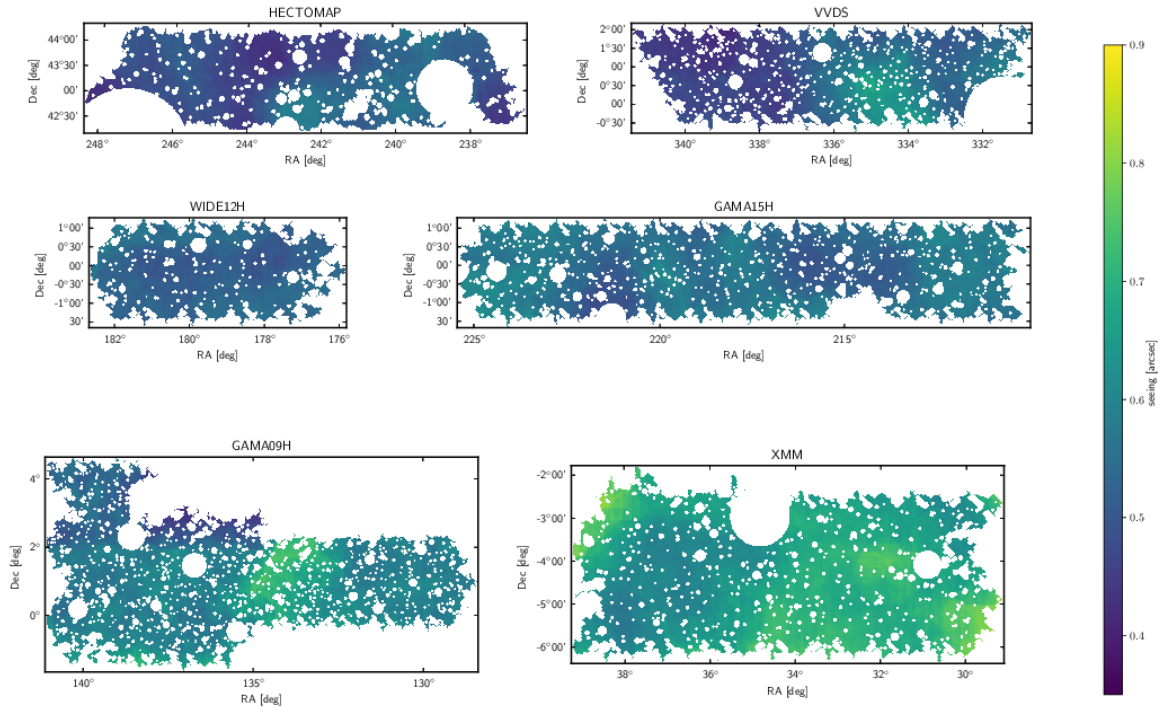}
\end{center}
\caption{Map of the $i$-band PSF FWHM across each field. The holes in area coverage are due to
  masking bright stars, while the other aspects of the area coverage are determined as described in
  Section~\ref{subsec:area}. 
}
\label{fig:seeing_map}
\end{figure*}

\begin{figure*}
\begin{center}
\includegraphics[clip,width=6.3in]{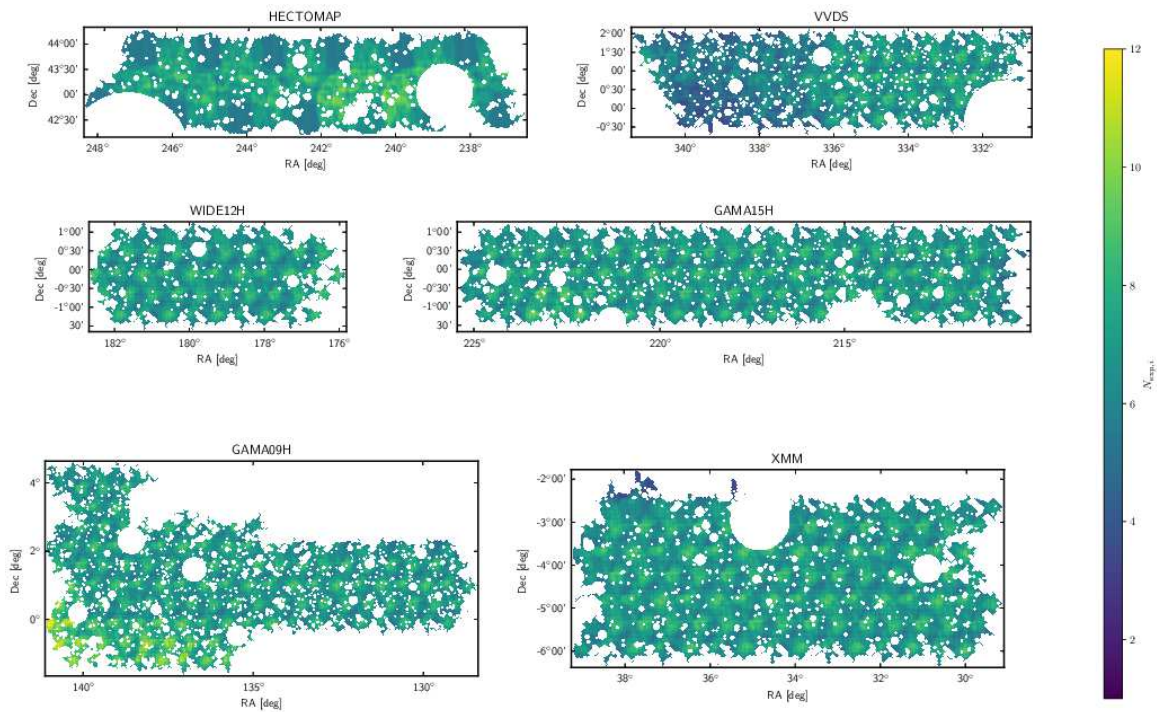}
\end{center}
\caption{Number of exposures contributing to the coadd in the $i$ band across each field.  The way
  exposures are tiled across each survey area results in the repeated pattern of overlap regions
  with more than the typical number of exposures\rmrv{; see \citet{SurveyOverview} for discussion of
    the tiling strategy.}}
\label{fig:nexp_map}
\end{figure*}

Among the 8-10m class telescopes, Subaru
is the one with by far the largest field of view.  Hyper Suprime-Cam
(HSC) takes advantage of the accessible field of view of the Subaru
telescope (1.5$^\circ$ diameter corresponding to 1.77~deg$^2$),
and thus has a survey power about 8
times larger than that of the previous camera, Suprime-Cam.


The focal plane includes a total of 116 Hamamatsu Deep Depletion
CCDs, each $2\,\rm K \times 4K$ pixels.  Four of the CCDs are used for
guiding and eight for automatically monitoring focus, leaving 104
science detectors with a circular-shaped field-of-view of 1.77~deg$^2$.
These chips, which are three-side buttable and have four
independent readout amplifiers, have excellent characteristics: low read
noise, excellent charge transfer efficiency, few cosmetic defects, and
most importantly, high quantum efficiency from 4000\AA\ to 10,000\AA.
The CCD pixels are
15~$\mu$m on a side, corresponding to $0.168^{\prime\prime}$ at the focal
plane.

\mtrv{In this paper we use the S16A internal release data of the HSC
Survey, which was released in 2016 August (see below for more details).}
The HSC weak lensing analysis
is based on the Wide layer data among the three survey
layers (the others are the Deep and UltraDeep layers).
The survey
fields are chosen based on the following considerations \mtrv{\citep[also see][for
  details]{SurveyOverview}:} The HSC survey footprint should overlap the Sloan Digital Sky Survey (SDSS)
   Baryon Oscillation Spectroscopic Survey \citep[BOSS;][]{2013AJ....145...10D} footprint,
    because {the BOSS data provide a
    huge spectroscopic sample of galaxies up to $z\sim 0.7$, which will
    be used to calibrate photometric redshifts via the cross-correlation method and as inputs to the cluster-finding
    algorithm, and for cosmological analyses \rmrv{that combine galaxy clustering and lensing statistics}.
The fields should be well distributed over a wide range of RA,
  such that fields are reachable at all times of the year.
The fields should overlap other
multi-wavelength datasets to maximize
      scientific outputs when combined with the HSC data.  The major
      datasets \rmrv{that} offer unique synergy with HSC data are the
      arc-minute-resolution, high-sensitivity CMB survey
 by the Atacama Cosmology Telescope (ACT; \citealt{Swetzetal:2011}) in
 Chile, and its polarization extension ACTPol \citep{Thorntonetal:2016};
 X-ray data from XMM-XXL \citep{XXL} and eROSITA\footnote{\url{http://www.mpe.mpg.de/eROSITA}};
 near-/mid-infrared imaging surveys (e.g., VIKING/VIDEO\footnote{\url{http://www.astro-wise.org/projects/VIKING/}}
and UKIDSS\footnote{\url{http://www.ukidss.org}});
 and deep spectroscopic surveys such as VVDS \citep{VVDS}, DEEP2\footnote{\url{http://deep.ps.uci.edu}},
zCOSMOS\footnote{\url{http://cosmos.astro.caltech.edu}},
VIPERS\footnote{\url{http://vipers.inaf.it/papers.html}},
GAMA\footnote{\url{http://www.gama-survey.org}}, HectoMap \citep{HectoMAP},
and AEGIS \citep{2007ApJ...660L...1D}.
Finally, the fields should be low in Galactic dust extinction and
as spatially continuous as possible, to enable cosmological analysis on large scales.

We designed the observation strategy of the HSC-Wide layer to better
control systematic errors inherent in the weak lensing (WL)
measurements. First, since the WL shear estimation uses the
$i$-band data, we took the $i$-band imaging data when the seeing
is better than $\sim 0.8^{\prime\prime}$, where
the on-site quick-look software \rmrv{\citep{quicklook:inprep}} was used to monitor the data quality with a lag of only a few
minutes.
Almost
all data used for the first-year science meet this
requirement, as we show below. Second, we employed a large-angle
dithering strategy (about one third of the HSC FoV radius\rmrv{, but without rotational dithering}) so that
objects appear in different positions of the focal plane in each
exposure, thus (at least partially) canceling out
\mtrv{various}
optical
and detector effects over the multiple exposures.  For each field we took a total of
20~minutes exposure time in the $i$-band, split into 6 exposures. Third, we separated the different exposures for
each field by
at least
a half hour
in order to have an independent sampling of the
atmospheric PSF. Finally, we maintained a high elevation for the observations
of each target field  in order to have
\mtrv{high}
atmospheric transparency; most of the data are taken at 60 degrees elevation
(airmass$\sim$1.2) \mtrv{or higher.}

\subsection{Area coverage of shear catalog}\label{subsec:area}

The data we use in this paper were taken during March 2014 through April 2016 with about
90~nights in total.  Note that the \rmrv{publicly-released} HSC DR1 data is based on data taken during March 2014 through
Nov 2015 with a total of 61.5~nights.  However, the same analysis pipeline was run on the
\mtrv{90 night}
dataset
used for the shear catalog described in this work \rmrv{(as described in the HSC DR1, which
  mentions this internal data release)}, ensuring consistency of all
aspects of image processing between the publicly-released subset of the data and the full catalog.
\rmrv{Maps of the $i$-band PSF FWHM and number of $i$-band exposures across this 90-night dataset
  are shown in figures~\ref{fig:seeing_map} and~\ref{fig:nexp_map}, and illustrate that the catalog
  covers six distinct fields that will hereafter be referred to as HECTOMAP, VVDS, WIDE12H,
  GAMA15H, GAMA09H, and XMM.}

For the weak lensing shear catalog, we make a number of well-motivated cuts on this dataset:
\begin{itemize}
\item Weak lensing full depth
\mtrv{and}
full color (WLFDFC) cut: We restrict ourselves to regions that reach
  the approximate full depth of the survey in all 5 broadband
filters ($grizy$), to achieve better uniformity of the shear
calibration and photometric redshift quality across the survey. This cut is non-trivial mainly
because of issues
  like chip gaps which
\mtrv{could}
result in
lattice-like
  features in the area coverage depending on how the cut is applied.
  In detail, this cut is imposed by requiring the number of visits
  within \texttt{HEALPix} pixels 
with {\tt NSIDE=1024} to be $(g,r,i,z,y) \geq
  (4,4,4,6,6)$ and $i_{\rm lim}>25.6$ \rmrv{(using a limiting magnitude definition described below)}.
\mtrv{We allowed}
the $i$-band number of
  exposures
to
be smaller than the ideal value \mtrv{(6)}
so as to avoid
  removing part of \rmrv{the VVDS field} where visits with excellent seeing (resulting in poor PSF
  modeling \rmrv{as described in the HSC DR1 paper}) were
  removed from the coadds.
\mtrv{Given} the very
  good seeing, the limiting magnitude can nevertheless meet our target.

The limiting magnitude is estimated as follows. First we obtain a limiting magnitude for each
patch\footnote{\mtrv{The HSC data is processed separately in equi-area rectangular regions on the
    sky. The regions, called \texttt{tracts}, are pre-defined as an iso-latitude tessellation, where
    each tract covers approximately $1.7\times 1.7$ deg$^2$. A tract is further divided into
    $9\times 9$ sub-areas, each of which is 4200 pixels on a side (approximately 12 arcmin) and is called a
\texttt{patch}.}}
from the database. This limiting magnitude is defined as the magnitude at which the PSF photometry has $S/N\sim5\sigma$
(for details, see the HSC DR1 paper).
However, we cannot immediately use this limiting magnitude because it fails in some patches \hmrv{due to the failure of forced measurements}. 
Instead, we perform a linear fit on the limiting magnitude as a function of seeing, which is again
obtained from the database, and the number of visits \rmrv{(ignoring the dependence on transparency)}, and use this linear fit for the limiting magnitude in all the patches. Note
  that this WLFDFC cut is defined differently from the full depth cut in the HSC DR1 paper\rmrv{,
    with the most important difference being that it is more inclusive in the VVDS field in
    regions where some exposures were removed}.
\item PSF model failures: as \rmrv{detailed} at the start of Section~\ref{sec:psf}, we eliminate regions with
  demonstrable PSF modeling failures in the coadd PSF (defined in Section~\ref{sec:pipeline}) according to a cut given in that section.
\item We remove disconnected regions created by the above two cuts
\mtrv{in the HEALPix pixelization,}
in order to obtain a contiguous survey area.
\item We require that the galaxies not lie within the bright object masks (which will be described
  in Section~\ref{sec:pipeline}).
\end{itemize}

After these cuts, the total area of the catalog is 136.9~deg$^2$.  As shown \rmrv{in figure~\ref{fig:seeing_map},} the best-seeing fields are HECTOMAP and VVDS,
while WIDE12H and GAMA15H are around the median value of seeing, GAMA09H has some areas that are
worse than the median, and XMM has clearly the worst imaging conditions.  Not surprisingly given the
imposition of cuts to achieve approximately full depth in all filters, the regions
all
\mtrv{have a}
fairly
similar
number of
contributing exposures \rmrv{(figure~\ref{fig:nexp_map})}.  The slight deficit in VVDS is a result of
data processing (removing exposures in which PSFs could not be modeled well, see introduction to
Section~\ref{sec:psf}) rather than observations.  Figure~\ref{fig:seeing_hist} shows the distribution of
$i$-band PSF FWHM values for the objects in the shear catalog.
\begin{figure}
\begin{center}
\includegraphics[clip,width=\columnwidth]{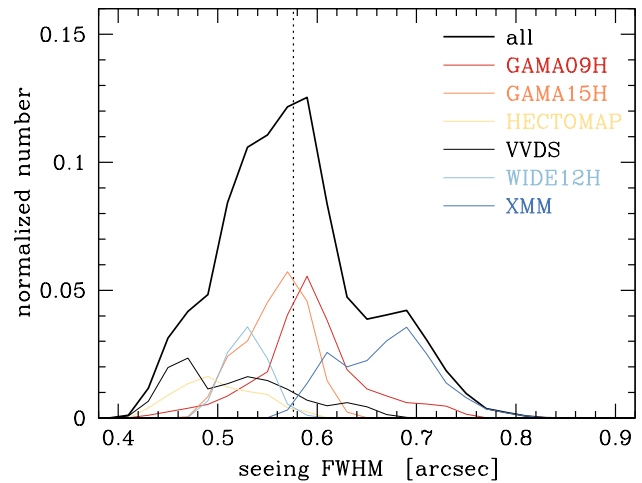}
\end{center}
\caption{Unweighted histogram of the $i$-band PSF FWHM values for galaxies in the shear catalog across each
  field and overall. The vertical dotted line indicates the average
  PSF FWHM value of $\sim 0.58$\arcsec.
}
\label{fig:seeing_hist}
\end{figure}

\subsection{HSC software pipeline}
\label{sec:pipeline}



The processing of single frame HSC images is described in detail in \citet{PipelinePaper:inprep}.
We only mention here the details \rev{of \texttt{hscPipe} that are} important for weak lensing measurements. \rmrv{Also, the software
  is constantly evolving; this paper describes a snapshot of it as of the time these data were
  processed\rev{, \texttt{hscPipe v4.0.2}, though the photometric redshifts rely on subsequent
    processing of the photometry from \texttt{v4.0.3} as described in \citet{2017arXiv170405988T}}.}

We utilize software being developed for
the Large Synoptic Survey Telescope (LSST; \citealt{2010SPIE.7740E..15A,2015arXiv151207914J}).   Basic routines are performed to remove the signature of
the instrument \mtrv{including flat-fielding, bias subtraction, correction of non-uniformity of plate scale, removal of bad pixels, and so on}.
Measurement and detection of objects occurs in two phases.  The first
phase only measures the brightest objects ($S/N \gtrsim 50$) to characterize the PSF
\rmrv{separately for each CCD} and do an initial
astrometric and photometric calibration.  From this initial bright object catalog, we select potential
star candidates for PSF estimation by looking at clustering in \rmrv{size}.  We use a k-means
clustering algorithm which iteratively assigns objects to the cluster with the closest mean.  We have found
that fixing the number of clusters to four and identifying star candidates as the cluster with the smallest
average size has worked reasonably well. We typically select $\sim 80$ star candidates per CCD.

\subsubsection{\rmrv{PSF modeling}}

The selected stars are fed into the PSFEx \citep{2011ASPC..442..435B} package to model the
position-dependent
PSF.  We altered PSFex so that it could be used as an external library in the LSST software, independent
of SExtractor.  Currently, we reserve 20$\%$ of the stars as a cross-validation sample and do not use them in
the modeling.  The \rmrv{PSF model is constructed in the native pixel basis} and we use a second order
polynomial per CCD for interpolation.
\mtrv{Using} a higher order polynomial \rmrv{is not worth the cost of the extra parameters, as it
  was found to produce only}
minor improvements in some CCDs at the \rmrv{focal plane edge}.

\subsubsection{\rmrv{The brighter-fatter effect}}

We also apply a correction to account for the brighter-fatter effect \citep{2014JInst...9C3048A,2015JInst..10C5032G}.
\mtrv{Charge}
that enters the detector
\mtrv{is}
deflected by a lateral electric field due to accumulated charge
in the pixel.  This alters the drift lines
\mtrv{pushing some of the}
charge to land in adjacent pixels, thus
causing bright stars to
\mtrv{appear}
slightly larger
than faint ones.  Ignoring this effect will cause the
PSF used for determining faint galaxy shapes to be incorrect.  We use an algorithm to revert this effect
\citep{BrighterFatter:inprep} where we assume the lateral electric field is curl-free and can
be written as the gradient of a kernel that is translationally invariant and proportional to the
accumulated charge.  We construct this kernel from flat-field images.  After applying this correction,
the dependence of PSF model size on magnitude is reduced below the scientific requirements
(see figure~\ref{fig:psf_residual_size_histmagcorr}, top right panel).

\subsubsection{\rmrv{Coaddition process}}
\label{subsubsec:coaddition}

For our first-year science analysis, all measurements of galaxies in HSC are performed on coadded images,
including shear estimation.  Coaddition is performed without PSF
homogenization \mtrv{(PSF matching of different exposures)};
instead we extend the ``StackFit'' approach of
\citet{2011PASP..123..596J} by computing the effective PSF on the coadd
at the position of each galaxy from the PSF models of the input
images rather than from the coadd itself \mtrv{\citep[see also][]{Annisetal:14}}.
Because convolution is a linear operation and we have applied the
same coordinate transformations and weights to the PSFs and the input images,
the PSF on this coadd (``coadd PSF'') can theoretically be predicted exactly (aside from astrometric
registration errors, which also affect non-coadd measurements), but will be
discontinuous in regions where the set of input images changes.  Because our
measurements assume that the PSF is constant over the scale of individual
objects, we do not attempt to include the effect of small-scale changes in
the input image set due to masked pixels on input images (e.g., cosmic rays or
bad columns) in the PSF.  Instead, we simply mask the regions of the coadd
for which the set of input images is not constant across detected objects, and reject galaxies that
overlap these regions from our sample.  Null tests of the PSF models are performed on the coadd to
ensure that the stacked PSF models adequately describe the coadd PSF (Section~\ref{sec:psf}).

This approach to coaddition also requires a strictly linear combination of
images (i.e., a weighted mean).  Using per-pixel robust estimation or
outlier rejection (such as a median or sigma-clipping) to combine images with
different PSFs would not merely invalidate our PSF estimation approach -- they
would ensure that the coadd does not have any well-defined PSF at all, because different input
images would be systematically clipped in the cores and wings of objects.  To
reject artifacts \mtrv{such as cosmic rays and saturated pixels}
from images, we instead identify the actual artifacts on the
input images, using the following procedure:
\begin{enumerate}
  \item We build a preliminary coadd with no outlier rejection.
  \item We build a second preliminary coadd using strong outlier rejection
  (mean with iterative 3-sigma clipping).
  \item We subtract the two preliminary coadds and threshold to find
\mtrv{pixels at which} they differ.
  \item We
\mtrv{monitor}
the input images for each such region \mtrv{in an automated way}, and keep only
  regions where one or two input images contributed to the difference.
  \item We expand each region on each of the input images that contributed
\mtrv{to}
it to
  include all simply-connected pixels above some threshold.
  \item We build a new coadd with no outlier rejection, with the input image
  regions defined above ignored and these locations masked on the final
  coadd. The \jbrv{coadded PSF models do not take into account pixels that are ignored in this manner and hence are subtly incorrect in these regions.}
\end{enumerate}
These coadd masks are then used to filter the galaxy catalog, ensuring that
it does not include any objects for which the PSF model \jbrv{is incorrect due to rejected pixels.}

\subsubsection{\rmrv{Detection and deblending}}

Coadds for different bands are built independently, and detections (above-threshold regions and
peaks within them) are identified separately in each
band.  We then merge the detections across all bands, merging co-located
peaks to eliminate redundant detections and computing new above-threshold
regions as the union of the per-band region.  Each simply connected region is
considered a family of blended objects.  We then independently deblend and
measure in each band.  This deblending takes the number of objects in each
family and their approximate positions as fixed from the previous step,
ensuring that these per-band measurements are broadly consistent, but we
currently do not otherwise ensure that per-pixel assignments of flux to
deblended child objects are consistent across bands.  \rmrv{This approach to deblending is likely to
change in future versions of the software pipeline.}

\subsubsection{\rmrv{Measurement algorithms}}

The measurement
algorithms run at this stage include centroiding, shape measurement, and PSF,
galaxy model, Kron, and aperture fluxes.  After measurement, we select a
``reference'' band for each object, using the $i $ band for all objects with
$S/N > 10$ and falling back to other bands as necessary in the order $r$,
$z$, $y$, $g$.  Measurements in the reference band are then used to drive
another round of per-band ``forced'' measurements, with centroids and shapes
held fixed at the values from the reference band.  Fluxes measured in this
forced mode are used to compute colors and photometric redshifts. 

\subsubsection{\rmrv{Bright object masks}}

Bright stars affect galaxy detection and photometry most notably through the luminous PSF pattern,
ghosting, and by altering the local sky background. We mask all stars that saturate the
typical exposures, \rmrv{accounting for the fact that} the saturation limit depends on the filter and on the observing
conditions (such as transparency and seeing).
\mtrv{To do so,}
 we use a sample of bright stars that is complete in all optical bands down to ${\rm
   mag}<17.5$. This conservative limit, estimated from PSF magnitudes in each HSC band, ensures that
 all saturated stars in our survey are properly masked (at the expense of a small fraction of
 non-saturated stars that are also masked). { We note that these masks are only used to remove
   sources from the shape catalog that are located near a bright object, however, they were not used
   at any other step during data processing.}




We first use the Tycho-2 star catalog \citep{2000A&A...355L..27H} to identify all
bright stars in our fields, and supplement this with the NOMAD catalog \citep{2004AAS...205.4815Z}
at fainter magnitudes ($10.0<{\rm minimum}(B,V,R)<17.5$), which is a compilation of a number
of
all-sky
star catalogs. The main caveat of the NOMAD catalog is \mtrv{that it is contaminated}
by a small fraction of bright
galaxies (visual inspection suggests that about 10\% of the objects are galaxies).
\mtrv{Given this contamination by galaxies,}
the
masks will be called \emph{bright object masks} from now on. We note that masking galaxies brighter
than $17.5$ has no impact on the galaxy shape catalog, composed of much fainter sources, however, we
warn that any other science
\mtrv{analysis}
making use of galaxies brighter than $17.5$ in any band should not use
the bright object masks. \rmrv{This caveat about inclusion of bright galaxies from the NOMAD
  catalog is valid for the bright object masks used
  here, which were an early version called `Sirius'.  The version described in \cite{2017arXiv170500622C}
  and called `Arcturus' are updated to remove the bright galaxies, and so are called `bright star' rather than `bright
  object' masks.  The updated version, which is suitable for use in a broader range of science cases, will be included}
in subsequent \rmrv{data} releases.

To build individual masks, we characterize, per bin of bright-object magnitude, the radial extent to which bright objects affect neighbouring source counts by measuring the two-point
cross-correlation function between bright stars and all HSC detected sources (without any cut on the
pipeline flags). At small radius, we observe a rapid decrease
\mtrv{in the density of detected sources.}
At large radius, however, we observe an enhancement of
\mtrv{detected sources;}
the luminous halo of the bright object
boosts the local background and leads
to an increased number of
\mtrv{noise}
detections.
This feature is more prominent for faint detections but occurs around the same radius,
\mtrv{for a given}
bright object magnitude, regardless of the detected source magnitude. We conservatively use the position of this feature to set the extent of the mask.
The size of the mask is kept fixed across the five HSC filters and the reference
bright-object magnitude ($m_{\rm BS}$) is chosen to be the
brightest of the optical magnitudes provided by each respective catalog, $B$, $V$, and $R$ from the
NOMAD catalog and the SDSS-emulated $g$, $r$ and $i$ \citep{2010PASP..122.1437P} from the Tycho-2
catalog. For each bright object, we build a circular mask whose radius depends on the bright-object magnitude ($m_{\rm BS}$) according to:
\begin{equation}
\label{eq:mask_star}
  r_{\rm{mask}} \, \mathrm{[arcsec]} =200\times10^{0.25(7.0-m_{\rm BS})}+12\times10^{0.05(16.0-m_{\rm BS})} \, ,
\end{equation}
%
where the parameters reproduce the measured radius at which the source detection near the bright
object starts
\mtrv{to feature}
an enhancement.
The total masked area due to bright objects in the WLFDFC region is 16\%.
We note that equation~(\ref{eq:mask_star}) 
diverges at bright magnitudes, and as a result
\mtrv{significantly}
overestimates the size of the required mask for a few very bright stars (mag$<4-5 $). \rmrv{This
problem is also updated in the mask version described in \cite{2017arXiv170500622C}}.

\subsection{Shear estimation algorithm}\label{subsec:shearest}

Since the initial development of algorithms for estimating shear
\mtrv{that correct}
for the effects of
the PSF on galaxy images using measurements of moments of the PSF and galaxy shape
\citep[e.g.,][]{1995ApJ...449..460K}, the field has seen tremendous development in
\mtrv{the} variety and accuracy of methods. This evolution
\mtrv{has been}
demonstrated through a series of community challenges
\citep{2006MNRAS.368.1323H,2007MNRAS.376...13M,2009AnApS...3....6B,2010MNRAS.405.2044B,2010arXiv1009.0779K,2012MNRAS.423.3163K,2014ApJS..212....5M,2015MNRAS.450.2963M}.
In the most recent of these, GREAT3, the types of methods used included those based on estimation of
per-galaxy shapes via measurements of moments, fitting parametric light profiles, decomposition into
basis functions, and machine learning, as well as methods that involve inferring ensemble shears without
per-galaxy shapes.  The majority of methods that are actively used for weak lensing science achieved
multiplicative bias in the $\lesssim 2$\% range\rmrv{, meaning that for a true ensemble shear $g$ the
  estimated ensemble shear $\hat{g} = (1+m)g$ for a multiplicative bias $|m|\lesssim 0.02$. As we will show
  in Section~\ref{sec:requirements}, these methods would} meet the requirements for first-year weak
lensing science with HSC, but not necessarily full HSC survey science requirements.

There are a number of important issues for shear estimation
\mtrv{methods that are}
based on averages of per-galaxy shape
estimates.  One of these is ``noise bias''
\citep[e.g.,][]{2002AJ....123..583B,2004MNRAS.353..529H,2012MNRAS.427.2711K,2012MNRAS.424.2757M,2012MNRAS.425.1951R},
wherein the pixel noise modifies the shape of the likelihood surface and causes the
maximum-likelihood estimator of per-galaxy shapes to be biased.  Another is ``model bias'': a number
of studies have convincingly demonstrated that when estimating shears
\mtrv{using a method}
that assumes a
particular galaxy model, the shears can be biased if the galaxy light profiles are not correctly
described by that model \citep[e.g.,][]{2010MNRAS.404..458V, 2010A&A...510A..75M}.  More generally, any
method based on the use of second moments to estimate shears cannot be completely independent of the
details of the galaxy light profiles, such as the overall galaxy morphology and presence of detailed
substructure \citep{2007MNRAS.380..229M,2010MNRAS.406.2793B,2011MNRAS.414.1047Z}.  Selection biases,
for example due to the selection criteria correlating with the lensing shear or PSF anisotropy, also
play an important role \citep[e.g.,][]{2004MNRAS.353..529H,2016MNRAS.460.2245J}.  In general, these
basic mathematical issues in estimating shear from per-galaxy shapes require either external
calibration using simulations, or some form of self-calibration
\citep[e.g.,][]{2017MNRAS.467.1627F,2017arXiv170202600H,2017ApJ...841...24S}.  An alternative is to
use a method that is designed to infer 
an unbiased estimate of the ensemble shear but not necessarily (or at all) an unbiased estimate of
per-galaxy shapes \citep{2014MNRAS.438.1880B,2015ApJ...807...87S,2016MNRAS.459.4467B}.

The first year shear catalog for HSC weak lensing was produced using a moments-based shape measurement method for which the shear is estimated by
averaging the shapes, described below; for this method, we use suites
of simulations (Section~\ref{subsec:great3like}) to remove the aforementioned forms of biases to which
the method is naturally prone.  A forthcoming paper \citep{Armstrong:inprep} will describe a
catalog produced using a Bayesian ensemble shear inference method \citep{2016MNRAS.459.4467B,2014MNRAS.438.1880B}.


Galaxy shapes are estimated on the coadded $i$-band images using the re-Gaussianization PSF
correction method
\citep{2003MNRAS.343..459H}.
This method was extensively used and
characterized for science in the Sloan Digital Sky Survey (SDSS;
\citealt{2005MNRAS.361.1287M,2012MNRAS.425.2610R,2013MNRAS.432.1544M}), and then incorporated into
the {\sc GalSim}\footnote{\texttt{https://github.com/GalSim-developers/GalSim}}
\citep{2015A&C....10..121R} open-source software package after further optimization for speed.  The
HSC pipeline relies on the {\sc GalSim} implementation of this algorithm.

The basic principle of galaxy shape estimation using this method is to fit a Gaussian profile with
elliptical isophotes to the image, and to define the components of the \rmrv{distortion}
\beq
(e_\mathrm{1},e_\mathrm{2}) = \frac{1-(b/a)^2}{1+(b/a)^2}(\cos 2\phi, \sin 2\phi),
\label{eq:e}
\eeq
where $b/a$ is the axis ratio and $\phi$ is the position angle of the major axis \rmrv{with respect
  to the equatorial coordinate system (sky coordinates)}.  In the course of
the re-Gaussianization PSF-correction method, corrections are applied to account for dilution of the
observed shape by the PSF, including the non-Gaussianity of both the PSF and the galaxy surface
brightness profiles \citep{2003MNRAS.343..459H}.  The ensemble average \rmrv{distortion} is then an
estimator for the shear $g$,
\beq
(\hat{g}_\mathrm{1},\hat{g}_\mathrm{2}) = \frac{1}{2\cal R}
\langle(e_\mathrm{1},e_\mathrm{2})\rangle,
\eeq
where ${\cal R}$
is called the `shear responsivity' and
represents the response of the \rmrv{distortion} (equation~\ref{eq:e}) to a small
shear \citep{1995ApJ...449..460K, 2002AJ....123..583B}; ${\cal R} \approx
1-e_\mathrm{rms}^2$, where $e_{\mathrm{rms}}$ is the RMS intrinsic
\rmrv{distortion} per component.  The equations used for the actual shear estimation process are given in
Appendix~\ref{app:shear}.  \rmrv{The notation used in this paper is that shear components called
  $(1,2)$ are in sky coordinates, while those denoted $(+,\times)$ are in a coordinate system
  defined for pairs of objects.  In that case $\hat{g}_+$ (tangential shear) is defined with respect to the vector connecting the pair and $\hat{g}_\times$
with respect to the axes rotated by $45^\circ$.}

It is useful to be able to apply selection criteria based on
\rmrv{how well-resolved the galaxy is compared to the PSF}.  For this purpose, we use the resolution factor $R_2$
which is defined using the trace of the moment matrix of the PSF
$T_\mathrm{P}$ and of the observed (PSF-convolved) galaxy image
$T_\mathrm{I}$ as
\beq\label{eq:def-r2}
R_2 = 1 - \frac{T_\mathrm{P}}{T_\mathrm{I}}.
\eeq
Well-resolved objects have $R_2\sim 1$ and poorly resolved
objects have $R_2\sim 0$.

The estimation of the
\mtrv{shear}
responsivity in light of shape measurement errors that make it difficult to
infer the intrinsic \rmrv{shape} dispersion, and the exact selection criteria to place on
the
resolution
factor
\mtrv{and}
other galaxy properties, are discussed in Section~\ref{sec:shear}.

\subsection{Systematics analysis software}


For our systematics tests we use the software Stile, Systematics Tests In LEnsing.  It is
open-source and publicly available\footnote{\texttt{https://github.com/msimet/Stile}}\rmrv{, and was
  developed specifically to enable easy calculation of a wide range of null tests for a large-area
  lensing survey such as HSC}.

The Stile code consists of a set of Python classes that perform systematics tests, plus
ancillary code for data handling.  The code used for this paper performs five main types
of tests. 
\begin{itemize}
\item The CorrelationFunctionSysTest. This acts as a wrapper for the software
TreeCorr\footnote{\texttt{https://github.com/rmjarvis/TreeCorr/}}~\citep{2004MNRAS.352..338J},
performing different kinds of correlation functions with appropriate parameter values.
\item The WhiskerPlotSysTest. This produces whisker plots of star and PSF ellipticities and their
residuals.
\item The ScatterPlotSysTest. This takes two columns of data and produces a scatter plot and
optionally a fitted trendline.
\item The HistogramSysTest, which produces histograms.
\item The StatSysTest.  This produces basic statistical quantities such as mean, median, and
percentile rankings.
\end{itemize}
The CorrelationFunction, WhiskerPlot, and ScatterPlotSysTests all have predefined systematics test
types (such as the $\rho$ statistics, equations~\eqref{eq:rho1}--\eqref{eq:rho5}, for the CorrelationFunctionSysTest and whisker
plots of the PSF-star residuals for the WhiskerPlotSysTest).  Additional tests can be added by making
subclasses of the existing tests, without much unnecessary overhead.  The existing tests handle details
such as forming residuals or other combinations of parameters if necessary, finding automatic bin
widths for the histograms, and fitting trendlines to scatter plots.  They also have plotting routines
if applicable, though we used specially-created plotting routines for the plots in this paper, rather
than Stile defaults (which do not currently allow us to plot multiple fields at once).

TreeCorr \rmrv{\citep{2015ascl.soft08007J}} is a fast correlation function code that uses a ball tree method (similar to a k-d
tree).  In this work we use its methods for point-shear correlation functions $\langle g\rangle$
(TreeCorr type `ng'), shear-shear correlation functions $\langle g\,g\rangle$ (TreeCorr
type `gg'), and scalar-scalar correlation functions $\langle\kappa\kappa\rangle$ (TreeCorr type
`kk').

The Stile tests operate on Python data arrays that can be accessed by column name.  A unified set of column names is expected, but the data can otherwise be in
any format.  
The Stile code also has the ability to interface with the LSST/HSC pipeline; we used this
functionality during early data processing but not in the final versions of the tests presented
here, which used versions of catalogs that have all weak lensing selection criteria applied.

More detailed information (including calculation details and usage instructions) can be found in the
package documentation on GitHub\footnote{\texttt{http://stile.readthedocs.io/en/latest/}}.

\section{Requirements}\label{sec:requirements}

Before presenting the results of systematics tests, we need to define the requirements for the
catalog to enable first-year HSC weak lensing science.  \rev{For the purpose of setting
  requirements, we consider two cosmological measurements as the goal of first-year HSC weak lensing
  science: the shear-shear correlation function
  using the entire shear catalog (no tomography), and the galaxy-shear correlation function (i.e.,
  galaxy-galaxy lensing) using the SDSS-III/BOSS \citep{2013AJ....145...10D} CMASS sample
  \citep{2016MNRAS.455.1553R}.  The latter will be combined with galaxy clustering measurements 
 to enable cosmological parameter constraints, as an alternate
  cosmological method to cosmic shear.} For this paper, the calculations for setting requirements
involve \mtrv{(a) estimating statistical errors in the fiducial weak lensing analyses for the
  first-year HSC data;} (b) assessing how systematic errors propagate into cosmological observables;
(c) comparing those systematic errors to the statistical
\mtrv{errors}; and
(d) requiring each
systematic error to contribute below some fraction of the statistical errors.  In principle, one can
set requirements by doing the full forecasting of how those systematics in the cosmological
observables propagate into systematics on cosmological parameter constraints
\citep[e.g.,][]{2016MNRAS.460.2245J}.  We did not adopt that approach in this paper, \rev{though we
  do present estimates of how our requirements, as currently calculated, propagate into cosmological
  parameter constraints}.  \rev{The overall process described here is broadly consistent with that
  used for several ongoing and future surveys  \citep{2008MNRAS.391..228A,2013MNRAS.429..661M,2016MNRAS.460.2245J,2017MNRAS.465.1454H}, differing only in small details of the implementation.}

\mtrv{In the following we will often assess the requirements in terms of
``multiplicative'' and/or ``additive'' biases in shear estimation, which are defined as
$\hat{g}=(1+m)g^{\rm true}+c$, respectively \citep{2015MNRAS.450.2963M}.}
Some
systematic errors, such as those that induce additive biases in shears, only contribute to
\rev{cosmic shear}.  Others, such as multiplicative biases in the shear, affect both 
\rev{cosmic shear and galaxy-galaxy lensing}.  In that case, we assess the requirements on the systematic separately for both
measurements, and adopt the more stringent requirement.  We will also consider additive and
multiplicative biases to be completely uncorrelated and hence constrain the two separately.  In
practice, selection biases or other effects could
\mtrv{couple}
these, but we expect this to be a
higher-order effect that is not of concern in first-year HSC analysis.

\rev{In this context, the word ``requirements'' has a very specific meaning: the requirements are defined
  such that if a systematic requirement is met, then the systematic is small enough that it can be
  ignored as part of the error budget in a science analysis.  If one of our systematic requirements
  is not met, then it does not mean science cannot be done; rather, it means that the systematic is
  not small enough compared to our statistical errors that we can safely ignore it, but rather needs
  to be explicitly accounted for (e.g., modeled and marginalized over) in a way that shows that it
  is a non-negligible portion of the error budget.  It is the goal of this paper to check for the
  various types of systematics that can affect weak lensing analyses, and classify them as ones
  meeting requirements (subdominant part of the error budget) versus those that are similar in
  magnitude to or exceeding our requirements, requiring further work to explicitly account for them
  in a science analysis.  For those systematics that can be detected using null tests, we carry out
  those tests; for others, we describe how the systematic is tested in a separate paper (if
  requiring external simulations or data).  Future science papers can then start with a focus on the
  small subset of problems that are identified as important for that science case in this work.}

In general, we require that systematic uncertainties be limited to $\lesssim 0.5\sigma$ of
the
statistical errors on the measurements.  Note that these are requirements on the systematic {\em
  uncertainties}, not {\em biases}.  Known biases should be removed from the measurements before
using them for science.  The choice of a $\lesssim 0.5\sigma$ threshold for defining requirements is
somewhat arbitrary.
  Ultimately for any science analysis, both statistical and systematic
errors must be reported and used to assess the relative importance of systematics.  If the science
case differs from the examples used in this section, a potentially lower $S/N$ could justify a
higher tolerance for systematic uncertainty.
However, we use this $0.5\sigma$ threshold as a
way of asserting that in general, for our highest-$S/N$ cosmological science cases, our goal is for
statistical error to dominate over systematic error. \rev{If statistical and systematic errors are
  independent and add in quadrature, then this $0.5\sigma$ threshold would mean that the systematic
  errors inflate the overall (total statistical $+$ systematic) error budget by 12 per cent.}

\rev{The connection between systematics requirements and statistical errors also means that the requirements become more stringent
  once more of the survey is complete.  A rough estimate of the final requirements (without
  accounting for details of survey edge effects) is that they will be a factor of $\sim 2.7$ times
  tighter than the ones presented in this work.  We do not compare our current estimates of
  systematic uncertainties with the tighter, full-survey requirements because the HSC pipeline
  as described in \citet{PipelinePaper:inprep} will be evolving in certain ways that are highly
  relevant to weak lensing, e.g., adoption of improved PSF modeling and shear estimation routines,
  so our current results are irrelevant to the full survey dataset.  However, this factor of
  $\sim 2.7$ can be helpful to bear in mind when identifying important directions for future work.}

Note that there are many sources of multiplicative bias in shear estimates.  When discussing
multiplicative bias in the subsections below, we will derive a constraint on the uncertainty in the
shear calibration (after correcting for known calibration biases) across all of these sources of
error: star/galaxy separation failure (i.e., stellar contamination \mtrv{of the galaxy sample}); PSF model size errors;
shear-related biases like model bias, noise bias, and selection bias; and photometric redshift biases.
In general, our constraints are
\mtrv{on}
the uncertainty in the component-averaged shear
calibration.  While many methods of shear estimation exhibit a slight difference in calibration bias
for the component that is along vs.\ diagonal compared to the pixel direction, any practical shear
measurement involves an average over those two components, so we place requirements on
\mtrv{the}
systematic uncertainty of the average.

\subsection{Covariances}

An important part of setting requirements is understanding the
statistical
\mtrv{errors}
in the relevant measurements.  Overly
optimistic statistical uncertainties due to neglecting  important error
contributions will result in overly stringent requirements on
systematics, which may be difficult to \rmrv{meet} in practice.  To account
for all important sources of error in shear-shear correlations
(including galaxy shape noise, cosmic variance, and super-sample
covariance) and in galaxy-shear correlations (including the above, plus
lens shot noise error terms) \citep{TakadaHu:13},
we use mock galaxy catalogs that include
various effects: properties of source galaxies, lensing effects on each
source galaxy due to the foreground large-scale structure (cosmic
shear), weights, and the survey geometry.  \rmrv{These are catalog simulations rather than image
  simulations, but the impact of shape noise and measurement error due to noise
in the images is included properly as described below.} Here we briefly summarize the
mock catalogs of HSC data, which we will use to derive
requirements on shape measurements for the science analysis \rmrv{with the shape catalog described
  in this paper}.  More detailed descriptions of mock catalogs will be presented in
\citet{Shirasaki:inprep}.

We first describe the mock shear catalogs of HSC source galaxies. We follow the
method developed in \citet{ShirasakiYoshida:14} \citep[also
see][]{2017MNRAS.470.3476S} to perform ray-tracing on a large number of \rmrv{cosmological} light-cone
simulations. We generate $48$ full-sky light-cones using outputs of different
box-size $N$-body simulations for the nine-year WMAP cosmology \citep{WMAP9}.
For details of the cosmological parameters and $N$-body simulations, see also
\citet{2017MNRAS.470.3476S}. We used the multiple-lens plane algorithm on the
sphere to simulate the light-ray path and lensing of a source galaxy by the
structure along its line-of-sight.
The ray-tracing simulations are designed to simulate the weak lensing
effect on source galaxies at different redshifts, over a full sky given
in the \texttt{HEALPix} format
of 0.43 arcmin pixel scale.  Although
the original $N$-body simulations have a higher resolution, we use this
pixel scale because we use this mock catalog to estimate the covariance
matrix of lensing observables.  The covariance at scales smaller than
this pixel scale is dominated by shape noise or shot noise, \rmrv{which we account for by
  preserving} pairs of two source
galaxies or lens-source galaxies at scales below the \rmrv{\texttt{HEALPix}} pixel scale \mtrv{in the galaxy-galaxy
lensing calculation.}

The simulation consists of 38 different source planes each separated by a
comoving separation of $\Delta \chi = 150\, h^{-1}{\rm Mpc}$, thus covering
source planes up to redshift $z_s\simeq 5.3$.  The angular resolution and the
redshift coverage are suitable for our purpose of creating mock catalogs for the
first-year data of HSC.  We incorporate our simulations with observed
photometric redshifts and angular positions of $real$ galaxies.  In brief, (1)
we insert each galaxy, taken from the real HSC catalog, into the nearest angular
pixel in the nearest redshift source plane, (2) we randomly rotate the
orientation of its \rmrv{shape} to erase the real lensing effect, and (3) we
simulate the lensing distortion effect on the source galaxy by adding the
lensing shear and the intrinsic shape, and (4) repeat the procedures (1)--(3)
for all the source galaxies.
Note that, for every realization, we randomly sample the source redshift from
the posterior \mtrv{probability} distribution of photometric redshift for each galaxy. Thus,
our mock catalogs include effects of properties of source galaxies (e.g.,
magnitudes, ellipticities and spatial variations in the number densities),
statistical uncertainties in photometric redshifts as well as the survey geometry.
Since the six HSC S16A regions are separated from each other \mtrv{as shown in
figures~\ref{fig:seeing_map} and~\ref{fig:nexp_map}}, we generate 21
different mock catalogs from 21 different rotations of the spherical coordinates
preserving the relative positional locations of the HSC S16A fields on the sky.
This allows us to generate 21 \mtrv{independent} mock catalogs from each full sky simulation
without
overlap. This results in a total of 1008 realizations of HSC mock
shear catalogs, generated from the 48 full-sky lensing simulations.

In what follows, we will also derive requirements on the level of residual
systematic errors for galaxy-galaxy weak lensing. For this purpose, we create
mock catalogs of CMASS galaxies, which are our primary targets for the
galaxy-galaxy lensing measurements, based on the halo occupation distribution
(HOD) approach.  To do this, we use the catalog of halos that are built from the
same $N$-body simulation outputs used for the ray-tracing simulations.  To
identify halos in each $N$-body simulation output, we used the software {\tt
Rockstar} \citep{Behroozi:2013} that identifies halos from the clustering of
$N$-body particles in phase space. Our $N$-body simulations allow us to resolve
dark matter halos with masses greater than a few times $10^{12}h^{-1}\, M_\odot$
with more than 50 $N$-body particles at redshifts $z\simlt 0.7$, which cover the
range of redshifts of CMASS galaxies.
We assign three-dimensional positions for all halos (angular position and
redshift) in the light cone depending upon their positions in the $N$-body
simulation output.

We populate galaxies in these dark matter halos with
an HOD that is
constrained based on the number density and spatial
clustering of CMASS galaxies in the redshift range $z\in[0.43, 0.55]$ and
$z\in[0.55, 0.7]$ but spanning the entire SDSS BOSS footprint. We constrain a
simple 5 parameter HOD \citep[see e.g,][]{White:2011} based on the analytical
modelling framework developed in \citet{vdB:2013}\footnote{We use the HOD
modelling code {\sc AUM} \citep{More:2013, 2015ApJ...806....2M} which is
publicly available at \texttt{http://www.github.com/surhudm/aum} .}. This HOD is used to
populate mock CMASS galaxies in the halos of the light cone. This galaxy catalog
in conjunction with the source catalog output from the same light-cone are then
used to perform 1008 mock measurements of galaxy-galaxy weak lensing.
These measurements are then used to compute a covariance matrix.

\subsection{Requirements for galaxy-galaxy lensing}

For galaxy-galaxy lensing (hereafter \rmrv{abbreviated} ``g-g''),
 it is possible to remove additive systematics in the shear via
cross-correlation with a random catalog that has the same area coverage as the lens sample
\citep[e.g.,][]{2013MNRAS.432.1544M}.  As a result, we are primarily concerned with placing
requirements on systematics that cause a systematic uncertainty in the overall amplitude of the
shear.  We will place two requirements in this section:
\begin{itemize}
\item an overall requirement on uncertainty in the shear calibration
(across all sources of multiplicative bias, as described in the introduction to this section) to achieve our
goal of having it contribute at $<0.5\sigma$ in all bins in CMASS galaxy-galaxy lensing with the
first-year shear catalog, and
\item a more specific and more stringent requirement that systematic uncertainty due to PSF modeling
  errors \rmrv{alone} contribute at
  $<0.25\sigma$.  Here the relevant PSF modeling errors are PSF size errors, which propagate into a
  multiplicative shear bias.
\end{itemize}

\subsubsection{Multiplicative bias in shear}\label{subsubsec:gglbiasreq}

To place a requirement on systematic uncertainty due to any source of \rmrv{uncorrected} multiplicative bias in the
shear for CMASS galaxy-galaxy lensing, we consider the expected fractional uncertainty in the
lensing signal $\Delta\Sigma$.  In particular, given a data vector $\vec{x}$ consisting of the
predictions for $\Delta\Sigma$ in bins of projected separation $r_p$ \mtrv{from the CMASS galaxies},
and the expected covariance matrix from the mock catalogs $\mathbf{C}$, we
can define the total SNR for CMASS galaxy-galaxy lensing as
\begin{equation}
    \text{SNR}_\text{g-g} = [\vec{x}^T \cdot \mathbf{C}^{-1} \cdot \vec{x}]^{1/2}.
\end{equation}
We obtain $\text{SNR}_\text{g-g}=29.9$ on scales from 0.5--30$h^{-1}$Mpc for
the sample of CMASS galaxies from $z\in[0.43, 0.7]$, including CMASS galaxies in BOSS regions
covering a wider area surrounding each HSC field. We require the overall
uncertainty in the shear calibration allowed for CMASS g-g lensing, $|\delta
m_\text{all-g-g}|$, to be below $0.5/\text{SNR}_\text{g-g}$, thus,
\begin{equation}\label{eq:mall-ggl}
    |\delta m_\text{all-g-g}| < \frac{0.5}{\text{SNR}_\text{g-g}} = 0.017\,.
\end{equation}



\subsubsection{PSF model size errors}\label{subsubsec:gglpsfmodelsizereq}

Here we want to place a more specific requirement on systematic uncertainty due to coherent PSF model size
errors, which can be considered a source of shear calibration uncertainty $|\delta m_\text{PSF-g-g}|$ that we
would like to be less than half the overall shear calibration uncertainty budget, $|\delta
m_\text{PSF-g-g}| < 0.5 |\delta
m_\text{all-g-g}|$ or $0.25/\text{SNR}_\text{g-g}$.  In other words, we need to leave some of the overall calibration budget for other
sources of systematics \rmrv{that cause multiplicative bias in shear besides the PSF model size errors}.  We then need to relate $|\delta m_\text{PSF-g-g}| \lesssim 0.5 |\delta m_\text{all-g-g}|$ to
some statistic that we can measure from the catalog, in order to place a requirement on that
statistic.

In this case, we can use the formalism from \cite{2004MNRAS.353..529H},
\mtrv{who}
show that the shear
calibration uncertainty for a given PSF model size error for a galaxy of resolution factor $R_2$ is
\begin{equation}
\delta m_\text{PSF-g-g}(R_2) = -\left(R_2^{-1} - 1\right) \frac{T_*-T_\text{PSF}}{T_\text{PSF}}.
\end{equation}
Here, $T_*$ and $T_\text{PSF}$ are the trace of the second moment matrix of stars and PSFs.
Well-resolved objects ($R_2$ near 1) are minimally affected by PSF model errors, while this
systematic becomes very large as $R_2$ approaches zero.  To relate this to quantities that we
measure, we first consider that the fractional error in the trace (which is a measure of area) is
double that of the linear size $\sigma$ which we use to quantify PSF model size errors.  We
furthermore place a constraint on the {\em average} PSF model size error, because stochastic errors
in PSF model sizes will average out and will not cause a bias in galaxy-galaxy lensing.  Hence
\begin{equation}
\langle\delta m_\text{PSF-g-g}(R_2)\rangle = -2\left\langle\left(R_2^{-1} - 1\right) \frac{\sigma_*-\sigma_\text{PSF}}{\sigma_\text{PSF}}\right\rangle,
\end{equation}
and asserting the independence of galaxy properties like $R_2$ and PSF model properties, we can
average over those properties separately:
\begin{equation}
\langle\delta m_\text{PSF-g-g}(R_2)\rangle = -2\left\langle R_2^{-1} - 1\right\rangle\left\langle \frac{\sigma_*-\sigma_\text{PSF}}{\sigma_\text{PSF}}\right\rangle.
\end{equation}

As will be described later, our catalog has a lower limit of $R_2=0.3$.  The weighted mean of
$R_2^{-1}-1$ over the whole catalog is $0.8$.  However, we would also like our requirements to be met by
sub-populations of the catalog selected by e.g., photometric redshift or magnitude, so we will
conservatively use a value of $\left\langle R_2^{-1} - 1\right\rangle = 1.0$ when placing the requirement.  So we will require
\begin{equation}\label{eq:mpsf-ggl}
\left\langle \frac{\sigma_*-\sigma_\text{PSF}}{\sigma_\text{PSF}}\right\rangle < \frac{|\delta
  m_\text{PSF-g-g}|}{2} \,\,\, < \,\,\,\frac{|\delta m_\text{all-g-g}|}{4} < 0.004.
\end{equation}


\subsection{Requirements for cosmic shear}

For cosmic shear, we care about both additive and multiplicative shear biases, including their
spatial correlation function.  We will place several requirements in this section:
\begin{itemize}
\item an overall requirement on shear calibration uncertainty (from all sources of multiplicative
  bias, as described in the introduction to this section) to achieve our goal of having it
  contribute at $<0.5\sigma$ in cosmic shear with the first-year shear catalog
  \rmrv{(Section~\ref{subsubsec:cosshearreq1}),}
\item a more specific and more stringent requirement that calibration uncertainties due to PSF size
  modeling errors contribute at $<0.25\sigma$
  \rmrv{(Section~\ref{subsubsec:psfmodelsize-ss}),}
\item a requirement that the systematic uncertainty due to overall additive biases from all sources
  (PSF \rmrv{shape} errors, selection biases, insufficient correction for PSF anisotropy) are
  sufficiently small to ensure that the resulting additive bias term contributes at $<1\sigma$ on
  all scales   \rmrv{(Section~\ref{subsubsec:additive-ss}),} and
\item a requirement that the uncertainty on the PSF \rmrv{shape} errors and their spatial correlation
  are sufficiently small to ensure that the resulting additive bias term contributes at $<0.5\sigma$
  on all scales (it can be removed, but we do not want to be making very large corrections to the
  signal) \rmrv{in Section~\ref{subsubsec:psfmodele-ss}}.
\end{itemize}

\subsubsection{Multiplicative bias in shear}\label{subsubsec:cosshearreq1}

Following
a similar methodology as in Section~\ref{subsubsec:gglbiasreq}, we place requirements on the
shear calibration uncertainty (from all sources of bias) for cosmic shear measurements.  We use the
same formalism to define an effective SNR on the shear-shear (``s-s'') correlations without tomography
($\text{SNR}_\text{s-s}$).  \rmrv{In this case we assume the observable quantity will be the shear
correlation functions $\xi_\pm(\theta)$.  Assuming we have per-object shear estimates $\hat{g}$
which for pairs of galaxies can be decomposed into components $\hat{g}_+$ and $\hat{g}_\times$, then $\xi_\pm$ can be
defined as}
\begin{equation}\label{eq:xipm}
\rmrv{
\xi_\pm(\theta) = \langle \hat{g}_+\hat{g}_+\rangle \pm \langle
\hat{g}_\times\hat{g}_\times\rangle.
}
\end{equation}
\rmrv{Hence our} data vector
$\vec{x}$ consists of the $\xi_+$ and $\xi_-$ values in $\theta$ bins, and $\mathbf{C}$ is its
covariance matrix. To avoid scales which could be affected by either baryonic
effects or theoretical uncertainties in the matter power spectrum, we use
$\xi_+$ and $\xi_-$ measurements on scales $\theta>4$ and $>40$ arcminutes,
respectively \citep{DES_CS_cosmology}.
For the maximum angular scales, we set $\theta_{\rm max}=285$~arcminutes.
\smrv{We have checked that the SNR is not very sensitive to values of
$\theta_{\rm max}$ above $50$~arcminutes, due to the limited sizes of our fields.}

With a multiplicative bias model that looks like
\begin{equation}
\hat{g} = (1+m_\text{all})g,
\end{equation}
where $g$ and $\hat{g}$ are the true and estimated shear, then in the limit that $|m_\text{all}|\ll 1$, this bias
primarily affects the shear-shear correlations as
\begin{equation}
\langle\hat{g}\hat{g}\rangle \approx (1+2m_\text{all}) \langle gg\rangle.
\end{equation}
We therefore require the overall value of $2 |\delta m_\text{all}|$ (systematic uncertainty in shear
calibration)  to be $\lesssim
0.5/\text{SNR}_\text{s-s}$, or
\begin{equation}
|\delta m_\text{all-ss}| \lesssim \frac{0.25}{\text{SNR}_\text{s-s}}.
\end{equation}
For the cosmic shear measurements, we obtain a SNR of $12.6$ \rmrv{from the simulations}, which results in
\begin{equation}
|\delta m_\text{all-ss}| \lesssim 0.020.
\end{equation}
 Given that
the SNR of the cosmic shear measurement is slightly less than half that
of the SNR of the CMASS g-g lensing measurement, the requirements for
the multiplicative bias uncertainty are slightly more stringent for g-g
lensing than for cosmic shear.  We adopt the requirement $|\delta m| <
0.017$ throughout and omit the subscripts ``all-ss'' and ``all-g-g''
hereafter.

\rev{So far the discussion has focused on the question of how systematics contaminate the
  observable quantities, such as the shear correlation function.  However, we can also check how
  $|\delta m| < 0.017$ translates into a bias on cosmological parameter estimates.  We do this for
  the case of cosmic shear over the adopted range of scales, when modeling statistical errors only
  -- i.e., without marginalizing over systematic uncertainties that are commonly marginalized over,
  such as intrinsic alignments and baryonic effects on the matter power spectrum.  In that sense,
  our results are somewhat conservative, since marginalizing over these other effects will increase
  the statistical error and thus our tolerance for systematic uncertainty in the shear signal.  We
  do not consider additional data, such as priors from the Cosmic Microwave Background data.  Our
  cosmological parameter set for this analysis is $(\sigma_8, \Omega_m, n_s, \Omega_b, \theta_*)$
  with wide top-hat priors, and we marginalize over the last three quantities while considering the
  best-constrained combination of $\sigma_8$ and $\Omega_m$, which is here denoted $S_8\equiv
  \sigma_8(\Omega_m/0.3)^{0.5}$.  We find the expected $1\sigma$ statistical uncertainty on $S_8$ is
  $\sigma(S_8)=0.029$.  The bias on $S_8$ as a function of $|\delta m|$ is found to be}
\begin{equation}
\rev{
|\Delta S_8| = (9.3\times 10^{-3}) \left| \frac{\delta m}{0.017} \right| = 0.32\, \sigma(S_8)\,\left| \frac{\delta m}{0.017} \right|
}
\end{equation}
\rev{Hence our requirements threshold corresponds to a bias in $S_8$ that is subdominant to the
  statistical errors, as expected.}


\subsubsection{PSF model size errors}\label{subsubsec:psfmodelsize-ss}

As shown in equation~(3.17) of \citet{2016MNRAS.460.2245J}, the
  additive shift in the shear-shear correlation function $\xi_+(\theta)$ due to PSF model size
  errors, $\delta\xi_+$, can be written as
\begin{equation}
\delta \xi_+ = 2\left\langle \frac{T_\text{PSF}}{T_\text{gal}}\frac{\delta T_\text{PSF}}{T_\text{PSF}}\right\rangle \xi_+(\theta).
\end{equation}
Here $T_\text{gal}$ is the trace of the second moment matrix of the galaxy itself (before PSF
convolution).
For the purpose of placing requirements, we use the theoretical model for $\xi_+(\theta)$ given the
HSC redshift distribution.  We also assert the independence of the first fraction (related to the
typical galaxy resolution) and the second fraction (related to the PSF model size errors).  As
before,
\mtrv{we set}
$\langle R_2^{-1}-1\rangle=1$ to place requirements, corresponding to
$\langle T_\text{PSF}/T_\text{gal}\rangle=1$, and again
\mtrv{set}
the fractional PSF model area error to
\mtrv{twice}
the
fractional PSF radius error.
This gives
\begin{equation}
\delta \xi_+ = 4 \left\langle\frac{\delta \sigma_\text{PSF}}{\sigma_\text{PSF}}\right\rangle \xi_+(\theta).
\end{equation}
We would like to constrain this potential additive error to be $\lesssim 0.5$ times the statistical
error, or
\begin{equation}
4 \left\langle\frac{\delta \sigma_\text{PSF}}{\sigma_\text{PSF}}\right\rangle \xi_+(\theta) \lesssim 0.5 \sigma_{\xi_+}(\theta).
\end{equation}
The resulting requirement on our observable quantity is
\begin{equation}
\left\langle\frac{\delta \sigma_\text{PSF}}{\sigma_\text{PSF}}\right\rangle \lesssim
\frac{ \sigma_{\xi_+}(\theta)}{8 \xi_+(\theta)}.
\end{equation}
However, we do not want our requirement to depend on our $\theta$ binning, as it would if we used
the above equation.  We therefore use $\xi_+(\theta)$ and its full covariance matrix to turn
$\sigma_{\xi_+}(\theta)/\xi_+(\theta)$ on the right-hand side into $1/\text{SNR}_{\rm s-s}$.  Using
the integrated signal-to-noise ratio leads to conservative requirements, assuming perfect correlations between the
bins.  Using a value of $\text{SNR}_\text{s-s}=12.6$ as above gives a requirement of
\begin{equation}
\left\langle\frac{\delta \sigma_\text{PSF}}{\sigma_\text{PSF}}\right\rangle
    \lesssim \frac{1}{8\,\text{SNR}_\text{s-s}} \approx 0.01\,.
\end{equation}
This is less stringent than the PSF model size requirement for g-g lensing (0.004, as given in
Section~\ref{subsubsec:gglpsfmodelsizereq}) and hence we adopt the g-g lensing-based requirement instead of this one.

\subsubsection{Additive bias in shear}\label{subsubsec:additive-ss}

In this section, we place a requirement on the overall coherent additive biases in the shear.  These
additive biases have multiple potential origins: PSF model \rmrv{shape} errors (which we will
consider more specifically in Section~\ref{subsubsec:psfmodele-ss}), inadequate removal of PSF anisotropy
in the shear estimation method, noise bias, selection bias, and more.  To place this requirement, we
consider coherent additive shears in the shear estimates as $\hat{g}=g+c$ \rmrv{(neglecting for this
purpose the multiplicative biases)}.  As demonstrated in
\cite{2015MNRAS.450.2963M}, insufficient removal of PSF anisotropy in the PSF correction method
typically results in $c$
\mtrv{having the form }
$a~e_\text{PSF}$, where $a$ is a prefactor;
however, the additive term has a
different impact if it is due to PSF modeling errors (Section~\ref{subsubsec:psfmodele-ss}).  In
\rmrv{the case of insufficient removal of PSF anisotropy},
the observed shear correlation function is $\langle \hat{g}\hat{g}\rangle = \langle gg\rangle
+ \langle cc\rangle$, under the assumption that the systematics do not correlate with lensing shear
\mtrv{(although the assumption could be violated, e.g.\ by selection bias).}
\mtrv{That is,}
we
place a requirement on any unknown
and therefore unremovable additive term in the shear correlation function, $\langle cc\rangle\equiv
\delta \xi_{+,\text{sys}}$.  This requirement will
depend on $\theta$, and is simply set by requiring it to be below $\sigma_{\xi_\pm}$ at each scale.
To avoid a binning dependence of this requirement, we use the same scheme
\mtrv{from Section~\ref{subsubsec:gglpsfmodelsizereq} of}
introducing a
conservative integrated SNR, giving a requirement on overall additive bias of
\begin{equation}\label{eq:additive-req}
\langle cc\rangle < \frac{\xi_+(\theta)}{\text{SNR}_\text{s-s}} = \frac{\xi_+(\theta)}{12.6}.
\end{equation}

In the case that this additive term is dominated by insufficient PSF correction, i.e.,
$c=a~g_\text{PSF}$, we can furthermore place a constraint on the typical value of $a$.  Alternatively,
we can require that this particular additive term be $<0.5 \sigma_{\xi_\pm}$ to allow some margin
for other sources of the additive term.  In this case,
the additive term looks like $a^2\langle g_\text{PSF}g_\text{PSF}\rangle=a^2
\langle g_* g_*\rangle$ \rmrv{(we use star shear estimates as a proxy for PSF model shear estimates)}.  For a known value of $a$ as quantified using simulations, we can remove
the term, leaving the uncertainty in the additive term due to any residual $\delta (a^2)$.  Hence we can write
\begin{equation}
   \delta (a^2) < \frac{0.5\sigma_{\xi_\pm}(\theta)}{\langle g_* g_*\rangle(\theta)}.
\end{equation}
Using the same (conservative) integrated SNR argument as in Section~\ref{subsubsec:psfmodelsize-ss} to avoid a
binning-dependence of the requirement, this becomes
\begin{equation}\label{eq:additive-psf-req}
   \delta (a^2) < \frac{\xi_+(\theta)}{2 \langle g_* g_*\rangle(\theta)\,\text{SNR}_\text{s-s}} = \frac{\xi_+(\theta)}{25 \langle g_* g_*\rangle(\theta)}.
\end{equation}
\rev{The scale-dependence of the star-star shape correlation (denominator) in the HSC survey is much flatter than that of the cosmic
  shear correlation function (numerator); hence the function in Eq.~\eqref{eq:additive-psf-req} is a declining
  function of scale, ranging from $2\times 10^{-3}$ at our minimum scale of $\theta=4$~arcmin to
  $3\times 10^{-4}$ at $\theta=1$~degree (beyond which we have little statistical power). Since we
  have already conservatively used the integrated SNR to define the prefactor in this equation, we
  do not need to conservatively choose the lowest $\delta (a^2)$ value as well; rather, we use the
  geometric mean of these values, requiring $\delta(a^2)<8\times 10^{-4}$.}

\rev{
Following similar methodology as in the end of Sec.~\ref{subsubsec:cosshearreq1}, we check how
cosmological parameter constraints (specifically $S_8$ as defined there) depend on $\delta (a^2)$.  We
find that}
\begin{equation}
\rev{
|\Delta S_8| = 0.06 \, \sigma(S_8)\, \sqrt{\frac{\delta (a^2)}{8\times 10^{-4}}} .
}
\end{equation}
\rev{The fact that the prefactor in this equation is $\ll 1$ means that this
  requirement based on integrated SNR is still relatively conservative.}

\subsubsection{PSF model \rmrv{shape} errors}\label{subsubsec:psfmodele-ss}

Finally, we need to place a requirement on how PSF model \rmrv{shape} errors propagate into an
additive term in the shear-shear correlations.  The expression for these additive terms is
found in
equation~(3.17)
in \citet{2016MNRAS.460.2245J}, and depends on the five $\rho$ statistics\rmrv{, two of which were defined in
\cite{2010MNRAS.404..350R} and the final three in \cite{2016MNRAS.460.2245J}.  If we define $\delta
g_\text{PSF}^* \equiv g_*-g_\text{PSF}$, evaluating the PSF model at the positions of the stars,
then the first two $\rho$ statistics are defined as}
\begin{align}
\rho_1(\theta) &\equiv \left\langle \delta g_\text{PSF}^*(\vec{r})\delta
g_\text{PSF}(\vec{r}+\vec{\theta})\right\rangle\label{eq:rho1}\\
\rho_2(\theta) &\equiv \left\langle g_\text{PSF}^*(\vec{r})\delta g_\text{PSF}(\vec{r}+\vec{\theta})\right\rangle.
\end{align}
\rmrv{Conceptually, $\rho_1$ is the auto-correlation function of PSF model shape residuals, while $\rho_2$
is its cross-correlation with the PSF shape itself. The remaining $\rho$ statistics involve $T_\text{psf}$, the trace of the second moment matrix of the
PSF.  They are used to estimate the systematic uncertainty in the shear correlation function due to
PSF modeling errors.}
\begin{align}
\rho_3(\theta) &\equiv \left\langle \left(g_\text{PSF}^*\frac{\delta
    T_\text{PSF}}{T_\text{PSF}}\right)(\vec{r}) \left(g_\text{PSF}\frac{\delta
    T_\text{PSF}}{T_\text{PSF}}\right)(\vec{r} + \vec{\theta})\right\rangle \\
\rho_4(\theta) &\equiv \left\langle\delta g_\text{PSF}^*(\vec{r}) \left(g_\text{PSF}\frac{\delta
    T_\text{PSF}}{T_\text{PSF}}\right)(\vec{r} + \vec{\theta})\right\rangle \\
\rho_5(\theta) &\equiv \left\langle g_\text{PSF}^*(\vec{r}) \left(g_\text{PSF}\frac{\delta
    T_\text{PSF}}{T_\text{PSF}}\right)(\vec{r} + \vec{\theta})\right\rangle  \label{eq:rho5}
\end{align}

We define a maximum tolerable
additive systematic on $\xi_+(\theta)$ due to PSF model \rmrv{shape} errors, by requiring that each
$\rho$ statistic contribute less than $0.5 \delta \xi_{+,\text{sys}}$.  In that case, our requirements on the $\rho$ statistics are
\begin{equation}
|\rho_{1,3,4}(\theta)| < \left\langle
  \frac{T_\text{PSF}}{T_\text{gal}}\right\rangle^{-2}\delta \xi_{+,\text{sys}}
\end{equation}
and
\begin{equation}
|\rho_{2,5}(\theta)| <
  |a|^{-1}\left\langle \frac{T_\text{PSF}}{T_\text{gal}}\right\rangle^{-1} \delta \xi_{+,\text{sys}}
\end{equation}
where $a$ and $\delta \xi_{+,\text{sys}}$ are defined as in Section~\ref{subsubsec:additive-ss}.  As in
Section~\ref{subsubsec:psfmodelsize-ss}, we adopt a value of $\langle T_\text{PSF}/T_\text{gal}\rangle=1$ for setting
requirements.  The value of $|a|$ is estimated using simulations in \citet{2017arXiv171000885M} as described in
Section~\ref{subsec:great3like} in this paper; we calculate its lensing-weighted average as approximately
0.02.
Finally, using the same (conservative) integrated SNR argument as in Section~\ref{subsubsec:psfmodelsize-ss} to avoid a
binning-dependence of the requirement, these \rmrv{requirements} become
\begin{equation}\label{eq:rho134-req}
|\rho_{1,3,4}(\theta)| < \frac{\xi_+(\theta)}{2\,\text{SNR}_\text{s-s}} = \frac{\xi_+(\theta)}{25}
\end{equation}
and
\begin{equation}\label{eq:rho25-req}
|\rho_{2,5}(\theta)| < \frac{\xi_+(\theta)}{2|a| \,\text{SNR}_\text{s-s} } = \frac{\xi_+(\theta)}{25|a|}.
\end{equation}

\subsection{Overall summary of requirements}

In Table~\ref{tab:requirements} we summarize the key requirements
\mtrv{in this section}.
In the case of
requirements
\mtrv{on the same quantity}
provided by galaxy-galaxy lensing and cosmic shear, we have always adopted
the more stringent requirement (from g-g lensing in our case).
\begin{table*}
\caption{Summary of key requirements from Section~\ref{sec:requirements}.}\label{tab:requirements}
\begin{center}
\begin{tabular}{ll}
\hline
Systematic uncertainty & Origin \\ \hline
Overall multiplicative bias & Galaxy-galaxy lensing,
Section~\ref{subsubsec:gglbiasreq}, equation~\eqref{eq:mall-ggl} \\
Multiplicative bias due to PSF model size errors & Galaxy-galaxy lensing,
Section~\ref{subsubsec:gglpsfmodelsizereq}, equation~\eqref{eq:mpsf-ggl} \\\hline
Overall additive bias correlation function & Cosmic shear, Section~\ref{subsubsec:additive-ss},
equation~\eqref{eq:additive-req} \\
Additive bias due to insufficient correction for PSF anisotropy & Cosmic shear,
Section~\ref{subsubsec:additive-ss}, equation~\eqref{eq:additive-psf-req} \\
PSF model \rmrv{shape} errors & Cosmic shear, Section~\ref{subsubsec:psfmodele-ss},
equations~\eqref{eq:rho134-req} and~\eqref{eq:rho25-req} \\ \hline
\end{tabular}
\end{center}
\end{table*}

\begin{figure}
\begin{center}
\includegraphics[width=3in]{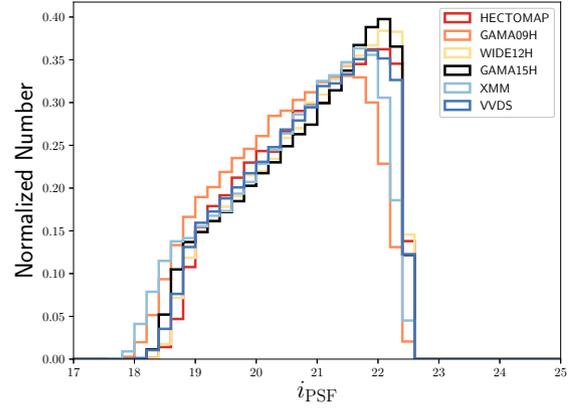}
\end{center}
\caption{$i$-band PSF magnitude distribution of the PSF star sample in each field.}
\label{fig:psf_mag_hist}
\end{figure}

\section{PSF modeling tests}\label{sec:psf}

In this section, we describe tests of
\mtrv{the fidelity of }
the PSF modeling.
\citet{PipelinePaper:inprep} show tests of PSF
modeling on individual exposures, including plots of PSF model size and \rmrv{shape} residuals as a
function of focal plane position, while this paper focuses
on tests of the
coadd PSF,
since  that
is the relevant
\mtrv{quantity}
for shear estimation.

We begin with a definition of the star
samples used for internal (to the data) tests of PSF models, and then describe the PSF model tests.
Note that no external simulations were used for tests of PSF models.  This is because the tests
internal to the data \mtrv{that we will show below}
suggest that PSF modeling errors are not a limiting systematic for first-year
HSC data.  In future years, when \rmrv{statistical errors are reduced due to the larger data volume
  and} other systematic errors are reduced through further study, PSF
modeling will require more scrutiny, \mtrv{perhaps} including external simulations.

The above statement about the impact of PSF modeling errors is valid only after some area
cuts were placed on the fidelity of the PSF model size.  Tests indicated that PSF models had serious
problems in very good seeing ($<0.45$\arcsec) as shown in more detail in
\citet{PipelinePaper:inprep}.

To bypass this problem, we introduced two
cuts\rmrv{, both of which were briefly mentioned in Section~\ref{subsec:area} but will be described
  in more detail here}.
The first cut removes visits in the VVDS field with PSFs which are too good to be modeled by PSFEx
properly before these visits go into the coadding process. In most cases such visits have seeing
better than $0.45^{\prime\prime}$
at the center of the field-of-view. This reduction is denoted as {\tt
  S16A.wide2}, with details described in
the HSC DR1 paper.

The second cut is to ensure that the PSF size is adequately modeled. For this purpose, we calculate the
fractional size residual of PSF stars (defined as described in Section~\ref{sec:pipeline}) which is
defined by $f_{\delta\sigma}=(\sigma_{\rm PSF}-\sigma_*)/\sigma_*$, where $\sigma_*$ and
$\sigma_{\rm PSF}$ are the determinant radius of PSF stars and PSF models (from the adaptive second
moments) reconstructed at the PSF
star positions, respectively. To suppress measurement noise, we average the fractional size
residuals $f_{\delta\sigma}$ within a \texttt{HEALPix} pixel with ${\tt NSIDE}=1024$, which corresponds to an
area of $\sim12\ {\rm arcmin}^2$. The number of PSF stars in a \texttt{HEALPix} pixel varies from $\sim10$ to
$\sim24$.
We then plot the fractional size residual $f_{\delta\sigma}$ as a function of seeing FWHM
(also based on the average size of PSF stars averaged within \texttt{HEALPix} pixels), as shown in
figure~\ref{fig:frac_size_cut}. In the GAMA09H field, we find a cloud of \texttt{HEALPix} pixels with very good
seeing ($\sim0.45^{\prime\prime}$) and large fractional size residual
($f_{\delta\sigma}\sim0.02$--0.2, with strictly positive values indicating a systematic bias). We
find that the inclusion of these regions degrades several of our PSF model-related null tests for
the GAMA09H field, so that it fails the requirements quite severely. After we remove such areas by
applying the $f_{\delta\sigma}<0.02$ cut, which removes $\sim4$\% of the total WLFDFC area (primarily
in GAMA09H but affecting HECTOMAP slightly, as well), $\rho_1$ goes down and meets the requirement.

This two-step approach was necessary because the first cut was intended to eliminate the
problem as originally noticed in VVDS, but later there were areas that were found to fail our
requirements in GAMA09H, necessitating a second cut that could be applied in postprocessing.

All tests described below use the final area of the shear catalog, after imposing these cuts on PSF
model quality.

\begin{figure*}
\begin{center}
\includegraphics[width=6in]{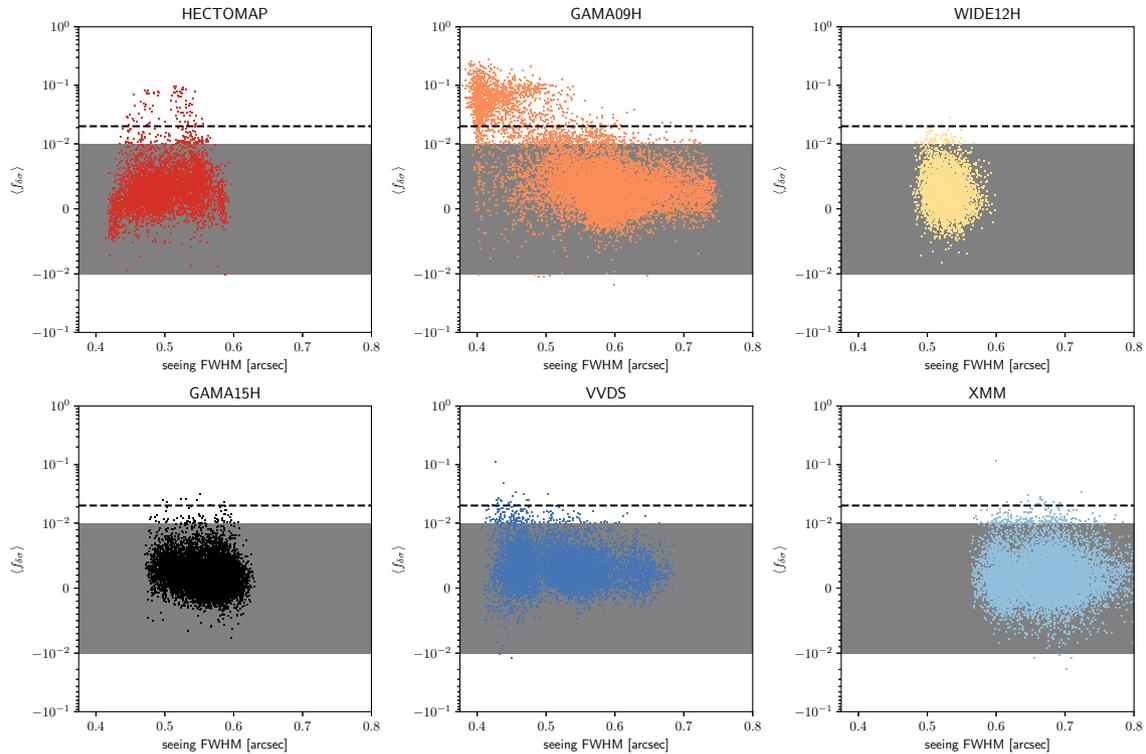}
\end{center}
\caption{The average fractional size residual $\langle f_{\delta\sigma}\rangle=\langle (\sigma_{\rm
    PSF}-\sigma_*)/\sigma_*\rangle $ between PSF stars and PSF models
  reconstructed at star positions, averaged over the PSF stars within \texttt{HEALPix} pixels with ${\tt
    NSIDE}=1024$, and shown \rmrv{for each \texttt{HEALPix} pixel} as a function of seeing. A 
symlog
scale is used to allow negative
  residuals to be shown. The dashed line shows the fractional size cut we apply\rmrv{, removing all
    points above the line and therefore eliminating }
  the cloud of points with large strictly positive $\langle f_{\delta\sigma}\rangle$ values \rmrv{that often but
    not always have very good
  seeing}. The best-seeing visits in the VVDS region were already removed before production of this
  figure. \hmrv{Regions with a dark grey background show the linear part of the symlog scale, with the rest
being logarithmic. }}
\label{fig:frac_size_cut}
\end{figure*}

\subsection{Star samples used}\label{subsec-starsamples}

As described in more detail in Section~\ref{sec:pipeline}, PSF star selection relies on clustering of
high-SNR objects in size, typically resulting in $\sim \rev{90}$ star candidates per
CCD chip.
Currently, $\sim$20$\%$ of the stars in a given exposure are
\mtrv{randomly chosen and}
reserved for cross-validation and are not
used for PSF modeling.  \rev{We found that the quality of the modeling was almost independent
of the number of stars as long as there were at least 20 stars used.  This was true for $\sim 99\%$ of 
the individual visists. Therefore, reserving 20$\%$ of the stars did not degrade the quality of the PSF models.}

We define two star samples for PSF modeling tests. The first includes stars used for the PSF determination by
PSFEx and the other includes bright, secure star detections that were not used for PSF.
Because the star sample used in PSFEx is derived on individual exposures, different
exposures will not necessarily select the same set of stars.
\jbrv{We label stars that were used on $\geq20\%$ of
the input visits as having been used in the modeling; because most of the Wide survey has six exposures in $i$, this typically requires a star to be selected for PSF determination in at least two exposures.}

The other set of stars we define are secure star detections that were not used for PSF
determination\rmrv{, but that were otherwise selected with the same flag and other cuts}.  The latter are important to use for testing in case over-fitting results in an
overly optimistic result on our null tests using PSF stars, and/or to pick up on PSF interpolation
failures.  For both catalogs, we first remove objects
\mtrv{whose photometry might be suspected due to a contamination}
at the edges of CCDs and pixels with saturation, interpolation, and cosmic rays, and with bad
centroid and shape measurement flags. We then select stars by
\texttt{iclassification\_extendedness=0} (a star/galaxy classifier based
on the
\mtrv{$i$-band}
images;
\citealt{PipelinePaper:inprep}) and restrict to a bright sample using $i$-band PSF
magnitude $<22.5$. \rmrv{As shown in figure~\ref{fig:psf_mag_hist}, this is quite close to the
  effective magnitude limit of PSF stars.} This choice is motivated by the finding (figure~13 of
the HSC DR1 paper)
 that
the star sample in this magnitude range is highly pure, but contamination by galaxies becomes more
important at fainter magnitudes.

The PSF star sample is
\mtrv{flagged by setting}
\texttt{icalib\_psf\_used=True}. Figure~\ref{fig:psf_mag_hist} shows the
distribution of $i$-band PSF magnitudes of this sample for different
fields. The magnitude of PSF stars ranges from $i_{\rm PSF}\sim18$ to
$22.5$. 

\subsection{Internal tests}
\label{subsec:psf-tests}


For weak lensing, we need to validate both the PSF model sizes and shapes.  Errors in the former
result in multiplicative biases in shear estimates, while errors in the latter result in additive
biases in shears.  For our tests, we use the effective stacked PSF (see
Section~\ref{sec:pipeline} for details)\rmrv{, and compare} with measurements of the stars in the coadd.
\citet{PipelinePaper:inprep} shows the result of PSF model tests on individual visits.  As
described in \rmrv{Section~\ref{subsec-starsamples}}, we carry out the majority of our tests
separately with \rmrv{two secure star samples: those that were and were not used to construct the
  PSF models.}

\begin{figure*}
\begin{center}
\includegraphics[width=3in]{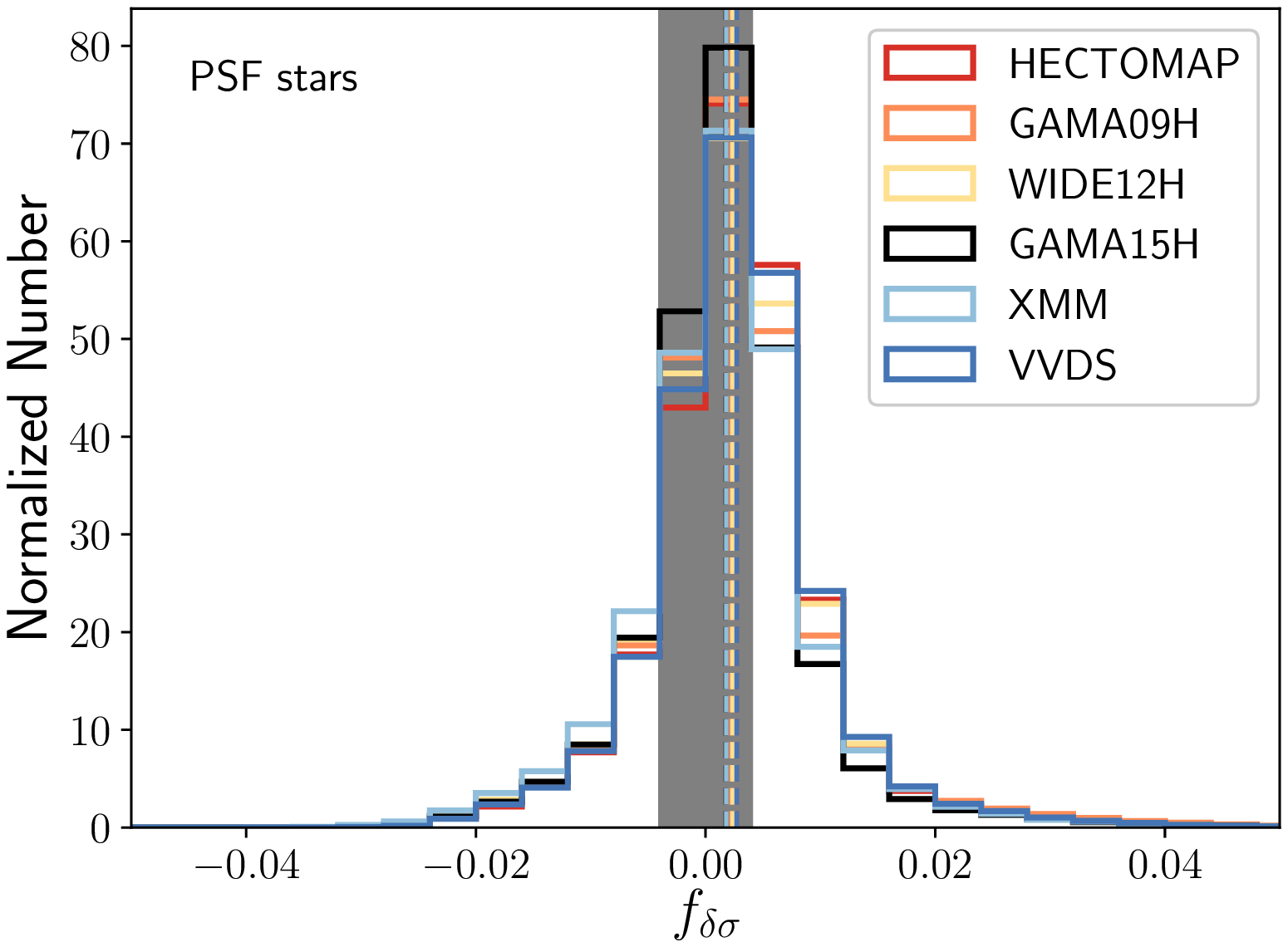}
\includegraphics[width=3in]{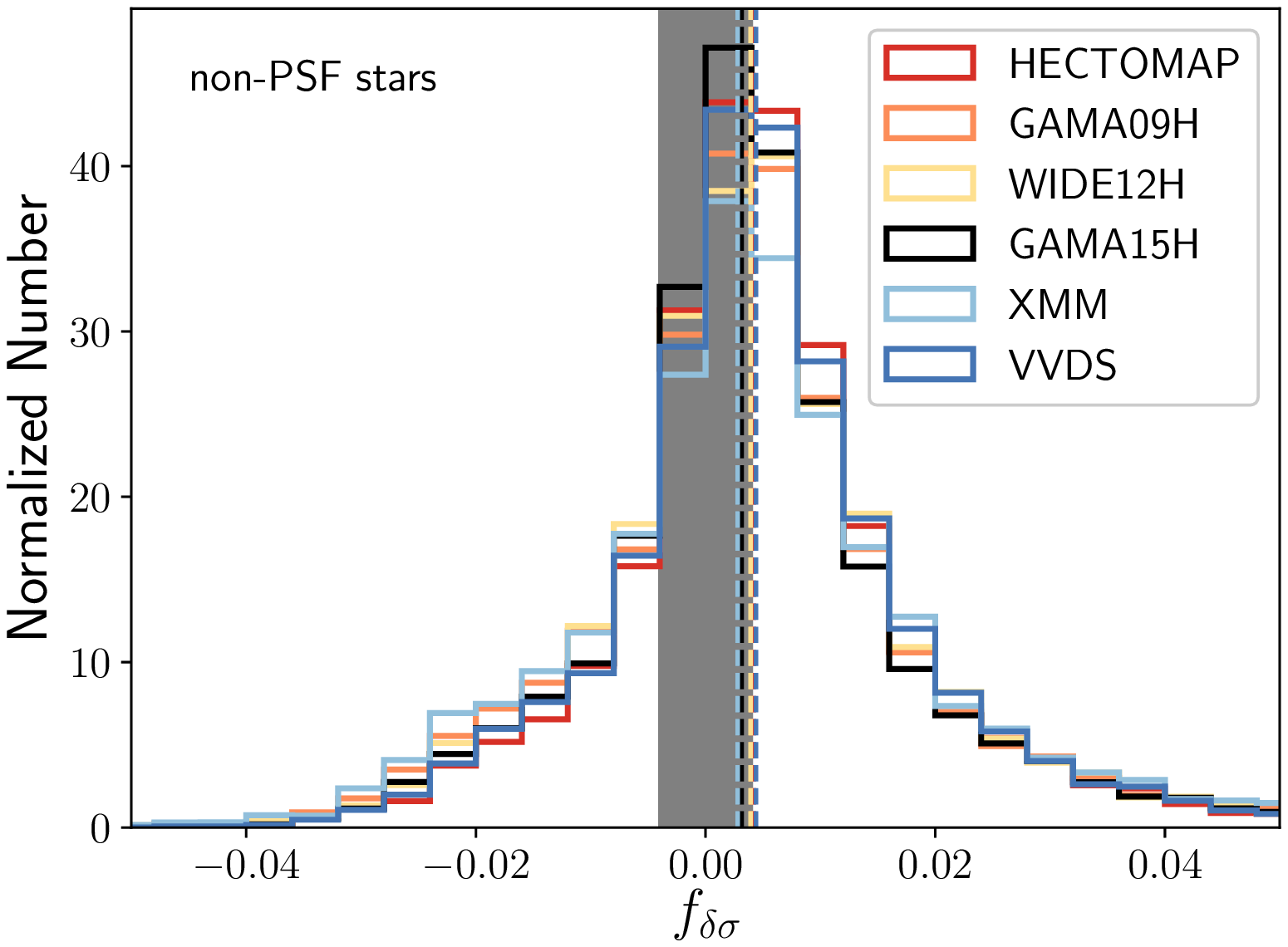}
\includegraphics[width=3in]{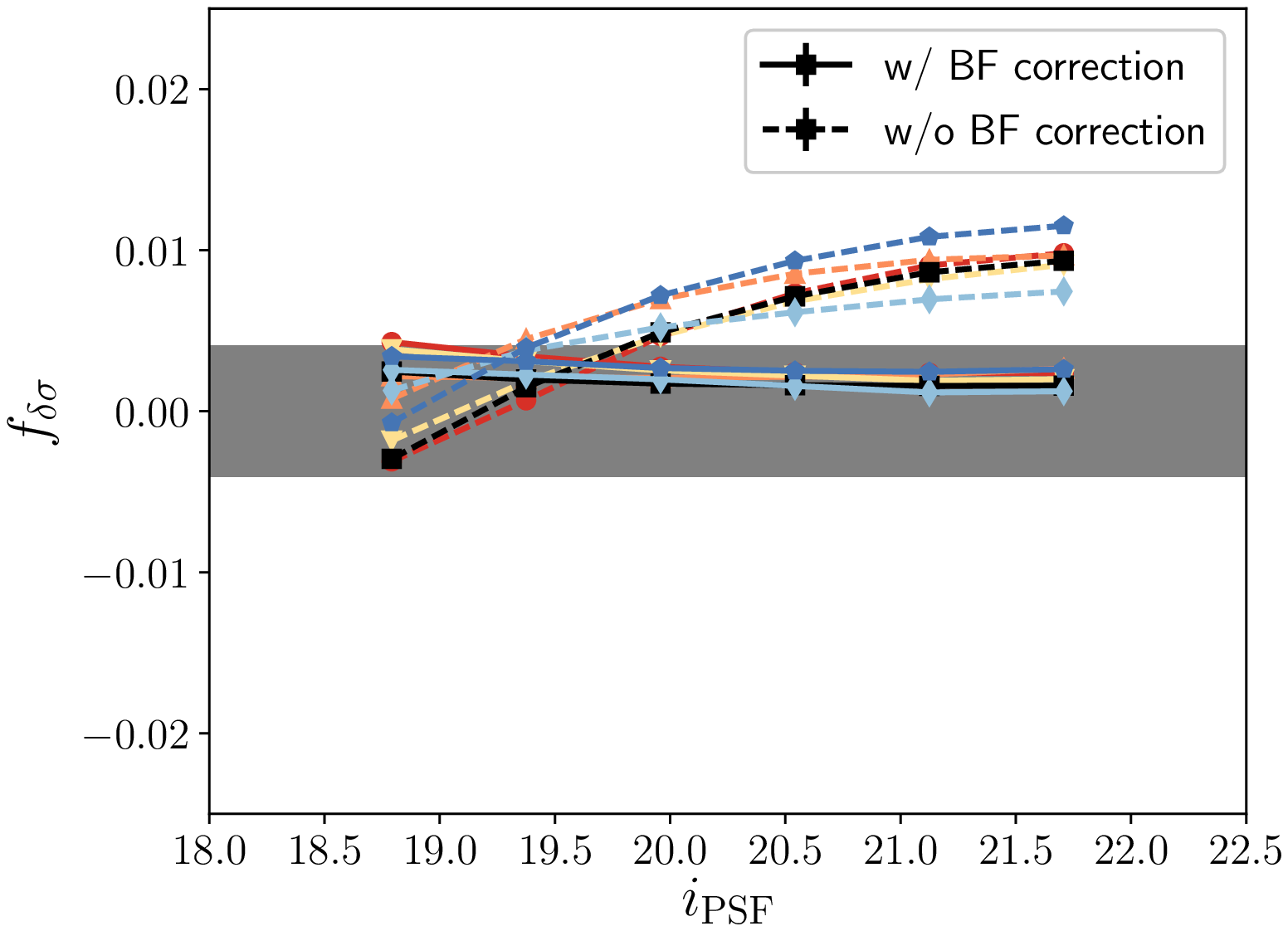}
\includegraphics[width=3in]{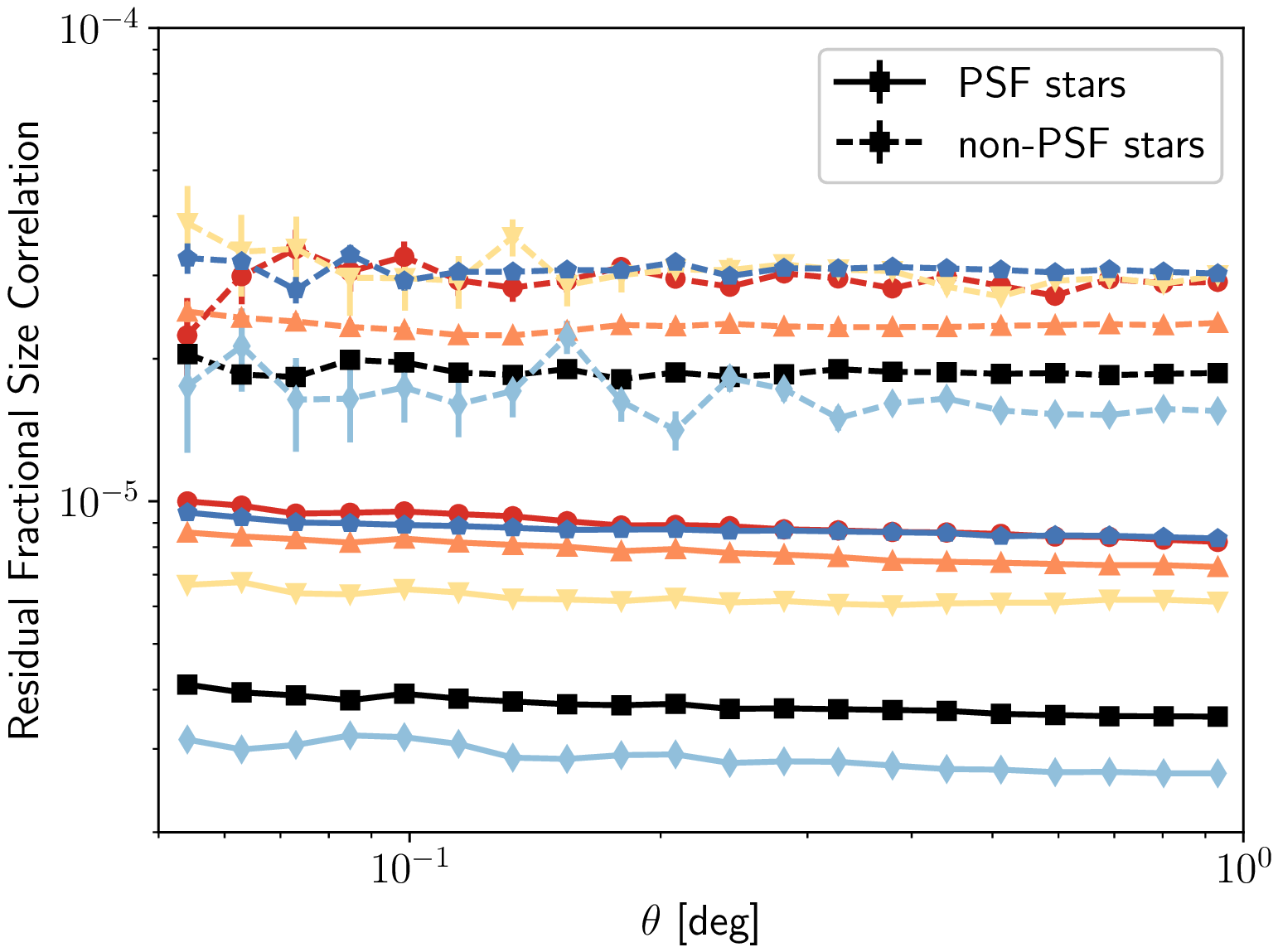}
\end{center}
\caption{{\em Top left:} Distribution of the fractional size residual
  of the PSF star sample in each field. The size is defined by the
  determinant radius. The gray shaded regions indicate the
  requirements on the mean residual for the first-year HSC survey
  data. The vertical dashed lines show the median of the fractional size residual of each field; the
  results are highly consistent across fields.
  {\em Top right:} Same as the top left, but for non-PSF stars\rev{, reweighted in order to match
    the SNR distribution of the PSF stars used to make the top left panel.}
  {\em Bottom left:} Solid lines are median and bootstrap error of
  residual size of the PSF stars after brighter-fatter correction, as
  a function of the $i$-band PSF magnitude. Dashed lines are without
  brighter-fatter correction, as calculated using \rmrv{the first 61.5~nights of data through November 2015} (hence the
  observed area is different from what went into the solid lines).  The gray shaded
  regions indicates the requirements on the mean residual for the
  first-year HSC survey data. {\em Bottom right:} The correlation function
  of the fractional size residuals as a function of separation
  $\theta$. Solid and dashed lines show results for PSF and non-PSF
  stars, respectively.}
\label{fig:psf_residual_size_histmagcorr}
\end{figure*}

\subsubsection{PSF model size}\label{subsubsec:psfmodelsize}

The results for all PSF model size tests are shown in
figure~\ref{fig:psf_residual_size_histmagcorr}.  For the size, we use the determinant radius
$\sigma$ \rmrv{for the
star images and the PSF model,
based on the adaptive second moments, to calculate the fractional size residual $f_{\delta\sigma}=(\sigma_{\rm
    PSF}-\sigma_*)/\sigma_*$}.

As shown in the top left panel of figure~\ref{fig:psf_residual_size_histmagcorr}, the distribution
of fractional PSF size residuals is narrowly peaked at 0 for all fields.  Nearly all stars used for
this test (PSF stars only) fall within the shaded region defined by our first year requirements.
The median values shown as vertical dashed lines fall within the first year requirements region.
The top right panel shows the same thing, but for non-PSF stars.  \rev{To make this plot, we have
  reweighted the stars to ensure that their SNR distribution matches that for PSF stars, to ensure
  that different levels of noise in the $\sigma_*$ estimates does not change the width of the
  distribution.}  Here the scatter is clearly
larger; the median values are close to the edge of our requirements, but still pass.  Given the
conservative assumptions made in defining the requirements, this is acceptable performance.
However, for the full survey, \rmrv{we may need} an improved PSF modeling algorithm
to achieve the more stringent resulting requirements.  \rev{Note that with this histogram alone, we
  cannot distinguish whether the additional width indicates overfitting in the PSF modeling process,
 or use of an insufficient PSF model interpolation scheme.  In \citet{PipelinePaper:inprep}, a similar test
 carried out with the single-epoch images that are used directly for PSF modeling produced results
 that are qualitatively similar to what we see here in the coadd.}

\rmrv{It does not seem obvious that the distribution of the quantity shown in figure~\ref{fig:psf_residual_size_histmagcorr}
  is the same as the quantity shown on the horizontal axis of figure~\ref{fig:frac_size_cut}.  The
  primary reason for this is that figure~\ref{fig:psf_residual_size_histmagcorr} shows the results
  for individual stars, which is substantially noisier than the results averaged within
  \texttt{HEALPix} pixels shown in figure~\ref{fig:frac_size_cut}.  For the earlier
  plot, our goal was to identify outlier regions, so we had to average over stars to reduce the
  noise that dominates the shape of the per-star histograms in the later plot.  A much more minor
  contributor to the differences between figures is that the quantity plotted has a different
  denominator in each case.  For the earlier plot, we used $\sigma_*$ because $\sigma_\text{PSF}$
  has significant errors in the outlier regions we had hoped to identify.  For the later plot, we
  used $\sigma_\text{PSF}$ to reduce noise.}

The lower left panel is a comparison of the average fractional size residual in bins of fixed
$i$-band PSF magnitude for two versions of the HSC pipeline: an older version without correction for
the brighter-fatter effect, and the version used for science including a brighter-fatter correction.
As shown, the version of the pipeline (dashed curves) without this correction shows the
characteristic signature of this systematic error: stars appear larger than the PSF model at the
brightest magnitudes, and smaller than the PSF model at fainter magnitudes. This trend is almost
completely removed by the brighter-fatter correction, giving PSF size residual curves that are
almost independent of magnitude.  The brighter-fatter
correction is necessary to ensure that we meet our requirements for first-year weak lensing
science.   This test again uses only the PSF star sample, however the
efficacy of the brighter-fatter correction should be the same for all stars at fixed magnitude, so
we do not expect
results to differ for the non-PSF star sample (which has the same magnitude range).

Finally, the bottom right panel of figure~\ref{fig:psf_residual_size_histmagcorr} shows the spatial
correlation function of the PSF fractional size residuals.  Note that curves are shown separately
for PSF stars (solid) and non-PSF stars (dashed).  All of these curves are nearly independent of
spatial separation and differ from field to field, unlike the distribution of fractional size
residuals, which are consistent across fields.  
We can also see that the fields with the best seeing, VVDS and HECTOMAP (see
figure~\ref{fig:seeing_map}), have the worst PSF model size residual correlation functions.  Finally, comparing the PSF stars
and non-PSF stars, we can see that the latter have a larger value of this correlation function,
again flat with scale.  The field-to-field trends
are similar \rmrv{to what we saw for} the PSF stars. This increase in the correlation function as we move from PSF
stars to non-PSF stars is suggestive of either problems with the PSF model interpolation or
overfitting.  However, the plot does not provide us with information to distinguish between these
options.

The important result of this section, however, is that while our PSF model size diagnostic for
non-PSF stars (which are likely more representative of PSF model fidelity at galaxy positions) is
worse than when we use PSF stars, the results are still within our first-year HSC
science requirements. 

One notable aspect of these results is the fact that $\langle \sigma_*-\sigma_\text{PSF}\rangle <
0$, that is, on average the stars on the coadd are slightly smaller than the stacked PSF model at
the 0.1\% level. In general, we expect that the impact of relative astrometric errors will be to
make the stars on the coadd \rmrv{appear} slightly \rmrv{broader} than the stacked PSF model.  Given the sign of the
average PSF size residual, we infer that the impact of relative astrometric errors \rev{cannot be responsible
for this difference.}

\begin{figure}
\begin{center}
\includegraphics[width=3in]{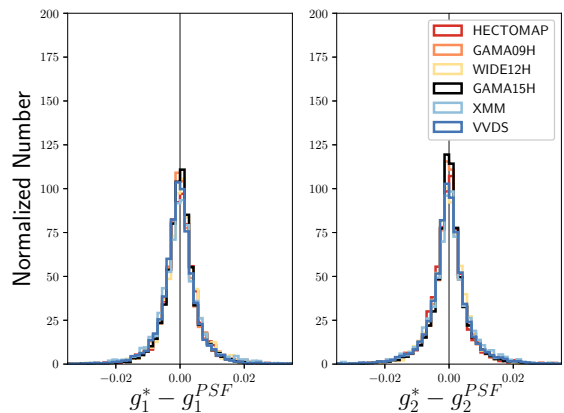}
\end{center}
\caption{Distributions of PSF residual \rmrv{shape} of the \hmrv{non-}PSF star sample in each
  field.}
\label{fig:psf_residual_shape_hist}
\end{figure}

\begin{figure*}
\begin{center}
\includegraphics[width=3in]{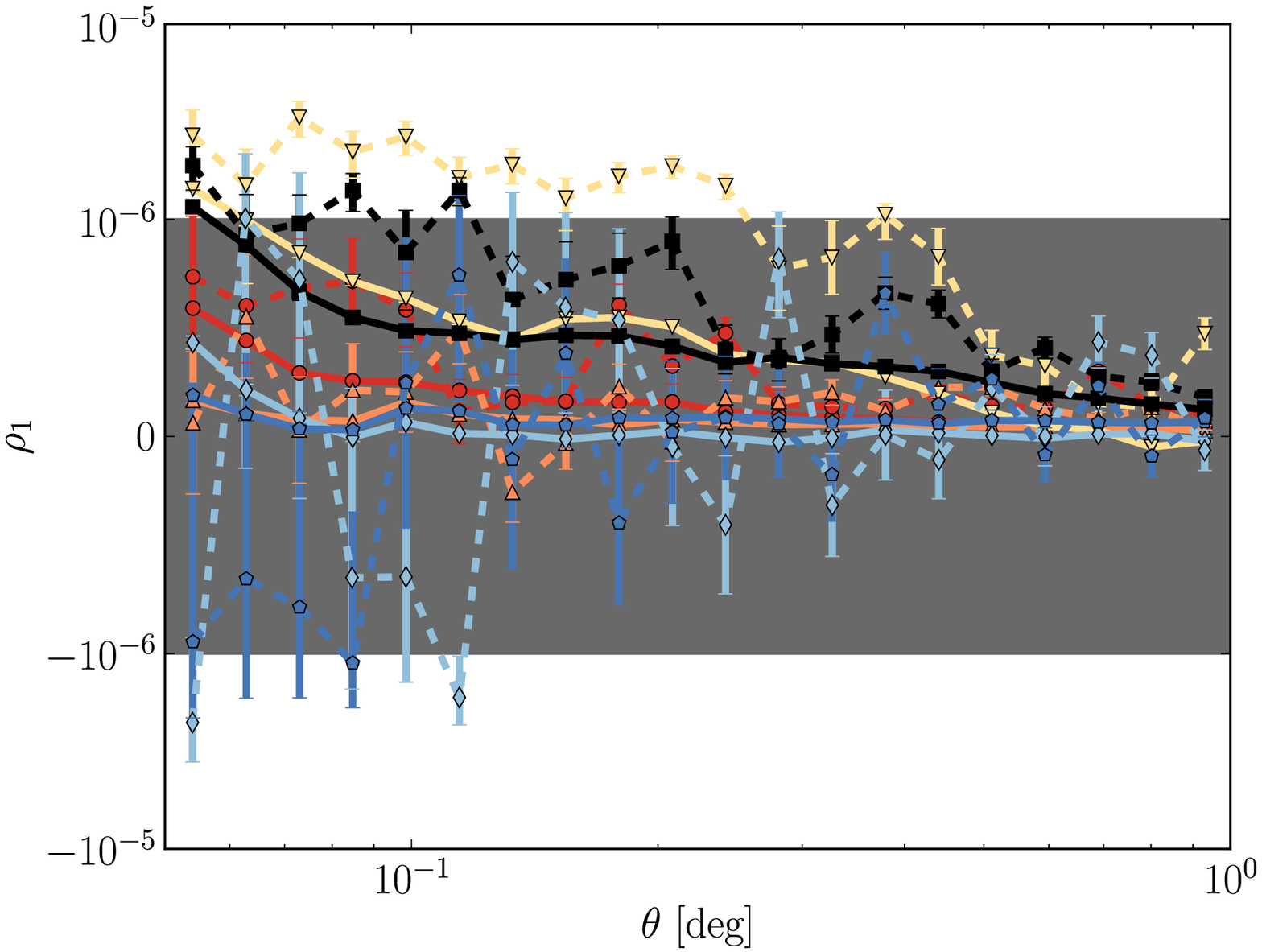}
\includegraphics[width=3in]{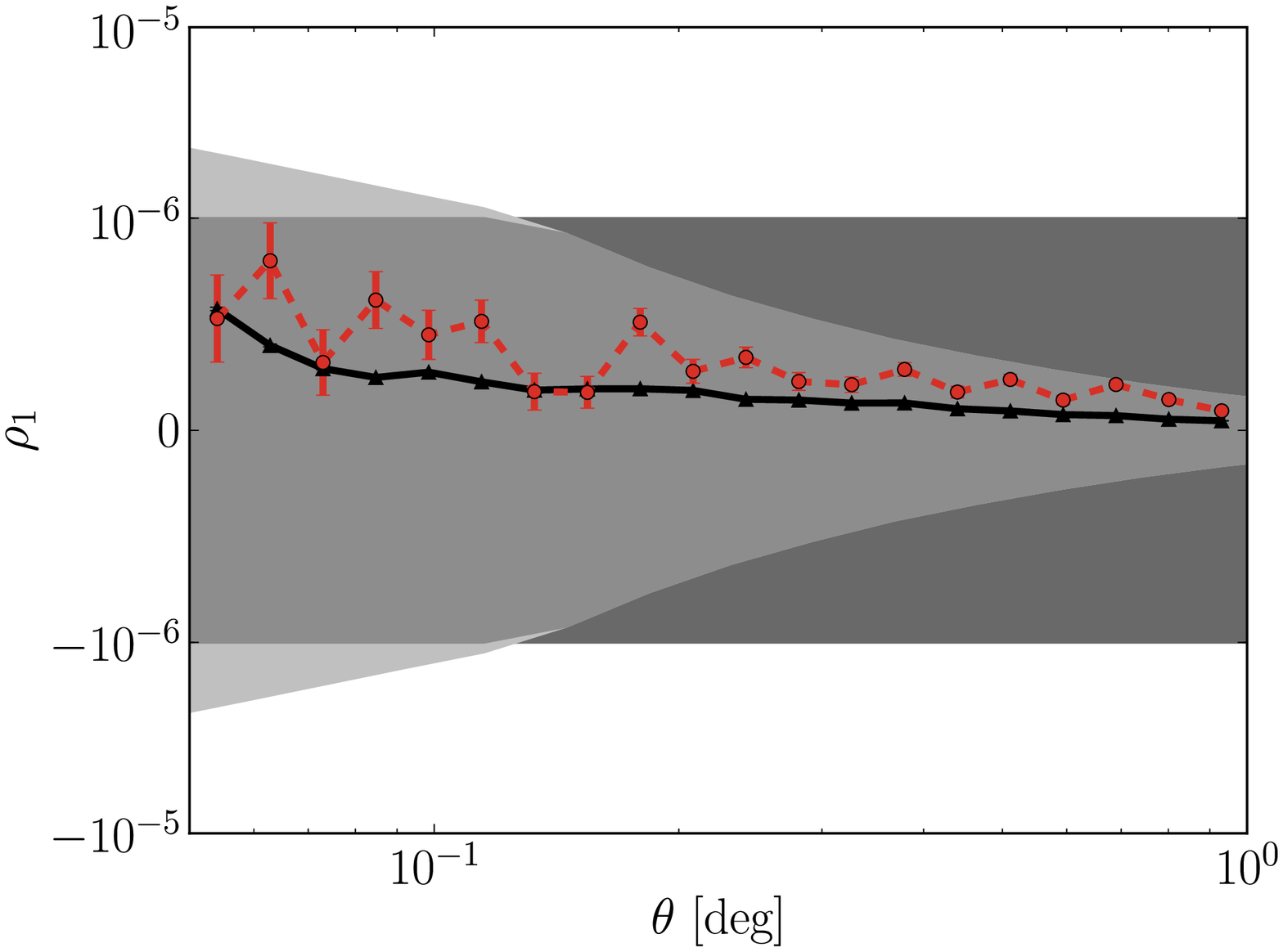}
\includegraphics[width=3in]{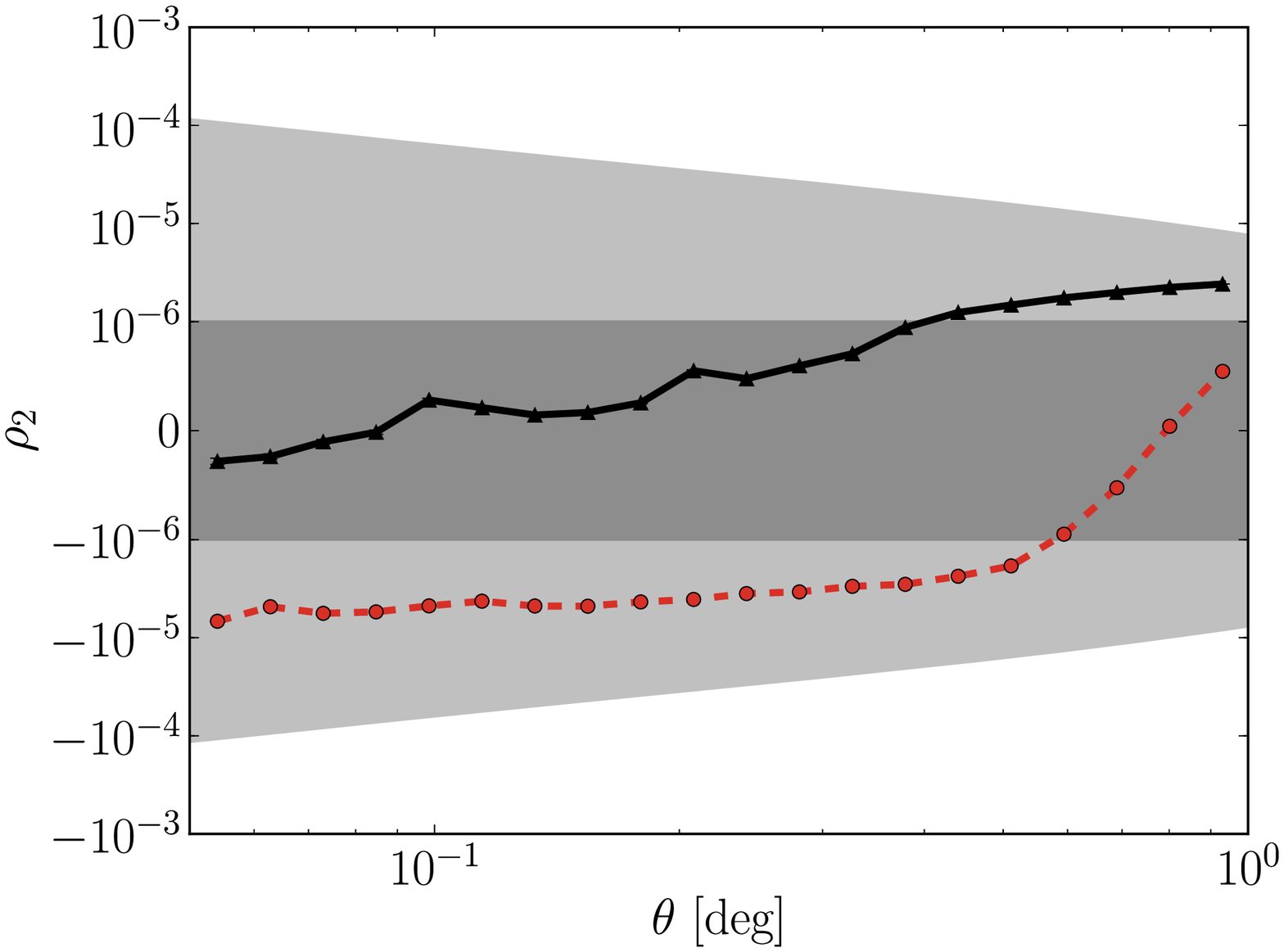}
\includegraphics[width=3in]{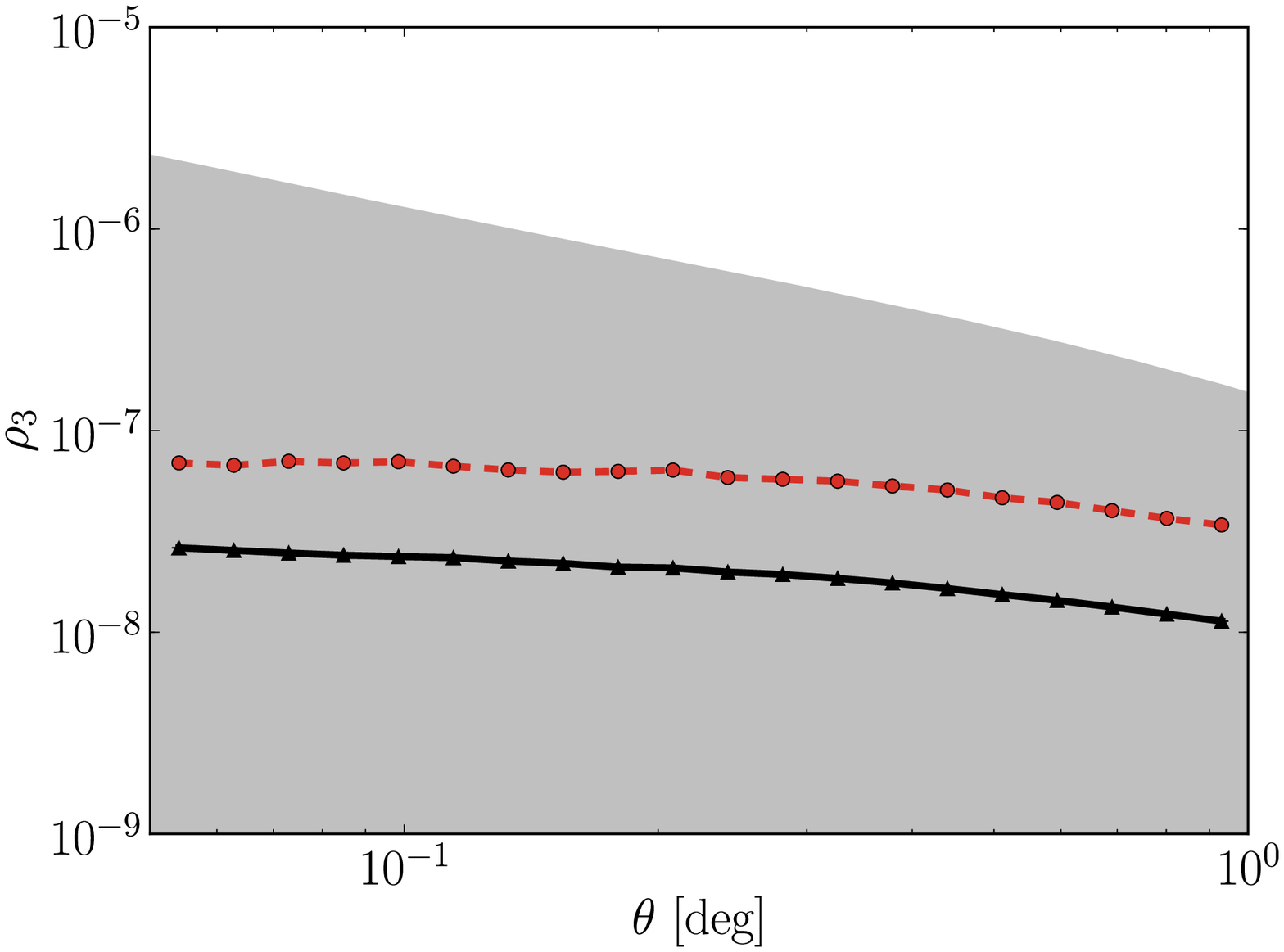}
\includegraphics[width=3in]{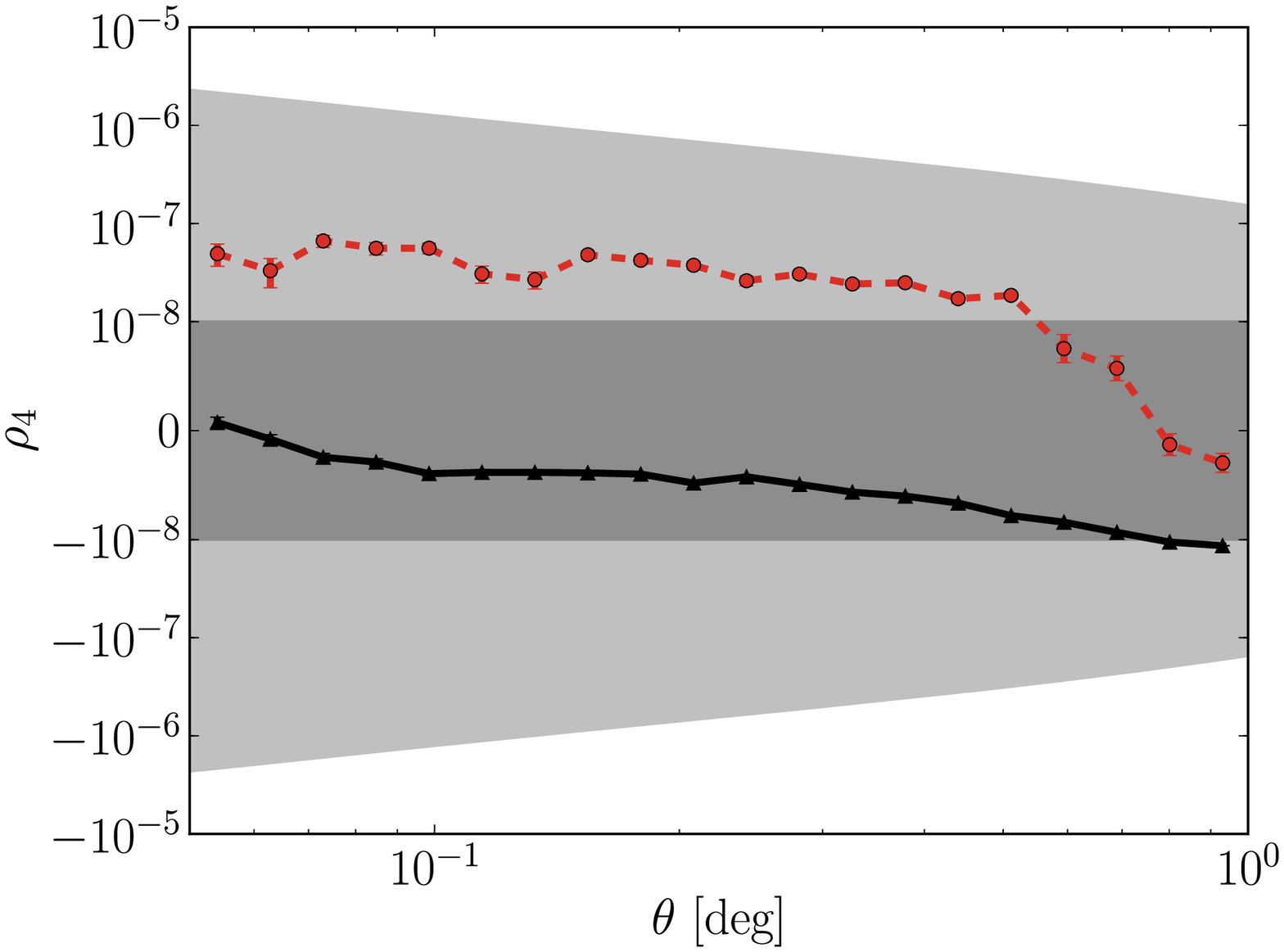}
\includegraphics[width=3in]{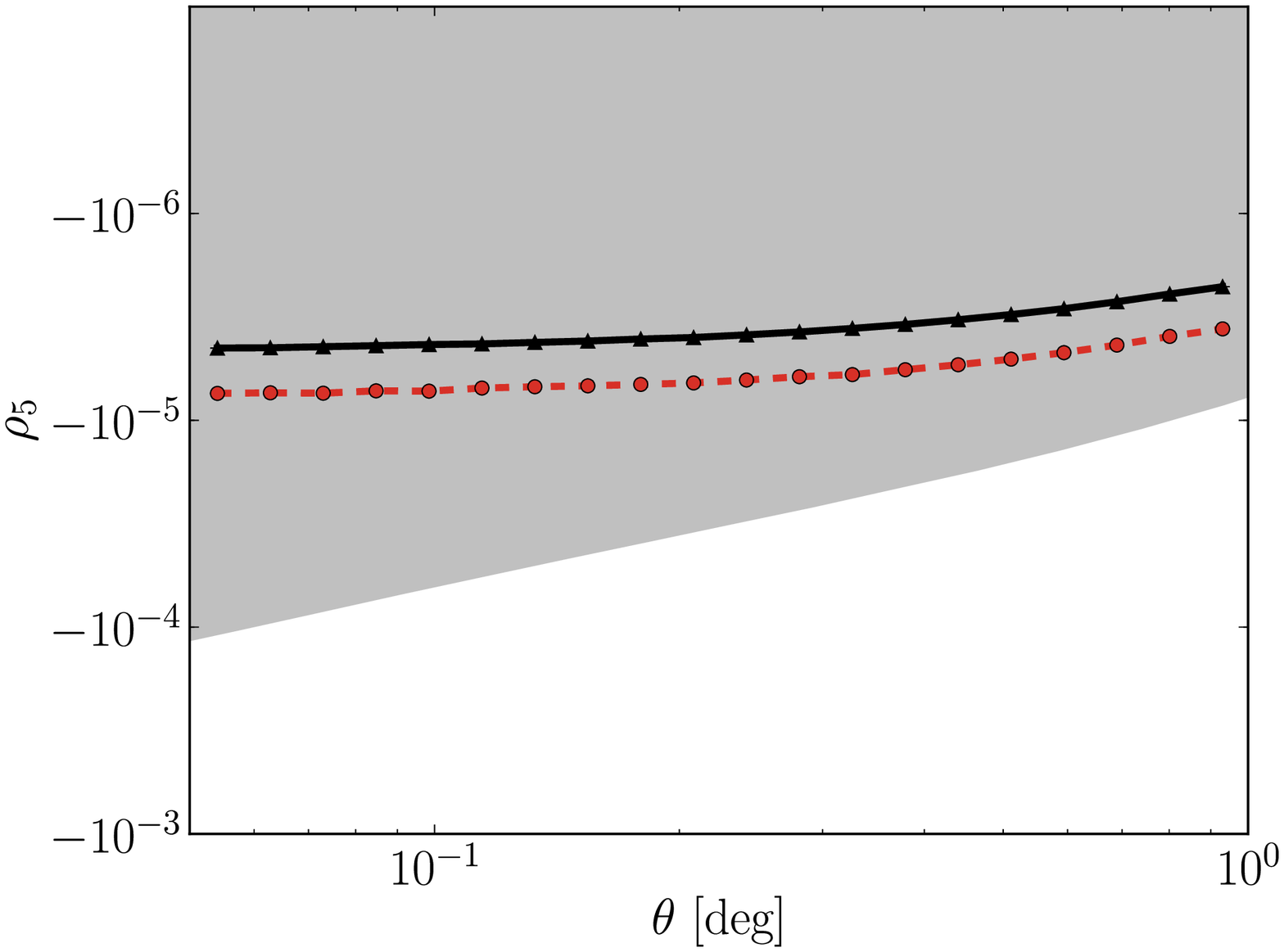}
\end{center}
\caption{PSF residual \rmrv{shape} correlations, or $\rho$ statistics,
  $\rho_1$ through $\rho_5$ \rmrv{(defined in Section~\ref{subsubsec:psfmodelshape})} as a function
  of separation $\theta$ on the sky.  The
  top left panel shows $\rho_1$ for each field to illustrate the
  significant field-to-field differences. The color scheme for the fields is the same as in
  figure~\ref{fig:psf_residual_shape_hist} (with an additional black curve showing the field-averaged
  results), while solid and dashed lines show results for PSF and
  non-PSF stars, respectively. The remaining panels show the $\rho$
  statistics averaged over the whole shape catalogue area along with our science
  requirements for first-year survey science as determined in Section~\ref{subsubsec:psfmodele-ss}. Here
  too the solid and dashed lines
  show results for PSF and non-PSF stars, respectively. Since the $\rho$ statistics can be negative,
  the vertical axes use a symlog
scale.  Regions with a dark grey background show the linear part of the symlog scale, with the rest
being logarithmic.  Regions with a light grey background are within the first-year survey
requirements for cosmic shear.}
\label{fig:rho_statistics}
\end{figure*}

\subsubsection{PSF model shape}\label{subsubsec:psfmodelshape}

The results for all PSF model shape tests are shown in figure~\ref{fig:psf_residual_shape_hist} and~\ref{fig:rho_statistics}.
While the shape definition returned by our shape measurement algorithms is \rmrv{the distortion
  defined in equation~\eqref{eq:e}}, we divide by
two to obtain shear estimates (this conversion is correct for stars and PSFs).

Figure~\ref{fig:psf_residual_shape_hist} shows the distribution of star $-$ PSF shears for the PSF star sample in each
field.  \rmrv{The typical width of the distributions is $<0.01$, with the means 
  quite close to zero}.

The panels in figure~\ref{fig:rho_statistics} show the $\rho$ statistics, \rmrv{defined in this
  paper in equations~\eqref{eq:rho1}--\eqref{eq:rho5} based on previous work
\citep{2010MNRAS.404..350R,2016MNRAS.460.2245J}.  These statistics  are constructed from spatial correlation functions of PSF model shape and size residuals.}
Requirements on the $\rho$ statistics were discussed in Section~\ref{subsubsec:psfmodele-ss}.

On all $\rho$ statistics, we show results for PSF stars (solid lines) and non-PSF stars (dashed
lines).  Clearly the results are not consistent in the two cases.  As shown in the top left panel of
figure~\ref{fig:rho_statistics}, there is clearly field-to-field variation in $\rho_1$; the same is
true for the other $\rho$ statistics.

The remaining panels of figure~\ref{fig:rho_statistics} show the survey-averaged $\rho$ statistics
compared with first-year survey requirements.  The results are consistently worse for the non-PSF
stars than for PSF stars, possibly due to inaccurate PSF interpolation or over-fitting; however, the
results are within our (relatively conservative) first-year requirements in all cases.  The
discrepancy between results with PSF stars and non-PSF stars is particularly striking for $\rho_2$
and $\rho_4$.  These are the only two of the $\rho$ statistics that incorporate a PSF model shape
correlated against a PSF model shape error.

We have the
least margin on $\rho_1$, which when measured with non-PSF stars, is close to our requirements
envelope for $\theta>0.3^\circ$.  This finding, combined with the mean PSF model size residual in
the previous subsection, suggests that an improved PSF modeling algorithm may
be needed to achieve our weak lensing science goals with the full HSC survey area.

\section{Shear catalog}\label{sec:shear}


In this section, we describe the shear catalog, including the
galaxy selection criteria (Section~\ref{subsec:selection}), its basic properties
(Section~\ref{subsec:basic-char}) and null tests
(Section~\ref{subsec:shear-tests}).  We
refer the interested reader to Appendix~\ref{app:catalog-def} for more details on the quantities
provided in the catalogs.

\subsection{Galaxy selection}\label{subsec:selection}

In this subsection, we discuss the galaxy selection criteria that are imposed after all area cuts
\rmrv{summarized} in Section~\ref{subsec:area}.  We describe these cuts and their purpose qualitatively, while in
Appendix~\ref{app:selection}, we give the exact flags and database columns to enable
reproducibility.  
Users who apply their own cuts that are less stringent than the ones given here should
be aware that the validation tests in this catalog were not performed for galaxies that failed the
cuts given below.

We begin with a set of cuts on HSC pipeline flags, where all of the below cuts are imposed in the
$i$ band \rmrv{and most of these are explained further in the HSC DR1 paper}.  First, we required that
\texttt{idetect$\_$is$\_$primary} be set; this flag is used to identify a single version of
  each astrophysical object, by rejecting duplicates due to overlap between different patches or tracts
  as well as still-blended parent objects (whose children are also measured). We also imposed  cuts  to avoid objects with problematic image processing, including the
following scenarios: extremely large groups of objects that the deblender skipped (usually caused by
bright stars); those with a failure in the centroiding algorithm; those too close to an image
boundary; those with a pixel very close to the object center that was flagged as interpolated,
saturated, a cosmic ray hit, otherwise bad, or just short of saturation (where nonlinearity
corrections are less certain); and those for which the shape measurement algorithm returned an error
code or a value of NaN for the shape measurement uncertainty.  Finally, we required that the object
have been classified as extended (not a point source) by the HSC pipeline.

In addition to flag cuts, we impose the following cuts on $i$-band object properties:
\begin{itemize}
\item The unforced cmodel $i$-band magnitude should be below $24.5$ after extinction correction.
  There are two reasons for this conservative cut.  First, our spectroscopic training sample used to
  calibrate photometric redshifts has very few objects fainter than this, which results in
  significant systematic uncertainty in the photo-$z$ calibration.  Second, the simulations that we
  use to calibrate our shear estimates are based on the \rmrv{HST} COSMOS survey, with galaxies with parametric
  model fits limited to $I<25.2$.  Given that the inserted galaxies have some scatter in their
  observed magnitudes, $24.5$ seems to be the practical limit before we start running out of
  galaxies due to the limitations of our parent sample.  In future years of HSC survey analysis, we
  hope to relax this cut and take a more full advantage of the depth of HSC data for our lensing
  shear catalogs.
\item The unforced cmodel\footnote{\rmrv{`cmodel' refers to composite model photometry that is
      estimated by fitting a linear combination of an exponential profile and de Vaucouleurs profile
      convolved with the PSF model to object light profiles
 \citep{Lupton:2001} as described in \citet{PipelinePaper:inprep}.}} $S/N$ should be $\ge 10$.  Note that given the magnitude cut
  \rmrv{above}, this cut is not very important in practice, as the \rmrv{$S/N$} distribution \rmrv{drops
    sharply} below
  \rmrv{$S/N=$20}.
\item Resolution factor (equation~\ref{eq:def-r2}) $R_2\ge 0.3$.  A completely unresolved
  object will have $R_2=0$, while a fully resolved one will have $R_2=1$.
\item Total \rmrv{magnitude of the distortion (after PSF correction) defined in
    equation~\eqref{eq:e} should satisfy} $|e|<2$.  Due to noise, the distribution of \rmrv{distortion}
  values extends into the non-physical $|e|>1$ regime.  Truncating the distribution too aggressively
  at $1$ leads to a negative shear bias; however, some truncation is needed to enable mean shear
  statistics to converge.
\item The catalog estimate of the shape measurement uncertainty due to pixel noise, $\sigma_e$,
  should lie in the range $[0,0.4]$.  This cut removes only a tiny fraction of highly anomalous
  objects, $<1$\% of those that pass the other cuts.
\item Multi-band detection cut, defined by requiring at least two other bands (out of $grzy$) to
  have a cmodel detection significance exceeding 5.  This cut removes a very small fraction of
  objects, $<1$\%, that pass our other cuts.  \rmrv{In addition to}
  ensuring enough color information to compute a photometric redshift, \rmrv{this cut} also helps remove
  junk detections, asteroids \citep{2017MNRAS.465.1454H}, and so on.
\item The blendedness parameter, \texttt{iblendedness$\_$abs$\_$flux}, which quantifies the
    relative contamination of the object light profile by the light from other nearby objects, should be less than
  $10^{-0.375}$.  \rev{As defined in section~4.9.11 in
    \citet{PipelinePaper:inprep}, the blendedness parameter would have a value of 0 for isolated
    objects by definition, while
  objects detected as being strongly blended would have a blendedness value approaching 1.  The cut
  value was determined in two stages.  First, an examination of the distribution of $R_2$ values in
  narrow bins in $\log_{10}{b}$ showed that above this value, the distribution of $R_2$ was
  skewed in an unphysical way towards very high resolution factor even for quite faint objects.
  Second, visual inspection of the objects above that cut value revealed that they} preferentially lie very near bright
  galaxies, and were either spurious detections or real detections of objects with completely
  unreliable photometry and shape measurement due to contamination by light from the nearby bright
  galaxy \citep{SynPipe:inprep}.  \rev{This cut value removes of order 1\% of the objects that would
  pass the other cuts, and should be considered a method of junk removal; it does not constitute an
  attempt to remove mildly blended objects from the catalog.} \rmrv{While a bug in the blendedness calculation was identified after
  introduction of this cut, as noted in \cite{PipelinePaper:inprep} and quantified in
  \cite{SynPipe:inprep}, the bug resulted in a very small remapping of the blendedness values that
  does not cause significant problems for our use of this cut value.}
\item  \jbrv{As noted in Section~\ref{subsubsec:coaddition}, the object must lie in
  a region where all overlapping exposures contributed to the coadd, so the
  coadded PSF model (which does not account for missing pixels within
  sensors) is correct. Due to a bug in \texttt{hscPipe}, this filtering was
  not complete; a small number of objects lying on CCD boundaries, sensor
  defects, or cosmic rays were not flagged by the pipeline and could not be
  removed in this cut. The internal PSF quality tests in Section~\ref{subsec:psf-tests}
  are sensitive to this problem, however, and demonstrate that
  its effects do not cause the PSF model errors to exceed our requirements.}
\end{itemize}

For the sake of scientific reproducibility, the publicly released shear catalog has a flag
indicating whether all lensing cuts are passed by each object.  Users do not have to separately try
to impose all flags and cuts on galaxy properties.

\begin{figure*}
\begin{center}
\includegraphics[clip,angle=-90,width=4in]{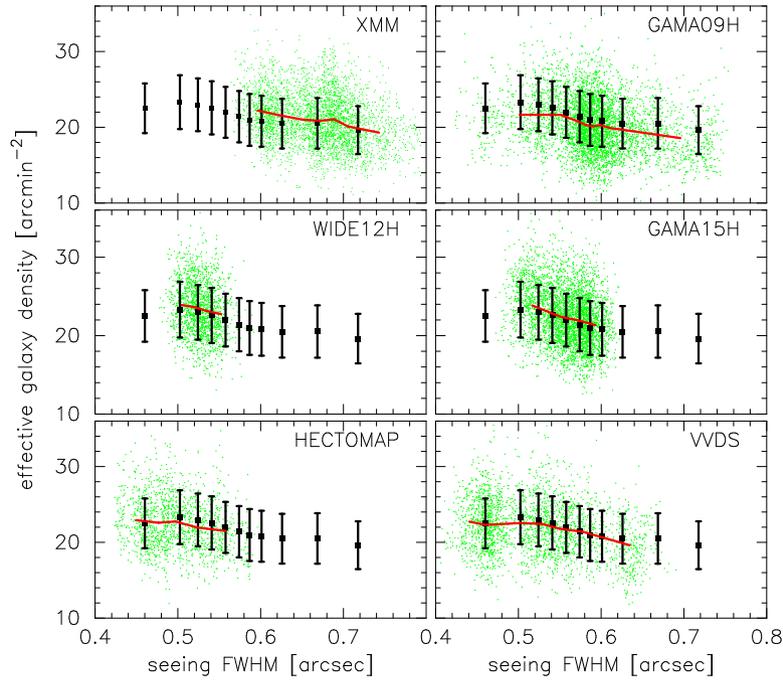}
\end{center}
\caption{The effective (weighted) galaxy number density as a function of seeing in each field.
  Green points are the density as computed on a regular grid with spacing of 0.5~arcmin on the
  tangent-projected sky using Gaussian smoothing with $\sigma\sim 1.06$~arcmin.  The red curves on
  each panel are the mean number density as a function of seeing in each field, while the black
  points with errorbars show the mean and standard deviation at fixed FWHM across the entire survey.}
\label{fig:seeing_ng}
\end{figure*}

\begin{figure*}
\begin{center}
\includegraphics[clip,width=6.3in]{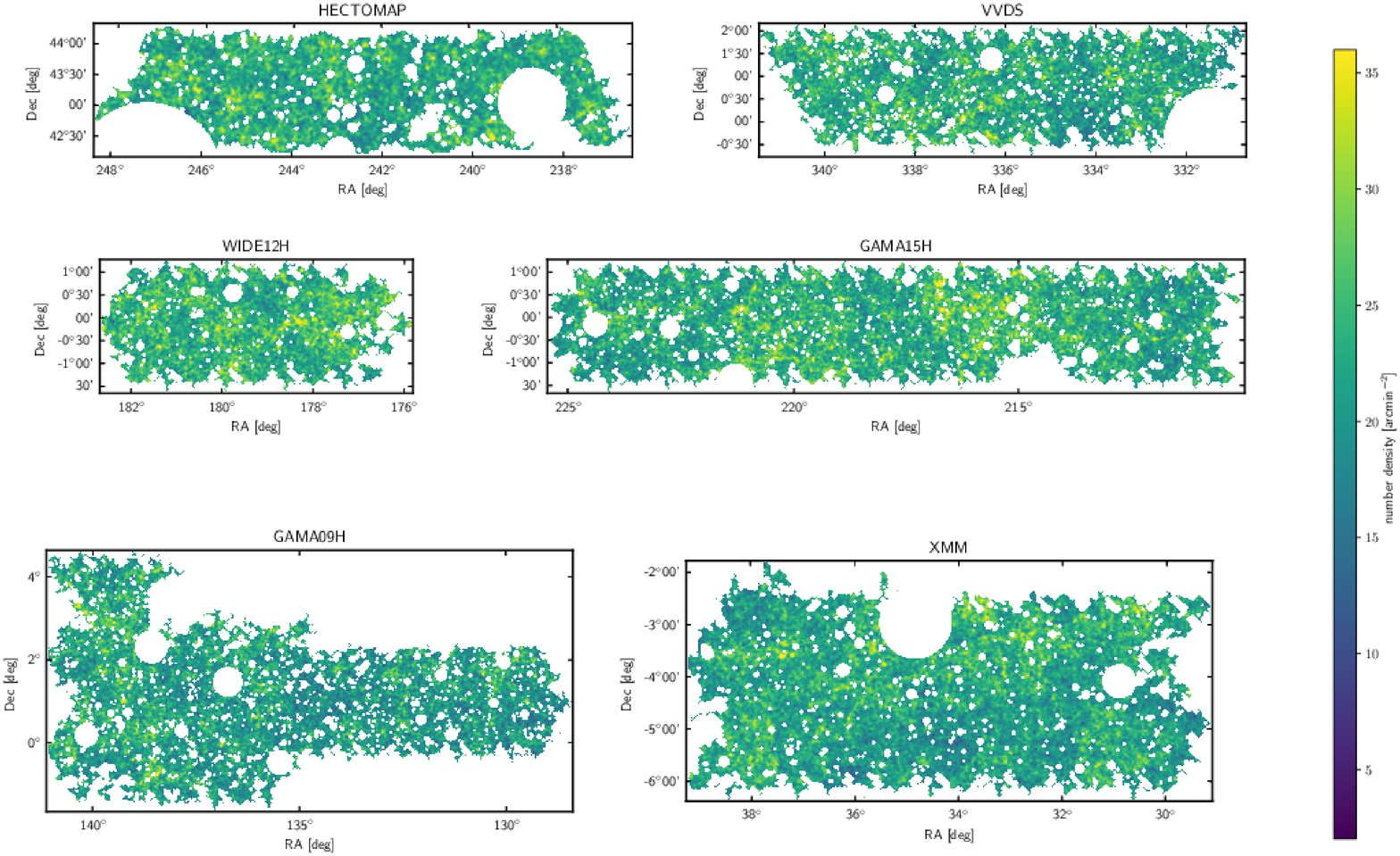}
\end{center}
\caption{Unweighted (raw) number density of sources passing all lensing cuts in each
  field.}
\label{fig:ng_map}
\end{figure*}

\begin{figure*}
\begin{center}
 \includegraphics[clip,width=6.3in]{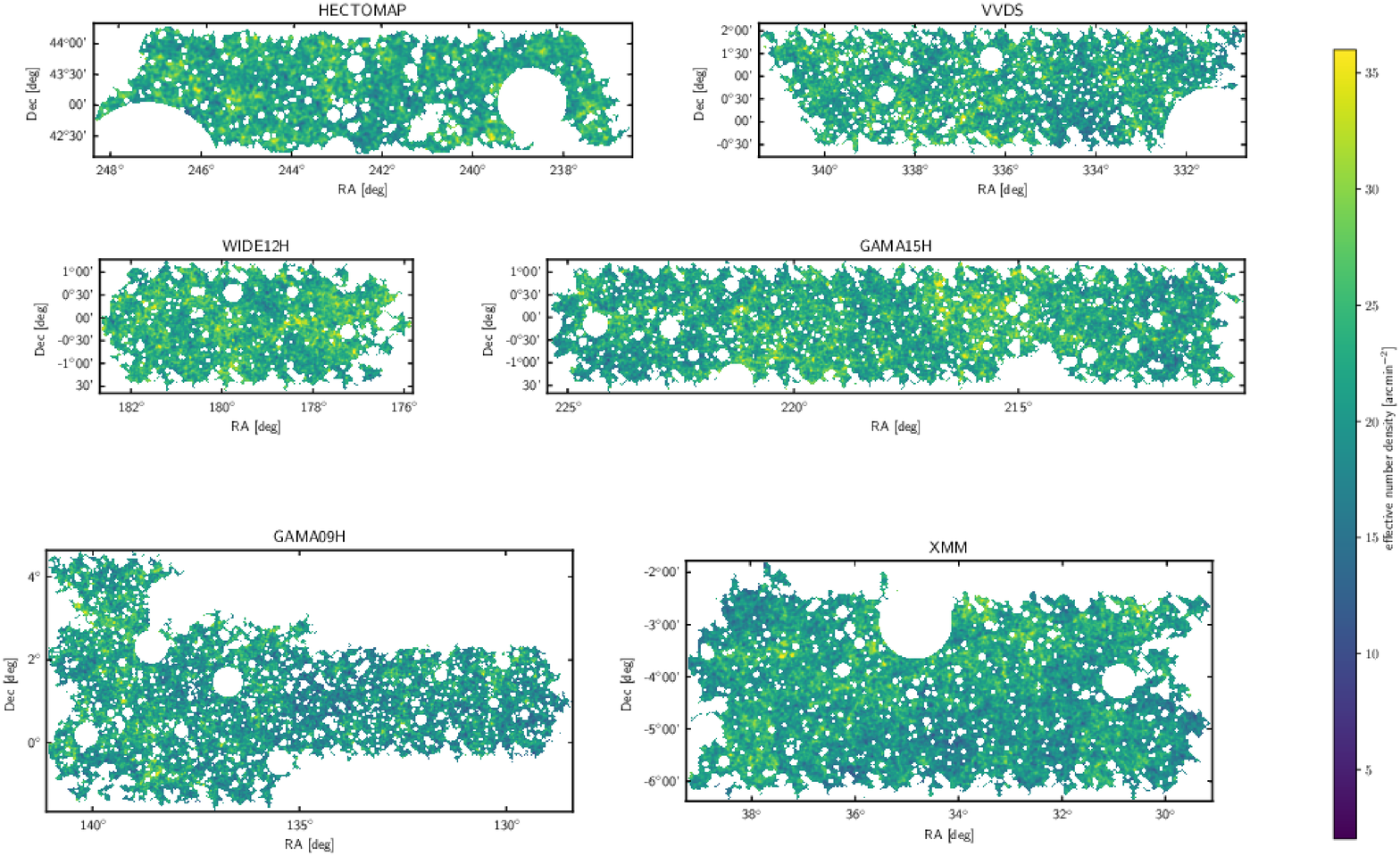}
\end{center}
\caption{Weighted number density of sources passing all lensing cuts in each
  field.}
\label{fig:ng_eff_map}
\end{figure*}

\begin{figure*}
\begin{center}
\includegraphics[clip,width=4.5in]{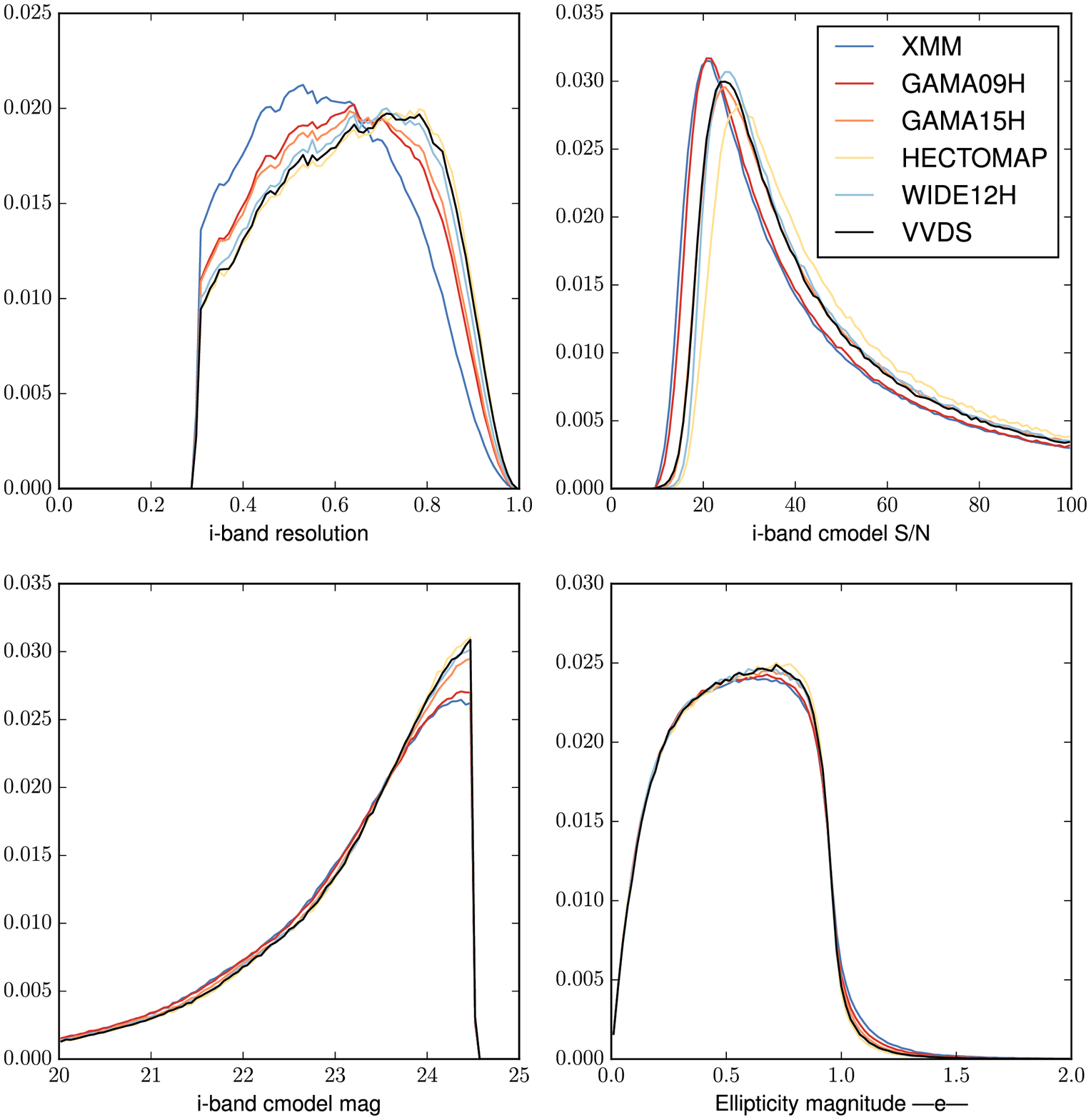}
\end{center}
\caption{Top left, top right, bottom left, and bottom right panels show the lensing-weighted distributions of galaxy
  $R_2$ values, $i$-band cmodel $S/N$, $i$-band cmodel magnitude, and total \rmrv{distortion} magnitude
  $|e|$ for each survey region after all selection criteria were imposed. \rmrv{The $|e|$
    distribution extends into the unphysical $|e|>1$ regime because $e$ is defined by taking the
    ratio of two noisy quantities, and thus noise fluctuations can create an unphysical tail.}
}
\label{fig:gal_prop}
\end{figure*}

  With $>12$ million galaxies in
the catalog after these cuts, and an area of 136.9~deg$^2$, the average source galaxy number density without any lensing weights is
24.6~arcmin$^{-2}$. Inclusion of lensing weights (described in Section~\ref{subsec:basic-char}) gives $n_\text{eff}=21.8$~arcmin$^{-2}$. Cuts on photo-$z$ quality or photo-$z$ values will reduce this number further,
but the level of reduction depends on the photo-$z$ algorithm.

\subsection{Basic characterization}\label{subsec:basic-char}

Here we present the basic characterization of the galaxy properties in
the shape catalog.

\begin{figure}
\begin{center}
\includegraphics[clip,width=3in]{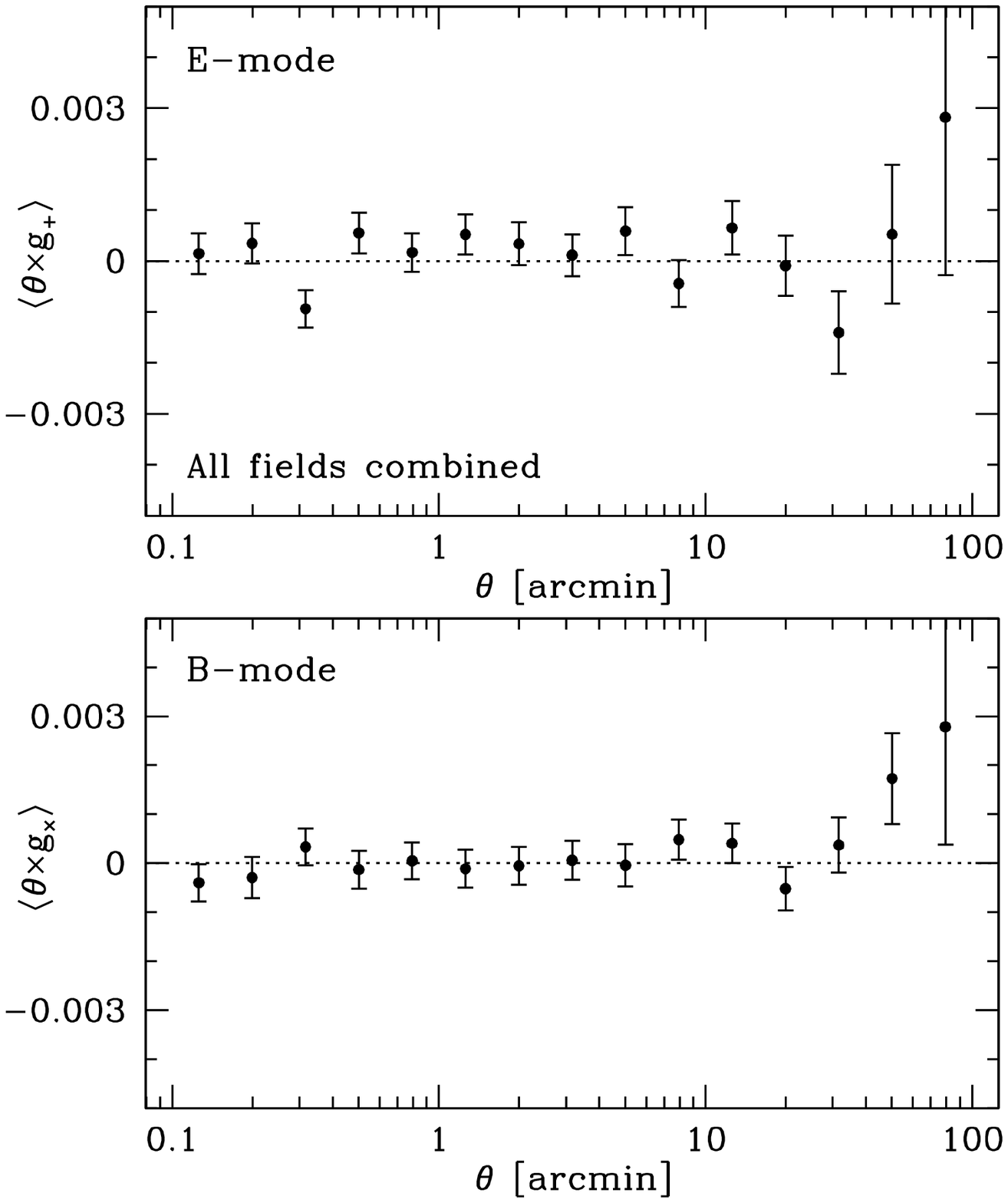}
\end{center}
\caption{Stacked tangential ({\it upper}) and cross ({\it lower})
  shear profiles around random points\rev{, averaged across the entire survey}. The number density of the random
  points is 50~deg$^{-2}$. Errors are estimated from mock catalogs
  which follow the same spatial distribution of galaxies as the HSC
  shear catalog and includes cosmic shear from ray-tracing simulations
  (see \citealt{Oguri:inprep} for more details). The $\chi^2$ and
  $p$-value for the null hypothesis \rev{for each field} are summarized in Table~\ref{tab:ranstack}.}
\label{fig:plot_ranstack_comb}
\end{figure}

Figures~\ref{fig:ng_map} and~\ref{fig:ng_eff_map} show the unweighted and weighted (respectively)
galaxy number densities after all selection criteria from Section~\ref{subsec:selection} are imposed on
the sample.  Here, the weight is the inverse variance weight based on the quadrature sum of shape
noise and measurement error, 
\beq
w = \frac{1}{\sigma_\text{SN}^2+\sigma_e^2}.
\eeq
Both quantities in this weight were estimated using simulations, see Section~\ref{sec:sims}.
Note that the weighted and unweighted number densities differ by only $\sim 12$\% because the shape
catalog does not go to very low $S/N$ detections.  The trend in number density as a function of
seeing is shown in figure~\ref{fig:seeing_ng} separately for each field.

Finally, figure~\ref{fig:gal_prop} shows the lensing-weighted distributions of several galaxy properties in each
survey field separately.  As shown, the distributions of resolution factor $R_2$ vary noticeably, in
a way that is consistent with per-field trends in seeing (figure~\ref{fig:seeing_map}).  For example,
the fields with the best (worst) seeing have the highest (lowest) resolution factors.  There are
similar trends for $S/N$, except that VVDS has only typical $S/N$ despite having nearly the best
seeing because of the omission of some exposures (figure~\ref{fig:nexp_map}).  However, the magnitude
and \rmrv{distortion} distributions are similar across all fields.



\begin{table*}[tp]
\caption{$\chi^2$ values and $p$-values for testing whether stacked $+$ and $\times$ shear profiles
  around random points \rev{(like figure~\ref{fig:plot_ranstack_comb} but also showing numbers separately for each field), $+$ and $\times$ shear profiles around
  bright stars in figure~\ref{fig:bright_star_shear}, and $\times$ shear profiles around
  CMASS galaxies in figure~\ref{fig:galaxy_cross_shear},} are consistent
  with zero signal. \rev{The last row shows $\chi^2$ values and $p$-values for all fields combined.
  The $\chi^2$ values for the first two columns have 15 degrees of freedom, while those for the last
three columns have 20 degrees of freedom.}}\label{tab:ranstack}
\begin{center}
\begin{tabular}{lccccc}
\hline
Field & Random $g_+$ & Random $g_\times$ & Bright star $g_+$ & Bright star $g_\times$ & CMASS $g_\times$ \\
 & $\chi^2$ ($p$-value) & $\chi^2$ ($p$-value) & $\chi^2$ ($p$-value) & $\chi^2$ ($p$-value) & $\chi^2$ ($p$-value) \\ \hline
GAMA09   & 22.80 (0.09) & 15.11 (0.44) & 32.24 (0.04) & 22.73 (0.30) & 27.17 (0.13) \\
GAMA15   & 14.34 (0.50) &  7.30 (0.95) & 21.41 (0.37) & 11.25 (0.94) & 49.81 (0.00) \\
HECTOMAP & 15.95 (0.39) & 15.41 (0.42) & 28.39 (0.10) & 17.91 (0.59) & 18.69 (0.54) \\
VVDS     & 24.25 (0.06) & 14.99 (0.45) & 14.32 (0.81) & 30.94 (0.06) & 12.58 (0.90) \\
WIDE12H  & 29.24 (0.02) & 15.61 (0.41) & 23.29 (0.28) & 16.21 (0.70) & 14.94 (0.78) \\
XMM      &  9.79 (0.83) & 11.09 (0.75) & 25.08 (0.20) & 27.15 (0.13) & 17.54 (0.62) \\
\rev{All fields combined} & \rev{20.04 (0.17)} & \rev{11.65 (0.71)} &
\rev{28.04 (0.11)} & \rev{27.35 (0.13)} & \rev{27.66 (0.12)} \\
\hline
\end{tabular}
\end{center}
\end{table*}

\begin{figure}
\begin{center}
\includegraphics[width=3in]{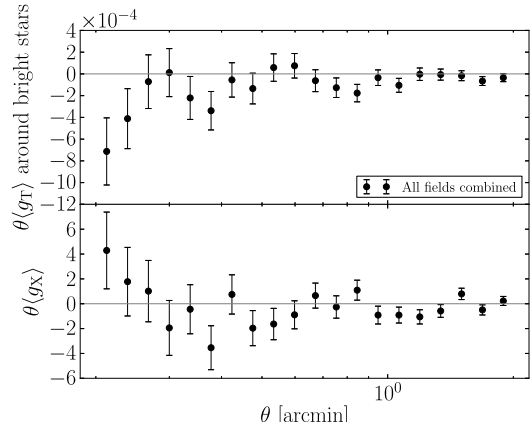}
\end{center}
\caption{Average tangential and $\times$ shear around bright stars \rev{across the entire survey}, defined as all stars with
  $i$-band magnitude $\le 22.5$ (and passing other cuts as defined in
  Section~\ref{subsec-starsamples}). We restrict ourselves to small scales here so that we can
  investigate the impact of nearby stars on shear estimates, for example due to their inducing
  errors in sky background estimation.}
\label{fig:bright_star_shear}
\end{figure}

\begin{figure}
\begin{center}
\includegraphics[width=3in]{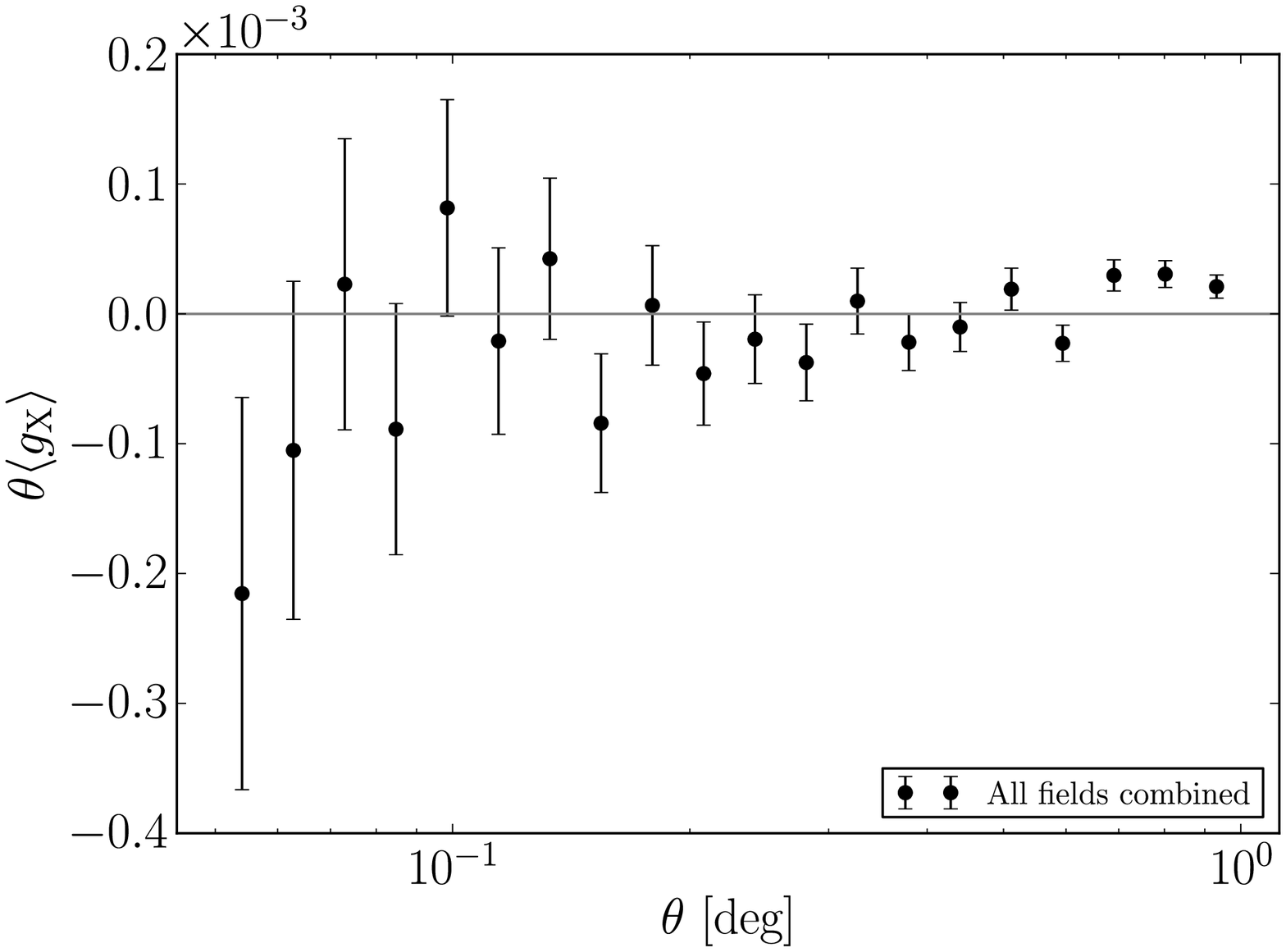}
\end{center}
\caption{Average cross shear around SDSS CMASS galaxies \rev{across the entire survey} at all redshifts. \rmrv{This quantity should
  be zero due to symmetry.}}
\label{fig:galaxy_cross_shear}
\end{figure}

\subsection{Null tests}\label{subsec:shear-tests}


In this section, we present a set of null tests for the galaxy shape catalog.   Note that all null
tests include statistical corrections for additive and multiplicative biases following a formalism
similar to Appendix~\ref{app:shear}.

\rev{As a first test, we calculate the mean shear estimates $\langle g_1\rangle$ and $\langle
  g_2\rangle$ within each of the six survey fields in sky coordinates, which are quite close to CCD
  coordinates for most of our fields. To ascertain the significance of any non-zero values, we must
  estimate errorbars including not only shape noise but also cosmic shear using many realizations of
  mock shear catalogs based on $N$-body simulations with  the same area coverage as the HSC survey.
Of the twelve numbers calculated (weighted mean values of two shear components in six fields), we
compute the $p$-value for a fit to zero signal.  Only one of the twelve $p$-values is 
below a nominal threshold of 0.05, and that one is still quite marginal ($p=0.03$), giving no reason
to suspect a systematic from this test.}

Figures~\ref{fig:plot_ranstack_comb} and~\ref{fig:bright_star_shear} show the stacked $+$ and
$\times$ shear signals as a
function of angular separation from random points and bright stars, respectively.  The first of
these goes to scales of 100~arcmin, to investigate large-scale systematic shear (which is revealed
as the angular scales get large enough that some of the annuli around the random points are
incomplete).  The second is on small scales only, and the goal in this case is to test for possible
apparent tangential or radial shears due to sky background misestimation near bright objects.
Finally, figure~\ref{fig:galaxy_cross_shear} shows the $\times$ shear component around BOSS CMASS
galaxies out to large scales. As shown, there
is no evidence for any statistically significant detection
for any of our survey fields, over any range of scales.  The $\chi^2$ and $p$-values for a
fit to zero signal are shown in Table~\ref{tab:ranstack}. 




\begin{figure}
\begin{center}
\includegraphics[clip,width=3in]{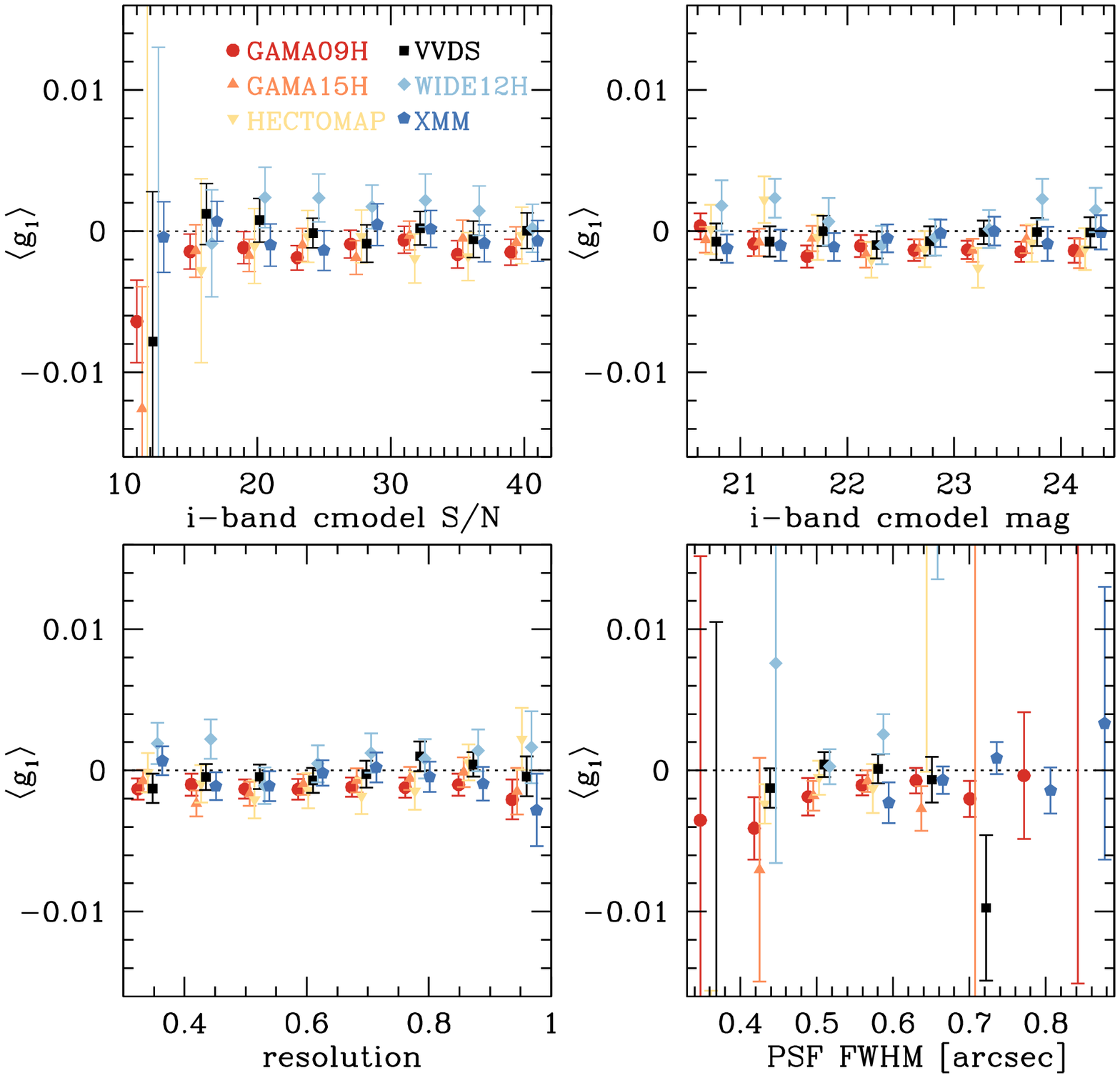}
\end{center}
\caption{Average $g_1$ values as a function of various parameters;
  $i$-band cmodel $S/N$ ({\it upper left}), $i$-band cmodel magnitude
  ({\it upper right}), the resolution factor, ({\it lower left}), and
  PSF FWHM ({\it lower right}).  Here $g_1$ is defined in the sky coordinates\rmrv{, which are very
    close to CCD coordinates for most of our fields}.
  As in figure~\ref{fig:plot_ranstack_comb}, errorbars are from mock
  shear catalogs and therefore include cosmic shear.
\label{fig:plot_eave_field_e1}}
\end{figure}

\begin{figure*}
\begin{center}
\includegraphics[width=3.2in]{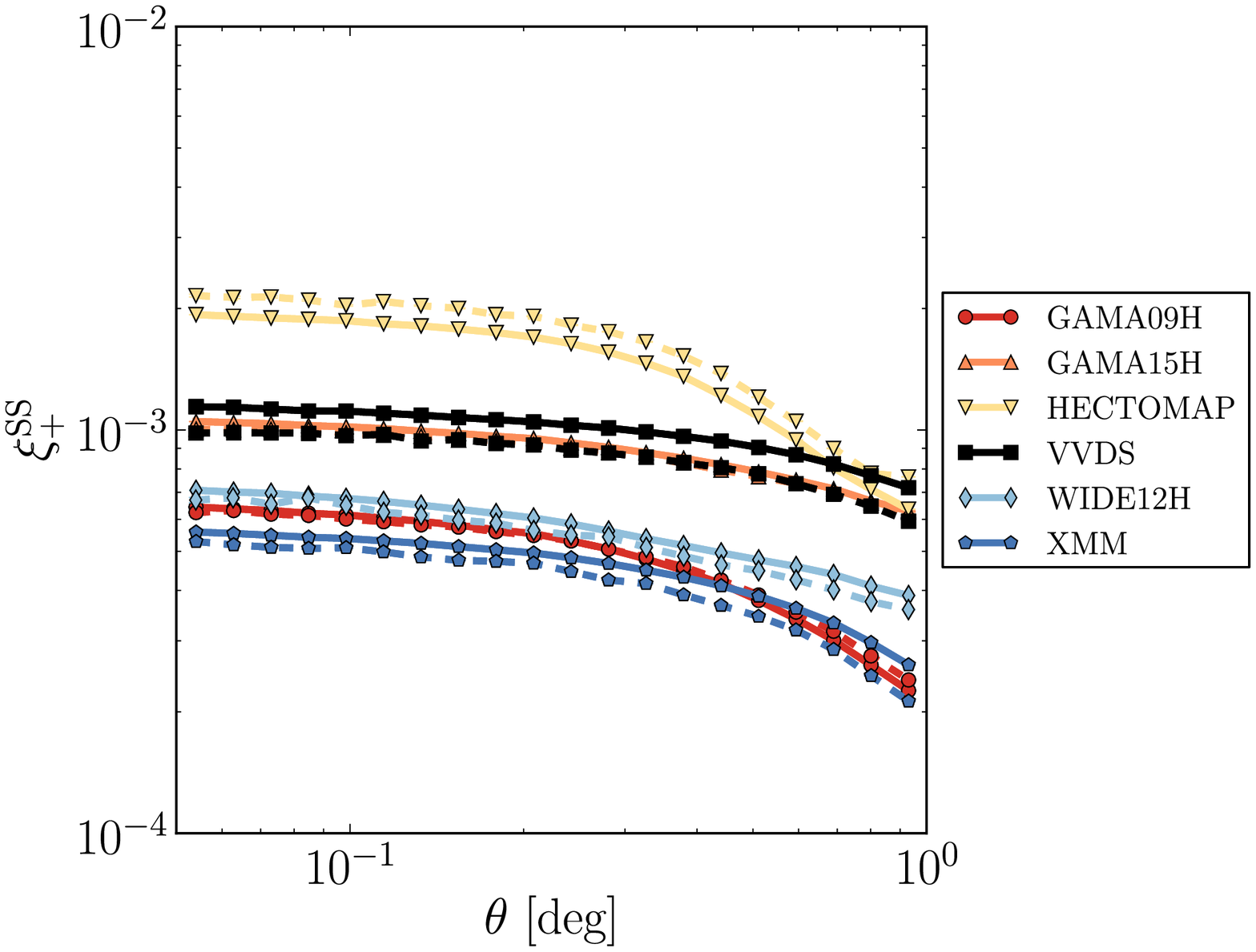}
\includegraphics[width=3.2in]{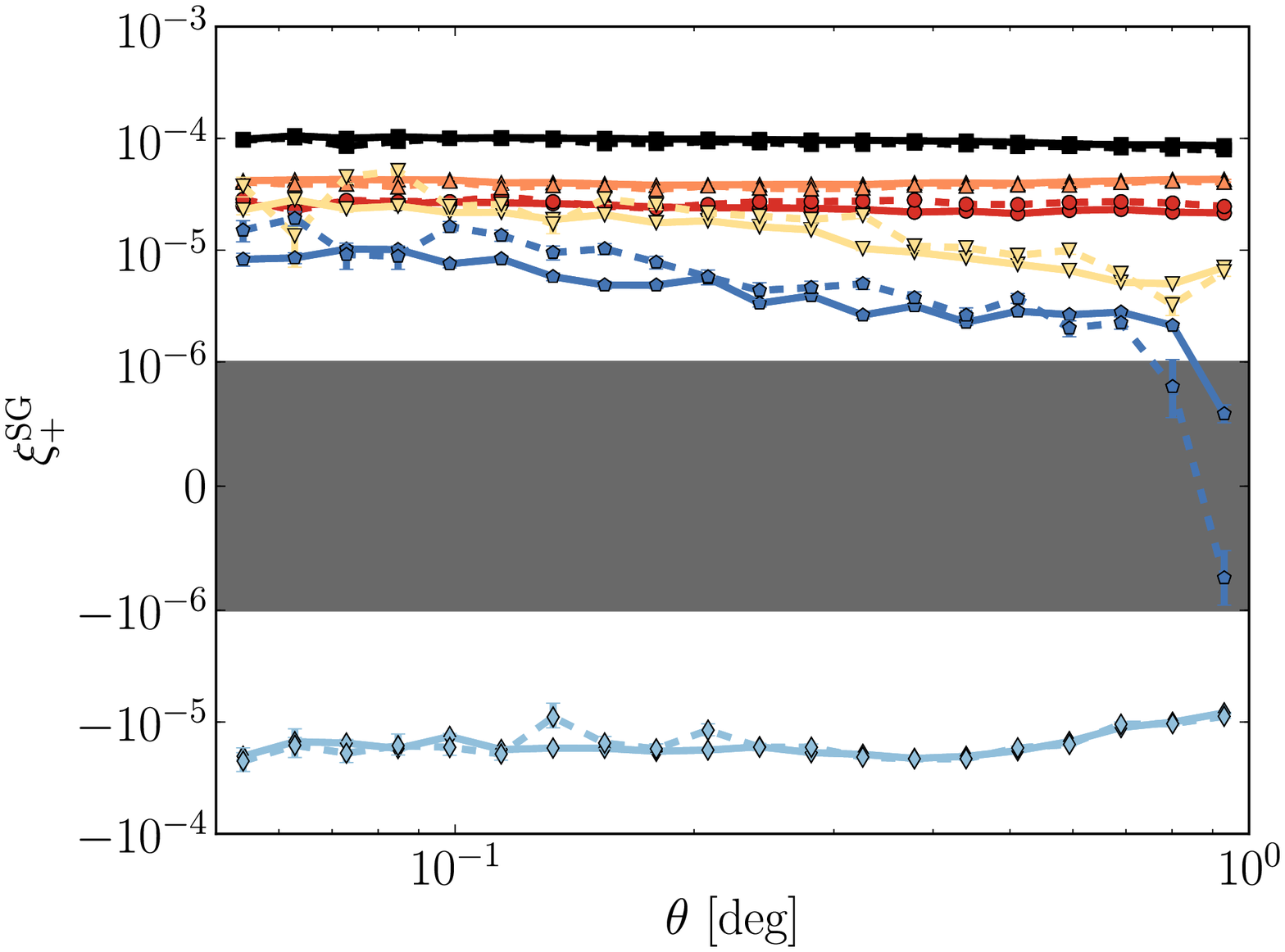}
\end{center}
\caption{Star-star (left) and star-galaxy (right) shear correlation functions $\xi_+$.  \rmrv{Solid
    and dashed lines were calculated using PSF and non-PSF stars, respectively.} Error bars are
    shown, but are generally smaller than the size of the plotted points.  \rev{For this plot, we
      have only shown $\xi_+$ because it carries nearly all the useful information.  However, we
      note that $\xi_-$ is consistent with zero for all fields on the scales on 
      which we have shown $\xi_+$}, except for HECTOMAP and
    VVDS.  HECTOMAP has a slight negative $\xi_-$ at scales $>0.7^\circ$: $\le 10^{-5}$ for the
star-galaxy correlation and $\le 10^{-4}$ for the star-star correlation, well below the $\xi_+$
    measurement.  VVDS has a slight negative $\xi_-$ of $\le 10^{-5}$ in the star-star correlation
    only at scales $>0.7^\circ$.}
\label{fig:ss-sg}
\end{figure*}

Figure~\ref{fig:plot_eave_field_e1}
shows $\langle g_1\rangle$ \rmrv{in sky coordinates} 
as a function of four properties of the $i$-band
images: cmodel $S/N$ and magnitude, galaxy size, and PSF FWHM.  \rmrv{Note that for most of our
  observations, sky coordinates are very close to CCD coordinates, so this plot has the potential to
reveal systematics that correlate with the pixel directions.} Results are similar for the
other shear component (not shown).
We find that the average shear values are consistent with zero, and the average shear values do not
show any strong dependence on galaxy properties explored here. In some fields (e.g., GAMA09H and
HECTOMAP), average shear values are persistently positive or negative in almost all bins.
This is most likely due to the cosmic variance (cosmic shear) which produces correlated residuals
between different bins of the galaxy properties.  Bin-to-bin correlation coefficients range from
0.3--0.6.  Other possible explanations for a nonzero mean shear signal, such as selection bias or
incomplete correction for PSF anisotropy, would typically result in some dependence on galaxy
properties -- inconsistent with what is shown in this plot.


As an indication of PSF anisotropy leakage into galaxy shapes (combined with PSF shape modeling
errors giving additive errors), figure~\ref{fig:ss-sg} shows the star-star and star-galaxy shape
correlation functions in each survey field, as measured using the PSF and non-PSF stars.  \rmrv{In
  both panels, the star shapes are used directly without correction for the PSF, because the goal is
to ascertain what fraction of the original PSF \rmrv{shape} (as traced by star shapes) leaks into the
galaxy shapes.} First we
consider the left-hand panel, the star-star correlations, where only $\xi_+$ (Eq.~\ref{eq:xipm}) is shown.  These curves
are fairly flat over separations of a degree, indicating that the PSF \rmrv{shape} exhibits slow
spatial variations in the coadd.  The magnitude of the curves, from $2\times 10^{-3}$ for HECTOMAP
down to $5\times 10^{-4}$ for XMM, reflects the typical PSF \rmrv{shape} magnitudes from 0.05 down to
0.02 in these fields.  \rev{The inverse correlation between seeing size and typical PSF shape may be
  caused by either increased contributions from optical distortions \citep{CameraPaper:inprep} in the very best seeing, or the
  fact that the amplitude of the atmospheric PSF ellipticity itself is inversely proportional to PSF
  size \citep{2013PASJ...65..104H}.}

One possible cause for a residual correlation between the shapes of stars and the PSF-corrected
galaxy shapes is use of an insufficient PSF correction method.  In the
simplest case, where the measured \rmrv{ensemble shear} is a linear \rmrv{combination} of the true \rmrv{shear} and the PSF shape
\rmrv{due to residual PSF anisotropy in the galaxy shapes},
$\langle \hat{g}_{\rm gal}\rangle = (1+m) \langle g\rangle +\langle a g_{\rm PSF}\rangle$, we should find a star-galaxy correlation that looks like
\begin{equation}\label{eq:star-galaxy-correlation}
\langle g_*\hat{g}_{\rm gal} \rangle = (1+m)\langle g_* g_{\rm true}\rangle
+a \langle g_*g_{\rm PSF}\rangle \sim a\langle g_* g_*\rangle,
\end{equation}
meaning that the star-galaxy correlation function should be simply a rescaled version of the star-star
correlation function with the same scale-dependence.  Examining the correlation functions presented in figure~\ref{fig:ss-sg},
however, we see that this simple rescaling does not hold.  Moreover, the relationship between the
amplitudes of the star-star and star-galaxy correlations changes from field to field (not only in
magnitude, but also in sign).  We therefore conclude that the prescription in
equation~\eqref{eq:star-galaxy-correlation} must be an incomplete description of the star-galaxy
correlations, with some other systematic error contributing.  One candidate is a contribution from
PSF modeling errors, which would give another term proportional to $\rho_2$ (shown in figure~\ref{fig:rho_statistics}).

Next, we \rev{carry out an empirical test for the possible impact of either PSF ellipticity
  modeling errors or residual PSF anisotropy in galaxy shapes}
 on cosmic shear two-point correlation function measurements.
In the same prescription as equation~\eqref{eq:star-galaxy-correlation},
a residual correlation caused by PSF anisotropy leakage
can be modeled by $\Delta  \langle g g \rangle \sim a^2 \langle g_\ast
g_\ast\rangle$, and can be evaluated by the combination of
\begin{equation}
\xi_{\rm
 sys} \equiv \frac{\langle g_*\hat{g}_{\rm gal} \rangle^2}{\langle g_* g_*\rangle}.
\end{equation}
\rev{PSF ellipticity modeling errors would also contribute additive terms, as discussed in
  Sec.~\ref{subsubsec:psfmodele-ss}.  This null test therefore detects both types of
  systematic errors combined.}
Figure~\ref{fig:xisys_field} shows $\xi_{\rm sys}$ for each field along
with the standard $\Lambda$CDM prediction of $\xi_+$\rmrv{, the galaxy shear correlation function.}
The amplitude of $\xi_{\rm sys}$ varies among fields as expected from
the above findings, and can be comparable to $\xi_{\rm gg}$ on degree
scales, suggesting that marginalization over a template for $\xi_{\rm sys}$ may be needed on those
scales.
As was shown above, the amplitude of PSF anisotropy leakage may
depend \rmrv{weakly} on galaxy properties \rmrv{(primarily resolution factor)}, suggesting that an appropriate galaxy
selection may reduce the amplitude of $\xi_{\rm sys}$.  However, selection bias itself can cause
some nonzero $\xi_{\rm sys}$, as can PSF modeling errors (illustrated through our small but non-zero
$\rho$ statistics).  \rev{Having identified this as a systematic that is likely to be important on
  the scales we would like to use for our analysis,  we defer the detailed} exploration of how to model this effect to reduce its impact on
cosmological \rev{lensing} analysis
to future work. \rev{However, note that standard template marginalization schemes are a promising
  candidate for mitigation of this effect, given that its scaling with $\theta$ differs strongly
  from the scaling of cosmological lensing signals with $\theta$.}

\begin{figure}
\begin{center}
\includegraphics[clip,width=3in]{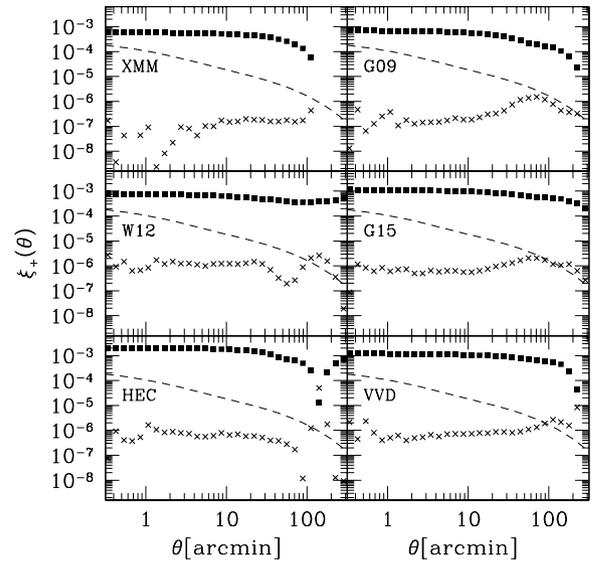}
\end{center}
\caption{Separate panels show  (for each survey field) the shape-shape correlation
  function $\xi_+(\theta)$ for PSF star shapes as points with \rmrv{errorbars}; the predicted cosmic shear
  correlation function with a WMAP9 cosmology using the $n(z)$ from HSC photometric
  redshifts without any correction for photo-$z$ errors (which illustrates the approximate magnitude
  of the expected cosmic shear signal) as dashed lines; and $\xi_\text{sys}$, defined as
  $\xi_\text{sg}^2/\xi_\text{ss}$, as crosses.}
\label{fig:xisys_field}
\end{figure}

\begin{figure*}
\begin{center}
\includegraphics[clip,width=4in]{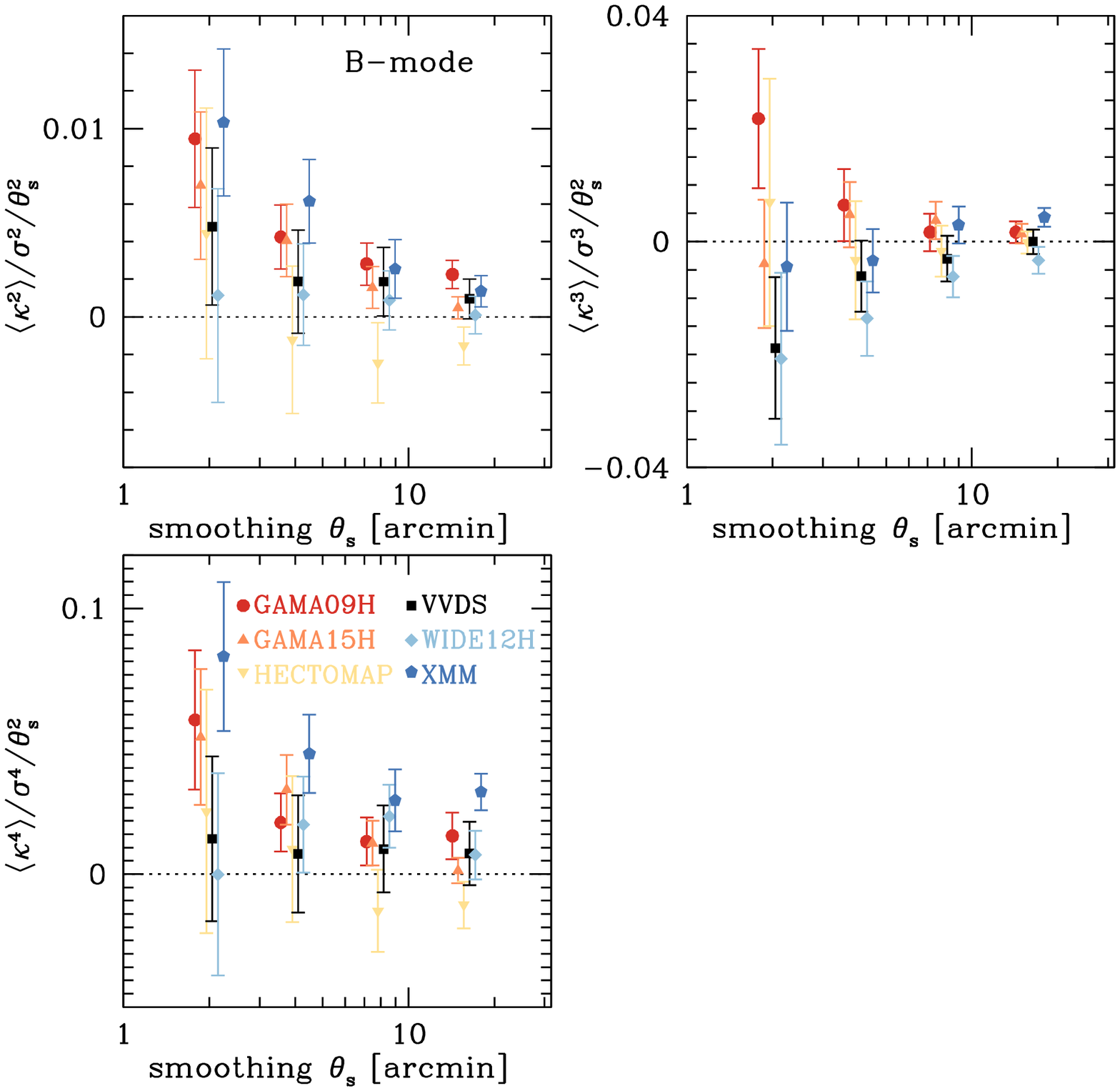}
\end{center}
\caption{Moments of B-mode mass map PDFs as a function of smoothing
  scale $\theta_s$. Mass maps are created using the standard \citet{1993ApJ...404..441K}
  inversion technique (see \citealt{Oguri:inprep} for
  more details). We show second ({\it upper left}), third ({\it
    upper right}), and fourth ({\it lower left}) moments, which are
  ``de-noised'' by subtracting moments originating from the shot noise
  (see, e.g., \citealt{2013MNRAS.433.3373V}). Since the boundary of the
  survey region also induces some B-mode in mass maps, we subtract the
  average moments from the mock catalogs from the observed moments
  to correct for the boundary effect \rev{(see text for more
    details)}. Errors are also estimated from the mock
  catalogs. \rmrv{In the absence of systematic errors, these
    quantities should all be  consistent with zero.}}
\label{fig:plot_mom_fid_bmode}
\end{figure*}

\rmrv{
 We can convert an observed shear field to a projected mass density
 (convergence) field because both are second derivatives of the
 gravitational potential \citep{1993ApJ...404..441K}. Weak
 gravitational lensing produces mostly ``E-mode'' convergence fields,
 whereas many potential systematic effects produce both ``E-mode'' and
 ``B-mode'' convergence fields. As a result, B-mode convergence fields
 (mass maps), which are produced by curl-like patterns in the shear
 field, are used to check for the presence of certain residual
 systematics in weak lensing shear catalogs \citep[e.g.,][]{2010RPPh...73h6901M}.
While not all systematics will produce a B-mode signature, this is nonetheless an important null
test.
 Here we present some results of our mass map analysis. Additional
 mass map analysis and further details about the mass map production is presented in
 \citet{Oguri:inprep}.
}

Figure~\ref{fig:plot_mom_fid_bmode} shows the second, third, and fourth moments of the B-mode
mass map probability distribution function (PDFs) as a function of smoothing scale, separately
for each survey field. These moments are
``de-noised'' by subtracting moments originating from the shot noise (see, e.g.,
\citealt{2013MNRAS.433.3373V}), and should therefore be consistent with
zero in the absence of systematics. 
\rev{It is known that boundary and masking effects also produce
  non-zero moments in the B-mode mass map PDFs, even after subtracting
  the shot noise contribution. We account for these effects using
  mock shear catalogs \citep{Oguri:inprep}. The mock shear catalogs
  have the same spatial distribution of galaxies as the HSC shear
  catalog, but their ellipticity values are replaced with simulated
  values that include the intrinsic ellipticity and cosmic
  shear from ray-tracing simulations \citep{2017MNRAS.470.3476S,2017arXiv170601472T}. 
  Since the mock shear catalogs have the same masking pattern and 
  spatial inhomogeneity as the HSC shear catalog, we remove the
  boundary and masking effects by subtracting the average moment values
  computed using the mock shear catalogs from those computed using the real HSC shear
  catalog.} 
As shown in this figure, the moments 
\rev{after the correction for the boundary and masking effects}
are mostly consistent with zero within $\sim 2\sigma$ level. This
means that the B-mode mass map PDFs are sufficiently close to Gaussian,
as expected \rmrv{in the case of no systematic effects}.
Small deviations from zero must originate
from PSF leakage as examined above, or PSF modeling errors as we will discuss below.


\rmrv{Mass maps can also provide a complementary check of the PSF
  leakage and PSF modeling errors.}
Figure~\ref{fig:plot_cc_fid_field} provides an alternative view of star-galaxy cross
correlations. We cross-correlated mass maps and ``star mass maps'' which are mass maps
constructed using star ellipticities \rmrv{\citep[for more detail, see][]{Oguri:inprep}}. We then quantify the correlation of two maps,
$\kappa_1(\boldsymbol{\theta}_i)$ and
$\kappa_2(\boldsymbol{\theta}_i)$ (\rmrv{which are normalized to zero mean, i.e.,}
$\langle \kappa_1\rangle=\langle \kappa_2\rangle=0$), using
the Pearson correlation coefficient $\rho_{\kappa_1\kappa_2}$ defined as
\begin{equation}
  \rho_{\kappa_1\kappa_2}
  =\frac{\sum_i\kappa_1(\boldsymbol{\theta}_i)\kappa_2(\boldsymbol{\theta}_i)}
  {\left[\sum_i\left\{\kappa_1(\boldsymbol{\theta}_i)\right\}^2\right]^{1/2}
    \left[\sum_i\left\{\kappa_2(\boldsymbol{\theta}_i)\right\}^2\right]^{1/2}},
\label{eq:pearson}
\end{equation}
\rmrv{where $\boldsymbol{\theta}_i$ specifies the sky position of each
pixel of the map, and the index $i$ runs over all the pixels of the map.}
For the star mass maps, we consider two cases, one with star ellipticities, and the other
case with star ellipticities after the PSF correction. If our PSF corrections are perfect,
we will have no residual star ellipticities after PSF corrections. Therefore, the latter case
explores \rmrv{the} potential impact of PSF modeling errors (see figure~\ref{fig:psf_residual_size_histmagcorr})
on weak lensing analysis. We find that the Pearson correlation coefficients are \rmrv{within
  $2\sigma$ of zero} for most cases. An exception is the correlation of B-mode
mass maps and B-mode star mass maps after PSF corrections, which show deviations larger than
$2\sigma$ in some cases.  This means that PSF modeling errors affect B-mode mass maps,
which is one of the sources of small deviations of moments of B-mode mass map PDFs shown in
figure~\ref{fig:plot_mom_fid_bmode}. We perform a more thorough analysis of cross-correlations
of mass maps and maps of potential sources of systematic effects in \citet{Oguri:inprep}.




\begin{figure*}
\begin{center}
\includegraphics[clip,width=4in]{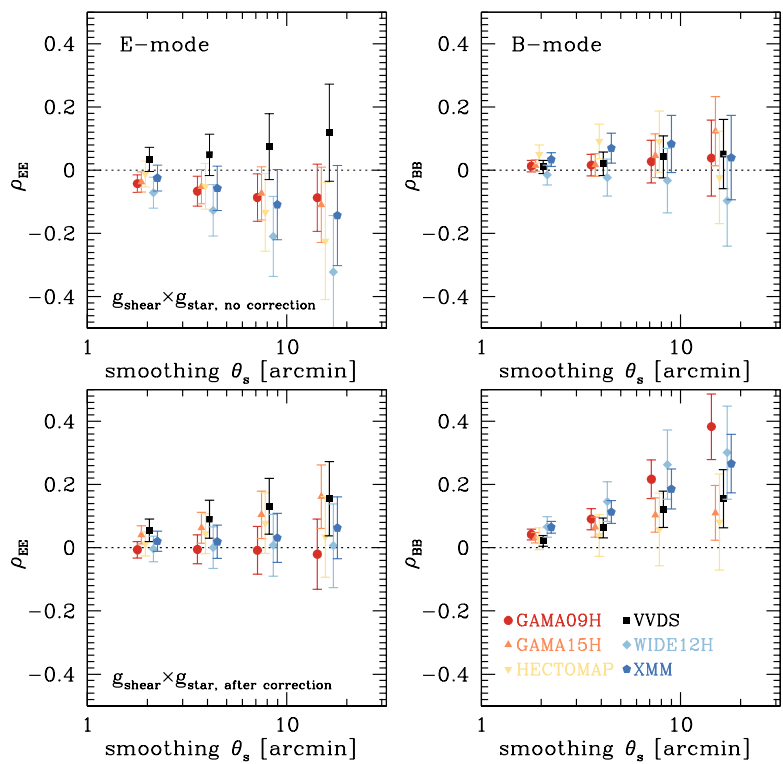}
\end{center}
\caption{Pearson cross-correlation coefficients (equation~\ref{eq:pearson}) of mass maps and
  star maps constructed using star ellipticities. For star ellipticities, we consider
  both cases with and without the PSF correction (which is done simply by
  subtracting the model PSF \rmrv{shape} from star \rmrv{shapes}, i.e.,
  $g_{\rm star}-g_{\rm PSF}$). Here we use all stars in both samples from
  Section~\ref{subsec-starsamples} for the analysis.}
\label{fig:plot_cc_fid_field}
\end{figure*}

\section{Simulations}\label{sec:sims}

In this section, we summarize the results of characterizing the shear catalog using simulated
\rmrv{images}.  Two \rmrv{image} simulation pipelines were used to address different problems, as
described in the subsections below.  \rmrv{These are completely distinct from the catalog-level
  simulations used to produce covariances in Section~\ref{sec:requirements}.  While the results
  described below primarily serve as short summaries of the conclusions of separate papers, they are
  important to mention in this paper because the image simulations are our primary means to answer
  several critical questions about the shape catalog that cannot be directly or cleanly answered
  through other means: What is the rate of \revrev{unrecognized} blends?   Do the measured shapes exhibit
  coherent alignments with bright objects due to image processing errors such as sky subtraction
  errors?  What are the primary sources of shear calibration biases, and how can we quantify and
  correct for them?  These are such critical elements of weak lensing science that we summarize the
  results of the simulation work and their main conclusions here.}

\subsection{Synthetic Object Pipeline}\label{subsec:synpipe}


The HSC Synthetic Object Pipeline\footnote{https://github.com/dr-guangtou/synpipe} (hereafter
\texttt{SynPipe}; \citealt{2017arXiv170501599H,SynPipe:inprep}) is a module that interfaces with
\texttt{hscPipe} and which can be used to insert synthetic objects into real HSC images. This is a
versatile simulation module that can be used for a variety of purposes such as evaluating the
completeness of a survey and testing the robustness of photometry measurements. Of relevance to this
paper, we use \texttt{SynPipe} to evaluate the overall performance of \texttt{hscPipe}, to estimate
the level of galaxy blending within the HSC Wide survey, to perform independent cross-checks on the
results from our GREAT3-like simulations (Section~\ref{subsec:great3like}), and finally to evaluate the number of background galaxies
lost to blending and masking effects around bright foreground galaxies \citep{Medezinski:inprep}.

The details of \texttt{SynPipe} are provided in \citet{2017arXiv170501599H}, and are briefly summarized
here.  \texttt{SynPipe} uses \texttt{GalSim} \citep{2015A&C....10..121R} v1.4 as a backend to
reliably create realizations of synthetic objects. These can be stars, galaxies described by single-
or double-{S\'{e}rsic} models, or galaxies described by the sum of a de Vaucouleurs and an
exponential profile. \texttt{SynPipe} is similar in spirit to the \texttt{BALROG} simulation package
\rmrv{used for the DES survey as} described by \citet{2016MNRAS.457..786S}, with two main differences. The first is that
\texttt{SynPipe} interfaces with \texttt{hscPipe} and can therefore be used to create synthetic
objects with properties measured in the same way as HSC data. Second, unlike \texttt{BALROG}, which
inserts synthetic galaxies into co-added images, \texttt{SynPipe} operates by injecting synthetic
objects directly into images at the single exposure level. Synthetic objects processed though
\texttt{SynPipe} therefore go through the same detection, stacking, and
measurement process as real stars and galaxies observed by HSC.

HSC is the deepest of \rmrv{the} existing wide field lensing surveys. This means that HSC source galaxies will
be subject to blending effects that may impact photo-$z$ and shear estimates and are hence a source
of systematic error. In \citet{SynPipe:inprep}, we use \texttt{SynPipe} to \rmrv{attempt to
quantify} the effects of galaxy blends for the HSC wide survey. We refer the reader to
\citet{SynPipe:inprep} for the full details of our study, and briefly summarize our main findings
here. 
Synthetic galaxies are classified into systems that are ``cleanly'' recovered versus those which are
subject to \revrev{unrecognized} blending (this occurs when \texttt{hscPipe} was unable to deblend a synthetic
galaxy from a neighbouring real galaxy \revrev{due to the failure to recognize multiple peaks in the
  object}). 
After imposing weak lensing cuts, we find that 
the \revrev{unrecognized} blend rate increases
because \revrev{unrecognized} blends have increased flux and apparent size, leading them to
scatter into our sample across the $i=24.5$ and $R_2=0.3$ cuts, which are the most relevant cuts
\rmrv{for}
defining our sample (see figure~\ref{fig:gal_prop}).


\rev{It is important to note that the simulations from Section~\ref{subsec:great3like} that are used
to quantify shear systematics (additive and multiplicative biases) include a realistic rate of
\revrev{unrecognized} blends.  This realistic \revrev{unrecognized} blend rate is achieved by including galaxies from
space-based imaging with all surrounding structures, whether galaxies or stars, and relying on the
HSC pipeline to detect, deblend, and select galaxies for weak lensing analysis.  The simulations
were then used to derive our multiplicative and additive shear bias corrections.  Hence the impact
of \revrev{unrecognized} blends on shear are already corrected for, while the impact on photometric redshift
estimates cannot be studied with either set of simulations described in this section (because they
are single-band only).  The impact of \revrev{unrecognized} blends on photometric redshift estimation for the
weak lensing sample is deferred to future work.}

In \citet{SynPipe:inprep} we also search for evidence of ``orientation'' bias. This is an effect in
which our detection algorithm may preferentially select galaxies with certain orientations in the
vicinity of neighbouring bright galaxies or in which neighbouring bright galaxies may create a bias
in our shape measurements. We find evidence for a strong orientation bias effect around bright
galaxies with $i<21$. The shapes of synthetic galaxies preferentially point towards the locations of
$i<21$ galaxies on scales below about 6 arcseconds (30~kpc at $z=0.4$), however this orientation
bias is not present on larger scales. Because galaxy clusters contain many bright galaxies, it is
possible that this orientation bias may affect cluster lensing studies, especially for lower
redshift galaxy clusters. 


\subsection{GREAT3-like simulations}\label{subsec:great3like}

To estimate the level of ensemble shear biases such as noise bias, model bias, and bias due to
intrinsic limitations of the re-Gaussianization method, and to estimate selection biases, we use a
set of simulations that are similar in spirit to those used for the GREAT3 challenge
\citep{2014ApJS..212....5M,2015MNRAS.450.2963M} \rmrv{and that do not involve injecting objects into
real data, unlike the \texttt{SynPipe} simulations}.  In these simulations, galaxies are placed on a
$100\times 100$ grid with the same lensing shear applied to each galaxy on that grid \rmrv{and
  analyzed using \rmrv{hscPipe}}.  Galaxies are
simulated in 90-degree rotated pairs to cancel out the shape noise to lowest order.  With a set of
many such grids, it is possible to test the recovery of shears, including both additive and
multiplicative biases.

\rmrv{I}n order
to include a realistic level of blending, parent galaxy samples were defined using HSC survey data
(with different observations that have different observing conditions) in the COSMOS field.  By
matching the HSC detections in that region against the HST-COSMOS images, and including a large chunk of
the actual space-based image without attempting to mask out nearby objects, realistic galaxy
morphologies, \revrev{unrecognized} blending effects, and contamination of light profiles by neighboring objects
are naturally included in this simulation despite its grid configuration.
\revrev{We found that the failure to include nearby galaxies in the simulations meant that (a) the
  simulated galaxy population looked very different from the real one (galaxies too small compared
  to those that are observed), and (b) the shear calibration was over-estimated by a very large
  amount, $\sim 9$\% when averaged over the entire sample.}

\rmrv{After defining this parent galaxy sample, }
the simulations were \rmrv{produced in a way that is} meant to represent coadded images.  Full coadd PSF images were drawn from random
locations in the survey, along with the sky variance at the same random locations.  Using that set of random locations, it was possible to
create a set of simulations with a distribution of observing conditions matching those in the real
HSC survey.  \rmrv{The
  parent galaxy sample from the HST-COSMOS images had the HST-COSMOS PSF deconvolved, then were
  sheared, convolved with the randomly-selected HSC PSFs, and resampled to the HSC pixel scale. Because of
the image resampling and combination process, the noise in the coadds exhibits pixel-to-pixel
correlations, but these correlations can be manipulated by adding a small amount of additional noise.  Instead of fully whitening the noise, the average noise correlation function in the HSC coadds was measured and used to produce the noise fields in the simulations.}

For more details of how these simulations were created, see \citet{2017arXiv171000885M}. As demonstrated \rmrv{there}, the observed distribution of galaxy properties after following this procedure for
producing the simulations looks remarkably similar to the distribution of object properties in the
real HSC data\revrev{, with the means of the $|e|$, $R_2$, magnitude, and SNR distributions in the
  simulations and data agreeing
at the 0.1, 1, 1, and 5\% level.}

These simulations were used for four main purposes, resulting in the development of fitting formulae
that were used to populate fields in the catalog and database:
\begin{enumerate}
\item Shape measurement error (i.e., statistical uncertainty in per-object shapes due to pixel
  noise): as demonstrated in, e.g., \citet{2012MNRAS.425.2610R}, the naive shape measurement errors
  from re-Gaussianization \rmrv{underestimate the real statistical errors} \revrev{by typically 30\%}.  Using the multiple noise realizations of the same
  galaxy, we determined a fitting formula for shape measurement errors and their dependence on galaxy
  properties.
\item Intrinsic shape noise: the response of the ensemble average ellipticities (distortions) to
  lensing shears depends on the intrinsic shape noise.  We used the shape measurement errors
  described previously, together with the observed shape dispersion in the real data, to produce a
  fitting formula for the intrinsic shape noise as a function of galaxy properties \revrev{with
    typical $\sigma_e\sim 0.4$ or $\sigma_\gamma\sim 0.24$.  These numbers increase at large $R_2$ likely due to
    the presence of unrecognized blends, which increase the intrinsic shape dispersion \citep{2016ApJ...816...11D}}.
\item Shear calibration bias: we tested the ensemble shear recovery using these simulations, and
  produced a fitting formulae for the average calibration bias as a function of galaxy properties.
  This can be used to correct the ensemble average shear \rmrv{estimated using} subpopulations of the catalog for the
  sources of calibration bias listed above.
\item Additive biases: we also produced a fitting formula for the additive bias due to residual PSF
  anisotropy, i.e., $\hat{g} = (1+m)g + a e_\text{PSF}$.  Using the formula for $a$ as a function of
  galaxy properties, and the PSF \rmrv{shape} as a function of position in the survey, we can
  generate estimates of the additive bias per shape component for each galaxy.  These numbers can be
  used as inputs to ensemble shear estimators.
\end{enumerate}

Selection biases due to the quantities used for selecting galaxies (Section~\ref{subsec:selection})
correlating with the shear and/or PSF anisotropy can be a highly significant source of selection
bias in ensemble shear estimation
\citep[e.g.,][]{2004MNRAS.353..529H,2005MNRAS.361.1287M,2016MNRAS.460.2245J}.  Fortunately, these
simulations make it possible to estimate which quantities used for selection can generate a
selection bias, and the magnitude of selection biases for subsamples of the catalog.  \revrev{There we
  ascertained using the simulations that without any additional selection criteria, the
  multiplicative selection
  bias due to our lower limit on resolution factor is $1.0\% \pm 0.3\%$, and presented a method for
  how to extend this result to subsamples of the catalog.  We also found that multiplicative
  selection biases due to other cuts such as our magnitude cut are statistically consistent with
  zero, as are additive biases due to any of our cuts.}  Note that
unlike the previously-discussed estimates, there is no per-object estimate of selection bias; by
definition, it is only defined for ensembles.  None of the plots in this paper are corrected for
selection bias; corrections will be applied for each science paper depending on the sample used.

\rev{The conclusions of that work are that when considering model bias due to realistic galaxy
  morphology, noise bias, the impact of light from nearby objects, \revrev{unrecognized} blends, and selection
  bias due to weights and selection criteria applied to the sample, additive biases are controlled
  at a level well below the requirements given in this paper.  The multiplicative bias corrections
  defined in that work were found to have a modest dependence on how the galaxy sample was defined, such
  that when applying additional cuts on photometric redshifts, the uncertainty in shear calibration
  may be as large as $\sim 1/3$ the requirements given in this paper.  \revrev{Factoring in these
    effects plus those mentioned above (sensitivity of the results to the galaxy population in the
    simulations), that work estimates a systematic uncertainty in the multiplicative shear calibration of
    0.01, which is within our requirement of 0.017.} As a consequence, 
  additive or multiplicative biases due to shear systematics originating from
the effects listed in this paragraph may be considered as controlled at the level of the
requirements given in this paper.}

\smrv{While performing cosmological analyses with the shear catalogs, we will
adopt a catalog level blinding approach which relies on manipulating the
fitting formulae used to correct for the various biases mentioned above. In
order to maintain this blinding, we do not provide the exact fitting
formulae here.}

\section{\rmrv{Other tests related to image} processing}\label{sec:earlystages}

In this section we show some additional tests of early stages of \rmrv{the image} processing that could
have an effect on shear systematics.

\subsection{Sky subtraction}

As shown in the SDSS \citep[e.g.,][]{2011ApJS..193...29A}, errors in sky subtraction around bright
objects (stars and galaxies) can cause systematic errors in the properties of faint nearby galaxies
such as those that dominate our source catalog.  This can result in inconsistent galaxy selection
criteria in dense regions vs.\ the rest of the survey, coherent tangential shear biases around
bright lens objects, and other systematic errors. \rmrv{Indeed, the HSC DR1 paper describes sky
  over-subtraction near bright galaxies, but the question of how this may result in coherent shear
  measurement systematics was not determined there.}

One test of sky subtraction errors comes from the already-computed tangential shear profile around
stars (after imposition of the bright star mask to remove regions that are badly affected by the
light from bright stars).  This was shown in figure~\ref{fig:bright_star_shear}, and demonstrates that
there is no coherent tangential or radial shear around bright stars detected down to scales of
$0.2$~arcmin.  As discussed in Section~\ref{subsec:synpipe}, there are signs of orientation bias
(coherent radial shear induced with respect to bright galaxies) in the source sample at separations
of $\le 6$\arcsec, which could be a sign of sky subtraction issues on even smaller scales.

\subsection{Relative astrometry}\label{subsec:astrometry}

One potential source of systematic errors in shear is the coaddition process.  Relative astrometric
errors should lead to \rmrv{an} additional blurring kernel in the stars and galaxies on the coadd
(effectively another Gaussian convolution in the PSF), but the coadd PSF does not include this term.
As shown in Section~\ref{subsubsec:psfmodelsize} we see no sign that the star images on the coadd are
larger than the coadd PSF model -- indeed, there is a $10^{-3}$-level discrepancy of the opposite
sign -- but nonetheless we can use the relative astrometric errors characterized in
the HSC DR1 paper
to estimate how important this effect may be for shear estimation.

As mentioned there, the internal astrometric accuracy is $\sim 10$ mas. Given a typical seeing size
of $0.6$\arcsec, and treating both the PSF and the blurring due to astrometric errors as Gaussians that can be added in quadrature, the effective PSF
would be larger than the coadded PSF by 1 part in $10^{-4}$ (in terms of the linear size).  Given
that the systematic bias in the shear due to this effective PSF model size error would be a similar
order of magnitude, we argue that this systematic error is subdominant to those discussed previously
in this paper.  This also explains why no signature of PSF broadening was seen in the tests of the
coadded PSFs.

\subsection{Star/galaxy separation}\label{subsec:stellar}

Failures in star/galaxy separation can cause two different types of systematic errors in weak
lensing.  First, if the sample of \rmrv{objects} used to determine the PSF includes some small galaxies (or
binary stars) then the PSF model \rmrv{will} be systematically biased to larger size (\rmrv{and/or gain} some spurious
\rmrv{shape}).  Figure~13 of the HSC DR1 paper
illustrates that the purity level of the
bright star sample is very high.  Here we rely on our direct empirical tests of PSF models
\rmrv{that
indicate that} their sizes are sufficiently accurate to meet our requirements.

Second, if the shear catalog includes some stars or binary stars misidentified as
galaxies, they will dilute the shear signal, resulting in a negative calibration bias.
We assess the level of stellar contamination in the shear catalog using the
HSC-COSMOS field, taking advantage of the  HST
observations also available in that region.
The HSC-COSMOS data used here correspond to the HSC Deep survey layer, and hence individual
exposure times are longer than in the Wide survey. Exposures with
different seeing values
were used to create Wide-depth stacks with effective seeing better than, around the median of,
and worse than the HSC Wide layer on average\footnote{\rmrv{After this work was completed, a problem with
  the median-seeing coaddition was identified, as described on
  \texttt{https://hsc-release.mtk.nao.ac.jp/doc/index.php/known-problems-in-dr1}, which resulted in
  that coadd being shallower than the others and than the actual Wide-layer coadds by about 0.16
  magnitudes.  Since we do not go near the detection limit given our $i<24.5$ cut, and since the
  results for the median-seeing coadd lie reasonably in between those with the best seeing and worst
seeing, we do not anticipate that this issue with the coadd causes a serious problem for the results presented in this paper.}}.  These stacks have seeing values of 0.5\arcsec,
0.7\arcsec, and 1.0\arcsec, as described in
Section~3.8 of the HSC DR1 paper,
and have fewer
exposures than in a typical Wide layer coadd due to the differences in exposure times.

We take as a reference the star-galaxy classification performed on the HST-COSMOS
data by \cite{2007ApJS..172..219L}. Because these data are higher resolution
than the HSC data, we regard the HST star catalog as the true, complete star
catalog for our purposes.  As shown in \citet{2007ApJS..172..219L}, the star-galaxy separation in HST-COSMOS is
reliable down to a depth of $i\sim 25$, which is fainter than our shape catalog.
 We cross-match the two datasets in the COSMOS region with a maximum matching radius of
$0.4$\arcsec, which is smaller than the HSC PSF and which can be clearly
distinguished as the distance at which spurious matches start to dominate the
cross-matched catalog.

Figure~\ref{fig:star_contamination} shows the fraction of
objects in \rmrv{the shear catalogs constructed for each of these stacks} that were classified as stars by
\cite{2007ApJS..172..219L}. For typical seeing conditions, the contamination is
below 0.2\% for $i$-band magnitudes $<$22, increasing to about 0.5\% for the
faintest sources included in the shape catalog. Even in the worst seeing
conditions (1\arcsec), which are highly non-representative of the shear catalog overall, there is only a 1\% contamination at the faintest
magnitudes.  For all three seeing conditions, the average over the catalog is at most 0.5\% (0.3\%
for median seeing).  We note that this low level of stellar contamination is due to a combination of
the pipeline star/galaxy classifier and other cuts, such as the resolution cut, that is designed to
remove objects (even galaxies) that are too poorly resolved for accurate shear estimation.
Such contamination is well within the levels required for current
science analyses, with the relevant requirement from Section~\ref{sec:requirements} being $|\delta
m|<0.017$.  Our most relevant estimate is somewhere between the best and median seeing, given the
seeing distribution of this catalog, and hence is about 15\% of that requirement. For the current
dataset we do not explore the stellar contamination
or its impact on scientific results any further.  We also note that the results are nearly
indistinguishable with and without lensing weights.

\begin{figure}
\begin{center}
\includegraphics[width=0.9\linewidth]{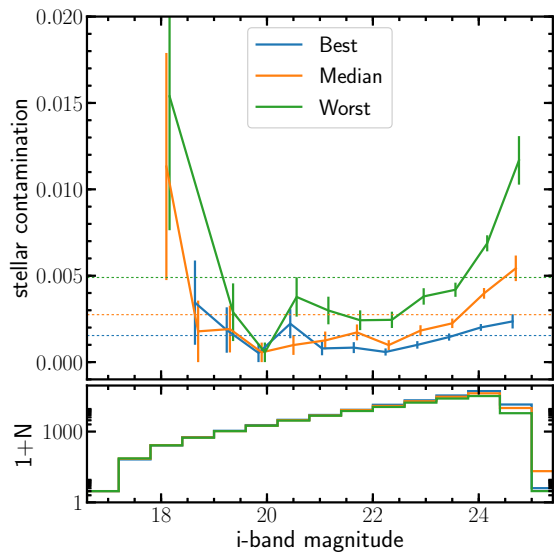}
\end{center}
\caption{Top: contamination of the shape catalog by stars incorrectly classified
as galaxies by the HSC pipeline, calculated in the COSMOS field by comparing to a
classification performed using HST data. We show the contamination as a
function of $i$-band magnitude for three different values of seeing which are
representative of good, typical and bad conditions across the HSC survey.
Errorbars show Poisson uncertainties. Dotted lines show overall averages. Bottom:
number of objects in each magnitude bin, for each dataset split by seeing.
Both panels show unweighted results, however the results are nearly indistinguishable when
incorporating the lensing weights.}
\label{fig:star_contamination}
\end{figure}

\section{Photometric redshifts}\label{sec:photoz}

Until now, we have focused exclusively on systematics related to shear estimation using the entire
sample of galaxies for which shapes could be measured.  In this section, we briefly comment on
photometric redshifts\rmrv{, which are discussed in much more detail in \cite{2017arXiv170405988T,Medezinski:inprep,Speagle:inprep,More:inprep}}. Our
simulations analysis \citep{2017arXiv171000885M} will address the question of whether applying
calibrations as a function of $S/N$ and resolution is sufficient to correct for redshift trends. 
In this paper, we focus on only the most basic aspects of photometric redshifts in the source
catalog.

The first question is how the requirement that there be a photometric redshift available reduce the
source number density.  For one typical photometric redshift catalog out of the several available
options, \texttt{mlz}\footnote{\rmrv{\texttt{http://matias-ck.com/mlz/}}} \rmrv{(machine learning
  and photo-$z$)}, the requirement that there be a value of photometric redshift results in a
12\% reduction in the source number density.  \rmrv{However, this statement is method-dependent as
  will be explored by \citet{Speagle:inprep,More:inprep}.}

Next, we consider how photometric redshift cuts modify the observed distributions of galaxies
properties, such as those shown in figure~\ref{fig:gal_prop}.  For this purpose, we \rmrv{divide the
source sample into three roughly equal subsamples by cutting at $z=0.6$ and $z=1$.}  Then
we plot the lensing-weighted distributions of galaxy properties for the entire sample of galaxies
with photo-$z$, and for those with photo-$z$ above $0.6$ and above $1.0$ (again using \texttt{mlz}).
The resulting distributions are shown in figure~\ref{fig:galprop-pz}.  As shown, the distributions of
resolution factor and \rmrv{distortion} magnitude are not strongly modified when placing photometric
redshift cuts, while requiring a higher photometric redshift skews the sample towards fainter
magnitudes and lower signal-to-noise ratio.  \rmrv{This suggests that shear biases such as noise
  bias will be the primary difference between source samples cut based on redshift, while effects
  that depend on the resolution factor distribution (e.g., the impact of PSF model size errors) will
affect the samples similarly.}
\begin{figure*}
\begin{center}
\includegraphics[clip,width=2.5in]{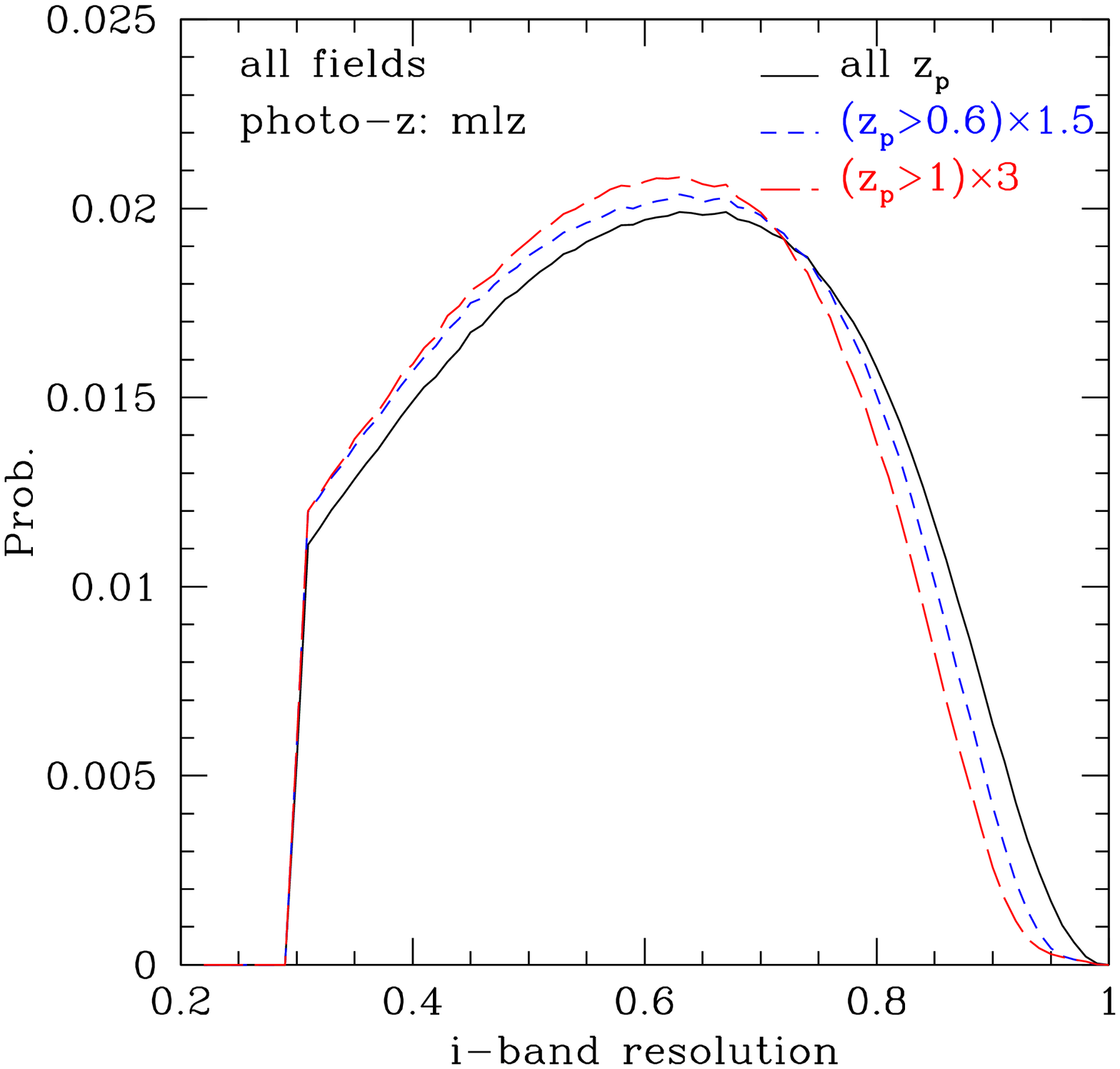}
\includegraphics[clip,width=2.5in]{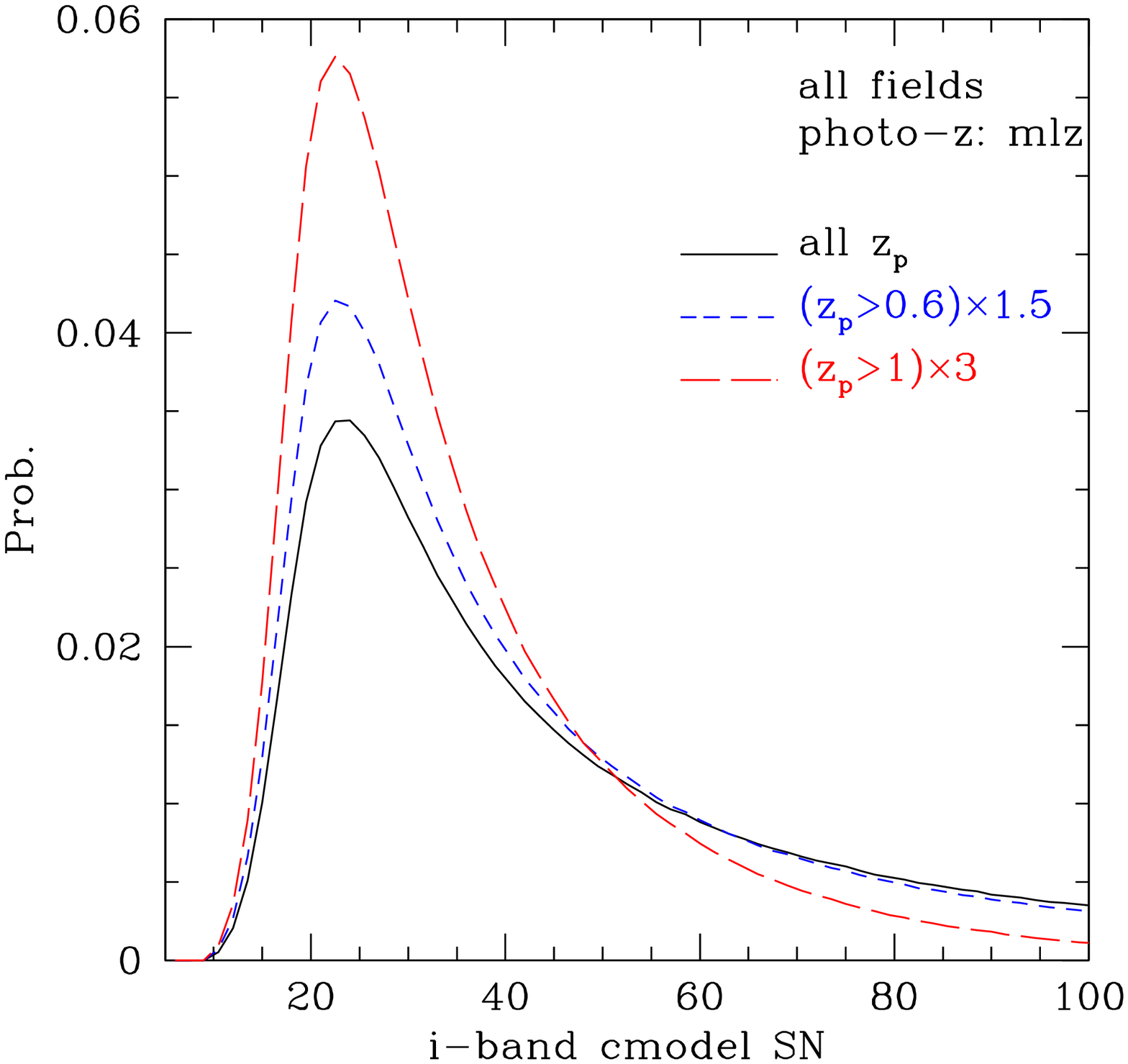}
\includegraphics[clip,width=2.5in]{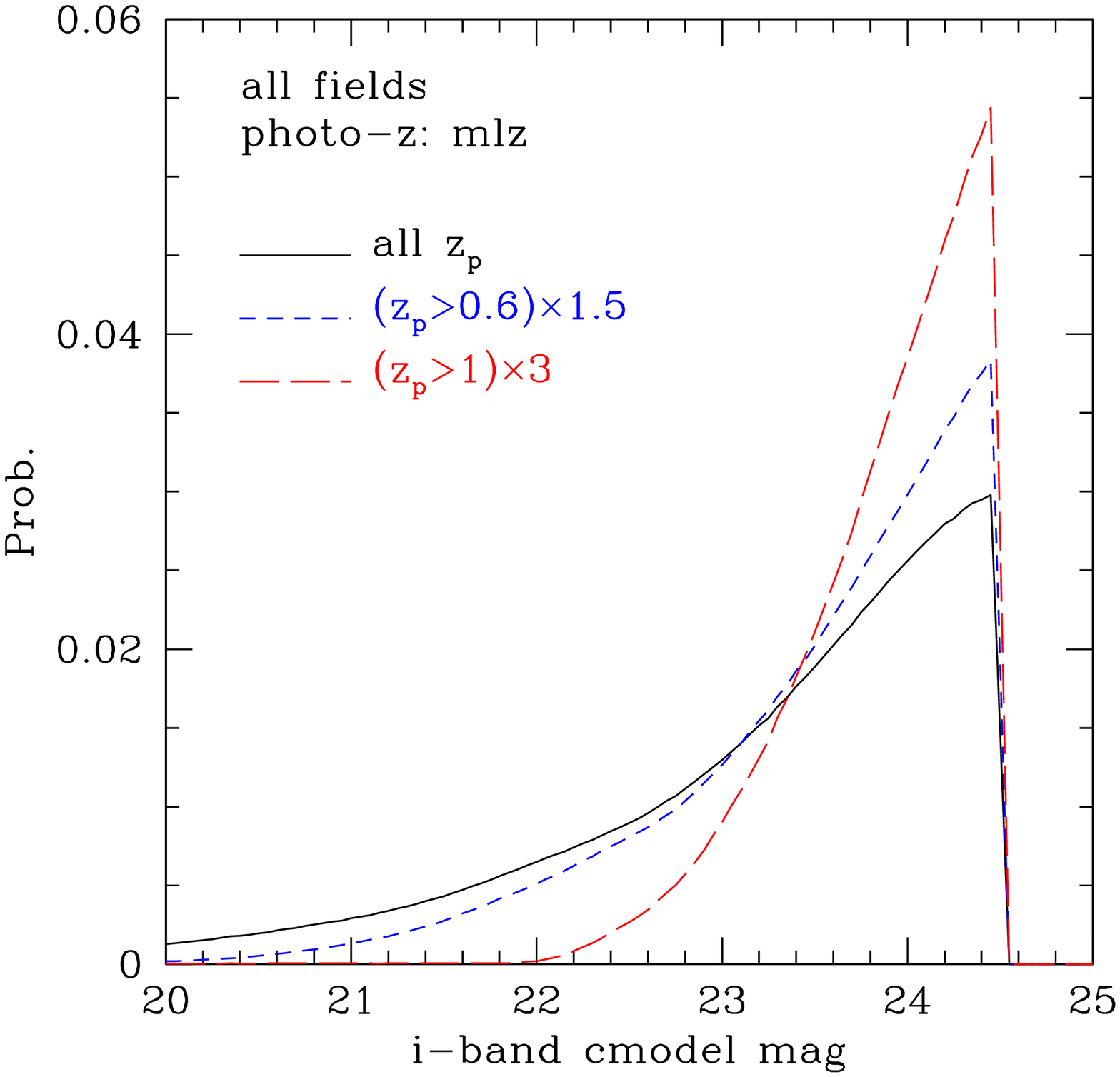}
\includegraphics[clip,width=2.5in]{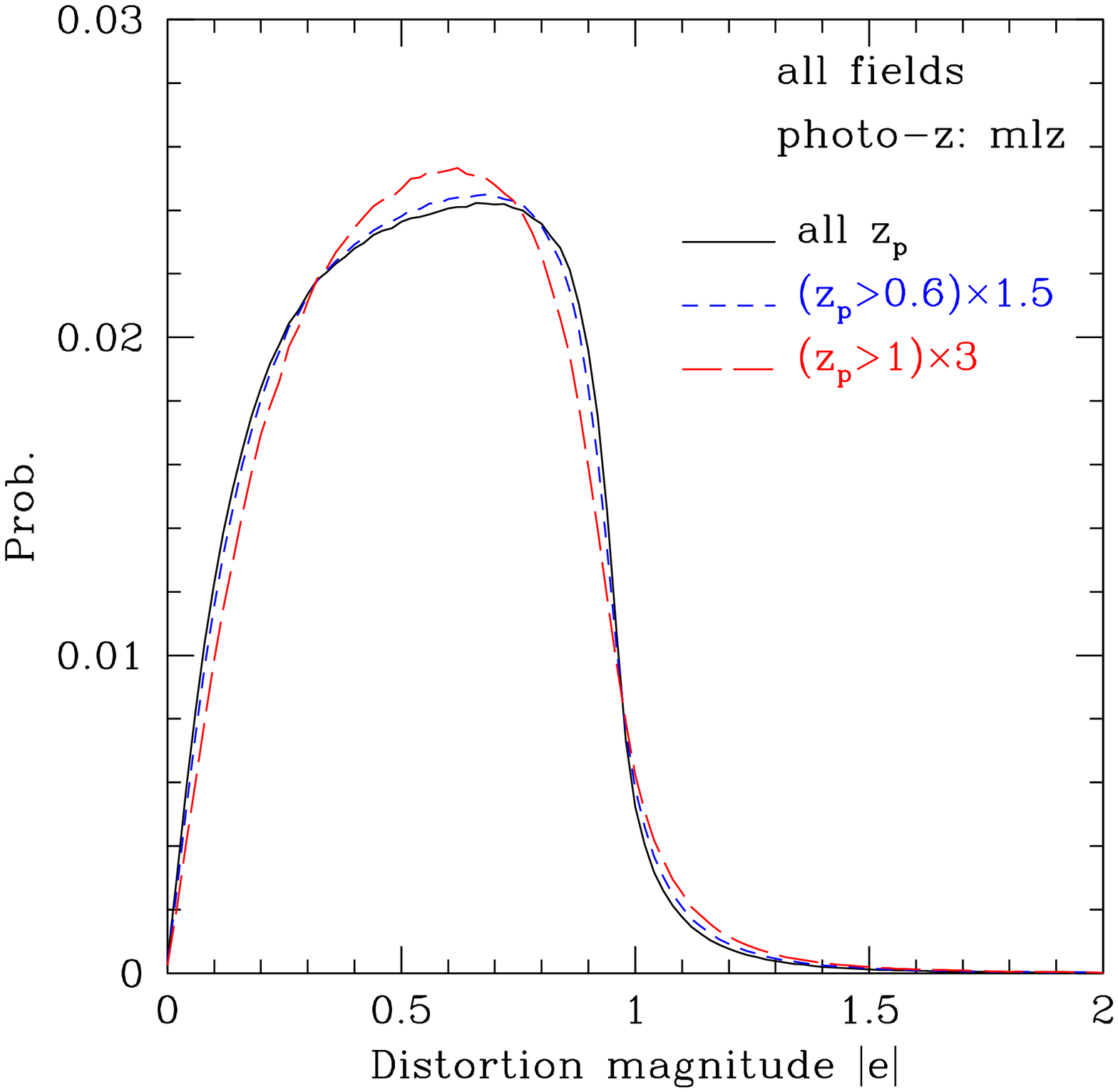}
\end{center}
\caption{We show how the lensing-weighted distribution of galaxy properties across all survey fields
combined changes when imposing lower limits on the photometric redshift from \texttt{mlz}.  The
resulting samples with these cuts are similar to those that might be used for tomographic analysis,
or for galaxy-galaxy lensing source samples.  Each panel shows a different galaxy property as
labeled on the panel itself, and all distributions are renormalized to integrate to the same total.
}
\label{fig:galprop-pz}
\end{figure*}

Finally, we note that we see evidence that the results of some null tests depend on the applied
photo-$z$ cuts.  Since this depends on the choice of photo-$z$ method and the cut itself, we defer
tests of this effect to science papers, which will be motivated by more specific photo-$z$ cuts and
will have null tests that depend on the science case.

\rev{While the impact of photometric redshift bias and scatter on the HSC weak lensing analysis is
  thoroughly quantified elsewhere \citep{Medezinski:inprep,Speagle:inprep,More:inprep}, and the biases in the signals depend on the adopted
  photometric redshift code, here we comment briefly on the current understanding of the residual
  systematic uncertainties (after correcting for known biases).  As shown in \citet{Medezinski:inprep}, when using a
  spectroscopic sample that has been reweighted to match the color and magnitude distribution of the
weak lensing source sample, the calibration of lensing signals for samples with lenses at $z>0.4$
has an uncertainty of $\pm 2$\%.  This uncertainty is primarily statistical, and is driven by the
limited size of spectroscopic samples covering the faint end of our source sample.  This slightly exceeds the
requirements given in Section~\ref{sec:requirements} for constraints on the calibration of the
lensing signals.  There, the number given was that shear calibration must be known to better than
0.017 for it to avoid contributing significantly to the overall error budget.  As this is a
requirement on calibration, a similar requirement applies to uncertainty in lensing signal
calibration due to photometric redshifts.  Hence the
results from \citet{Medezinski:inprep} suggest that our uncertainty in the calibration of the lensing signals due to
photometric redshift uncertainties is a systematic that must be explicitly added to the error budget
for cosmological weak lensing analysis with first-year data, due to it no longer being completely
negligible compared to our statistical uncertainties.  We caution that the specific values of this
systematic uncertainty in
\citet{Medezinski:inprep} are specific to the adopted $p(z)$ method and cuts in that work;
for cosmological galaxy-galaxy lensing analyses, the same methodology will be applied to derive
numbers that apply to that analysis
\citep{More:inprep}. }

\rev{Ongoing spectroscopic campaigns such
as the one described in \citet{2017ApJ...841..111M} will improve spectroscopic redshift coverage in
areas of color and magnitude space where existing samples are relatively sparse, and thereby reduce this systematic uncertainty for
future analyses.  In addition, once a larger HSC survey area is available it will be possible to use
clustering cross-correlations to reconstruct ensemble redshift distributions for photo-$z$-selected
samples, as is used by the Dark Energy Survey as a cross-check to the spectroscopic sample reweighting presented in \citet{2017arXiv170801532H}.}

%
%
%

\section{Summary}\label{sec:summary}

\rev{
In this paper, we have characterized the catalogs to be used for weak lensing science with the first
data release of the HSC survey.  These catalogs were produced using the moments-based
re-Gaussianization method of PSF correction as applied to a linear coadd, for which the PSF model
was constructed \rmrv{as a linear combination of the single-epoch PSF models}.  The tests \rmrv{of
  the quality of these catalogs} presented in this paper are predominantly internal tests such as 
the calculation of standard weak lensing null tests.  Some systematics cannot be assessed
  using null tests; these were only briefly summarized here, and are evaluated in detail in other work:}
\begin{itemize}
\item \rev{\revrev{Unrecognized} blend rate as a function of galaxy properties, and spurious shear due to bright
  objects, are quantified using \texttt{SynPipe} in \citet{SynPipe:inprep}.}
\item \rev{Multiplicative and additive biases due to the PSF correction algorithm used here, including
  model bias, noise bias, the impact of \revrev{unrecognized} blends, and selection biases, are quantified using
  simulations in \citet{2017arXiv171000885M}.}
\item \rev{The impact of photometric redshift errors \citep{2017arXiv170405988T} on weak lensing cosmology analyses are
  quantified in \citet{Medezinski:inprep,Speagle:inprep,More:inprep,CosmicShear:inprep}.}
\end{itemize}

\rev{ Finally, null tests carried out at the map level in \citet{Oguri:inprep} by correlating the
  lensing mass maps against maps of quantities that can induce systematic error (e.g., PSF shape and
  size) produced null detections.  These are nicely complementary to the null tests carried out in
  this paper at the level of 1- and 2-point correlation functions.}

\revrev{We emphasize that this work represents the technical underpinnings of HSC weak lensing
  science papers using this catalog only.  Future data releases will cover more area and will be
  processed by a different software pipeline.  As a result, the performance requirements become more
  stringent, but the results of the null tests and other tests to characterize performance of the
  catalog will also change.  We particularly highlight the fact that several of our PSF
  model-related null tests (specifically the PSF model size residuals and $\rho_1$) meet our requirements with
  the current catalog, but would not meet the requirements for the full HSC survey area, which are
  a factor of $\sim 2.7$ more stringent.  Of the improvements mentioned below in
  Section~\ref{subsec:future}, the PSF model improvements are most clearly motivated by our current
  failure to meet the full-survey requirements on PSF model fidelity.}

\revrev{  However, while our shear
  estimation method is not demonstrably producing multiplicative shear systematics that would exceed
the full-survey requirement of $6\times 10^{-3}$, it seems unlikely that our simulation-based
calibration method alone can reduce the shear calibration uncertainty much below our current uncertainty of
$10^{-2}$.  Hence, the difficulties in precise estimation of the shear calibration using simulations (rather than 
clearly evident systematics exceeding the full survey requirements) have motivated several of the
shear methodology updates listed in Section~\ref{subsec:future} for future data releases. These
methodology updates will result in us having two or three independent means of validating our shear
calibration, providing the multiple cross-checks that will be needed to reduce that part of the
systematic error budget.}

\subsection{Systematic error budget}\label{subsec:syserrorbudget}

\rev{In this section we present a brief quantitative summary of the elements of the systematic
  error budget, for which the key components are listed in Table~\ref{tab:requirements} and the
  requirements themselves were derived in Section~\ref{sec:requirements}.  For each
  row in the table, we discuss what systematic effects contribute and at what level.}

\rev{The first row in the table is the uncertainty in the overall multiplicative bias in the
  galaxy-galaxy lensing signal.  While the discussion in Section~\ref{sec:requirements} focused
  primarily on multiplicative biases in shear, in principle any effect that causes a multiplicative
  bias would contribute to this row.  The relevant effects discussed in this paper\footnote{\rev{PSF
      model size errors would also contribute, but are excluded from this list because requirements
      are placed on them separately.}} are (i) stellar contamination (Section~\ref{subsec:stellar}),
  (ii) astrometric errors (Section~\ref{subsec:astrometry}), (iii) photometric redshift errors
  (Section~\ref{sec:photoz} summarizing results from other work), and (iv) a range of shear-related
  multiplicative biases (Section~\ref{subsec:great3like} summarizing results from other work).  Of
  these, (i), (ii), and (iv) were found to be below our requirements, but (iii) exceeds our nominal
  requirements and hence must be tracked as a separate, significant
  component of the systematic error budget.}

\rev{The second row in the table is the uncertainty in the shear calibration due to uncertainty in
  PSF model size errors.  As shown in Section~\ref{subsec:psf-tests}, the mean of the PSF fractional
  size error (evaluated with the non-PSF star sample) is just barely within our requirement of
  $4\times 10^{-3}$.  Hence this systematic error component should in principle be tracked as a
  component of the systematic error budget.  However, as mentioned above, we already need to track
  the photometric redshift error contribution to the systematic error budget, and that contribution
  is several times the size of the PSF model size error, so when added in quadrature, the PSF
  model size error is relatively unimportant.}

\rev{The remainder of Table~\ref{tab:requirements} relates to additive biases.  As discussed in
  Section~\ref{subsec:great3like}, simulations have shown that additive bias due to insufficient
  correction for PSF anisotropy can be removed effectively to well within the requirements.  The
  other major component of the additive biases can come from PSF model shape errors, for which we
  placed requirements on the five $\rho$ statistics.  As shown in
  Section~\ref{subsubsec:psfmodelshape}, the $\rho$ statistics in this catalog do satisfy our
  (conservative) requirements, though $\rho_1$ is quite close to the requirements.  This is likely
  the source of the spurious star-galaxy shape correlation function presented in
  Section~\ref{subsec:shear-tests}, which on most scales was subdominant to the expected cosmic
  shear signal, but may need to be modeled out on scales exceeding $1^\circ$ (which fortunately do not dominate our
  cosmological constraining power, due to the cosmic variance errorbars).}

\subsection{Areas for future improvement}\label{subsec:future}

\rev{In the near future, we
anticipate that another shear catalog using a very different method of shear inference (an
implementation of the Bayesian Fourier Domain or BFD algorithm from \citealt{2016MNRAS.459.4467B})
will be produced and cross-comparisons will be made with the catalog described in this paper,
providing a complementary cross-check on the first year HSC shear estimation process with \rmrv{the older re-Gaussianization}
method.  This will be valuable given the very different assumptions behind the two methods.}

\rev{The results of the tests in this paper and in \citet{PipelinePaper:inprep} suggest several
  avenues for improvement in future data releases, where \rmrv{contiguous regions of at least twice
    the size of the regions in DR1} will necessitate a better handling of systematics.  We enumerate
  the highest priority plans here:  }
\begin{enumerate}
\item \rev{We would like to have at least two shear estimation
  algorithms available concurrently rather than in succession; work in other surveys
  \citep[e.g.,][]{2016MNRAS.460.2245J} has demonstrated the power of ensemble shear
  cross-comparisons between two methods applied to the same dataset.}  
\item \rev{At least one method
  should operate at the level of individual exposures rather than coadds.  }
\item \rev{Rather than
  assessing shear calibration only via simulations, which \rmrv{are limited} due to imperfect
  knowledge of the galaxy population, future data releases should use some implementation of the
  metacalibration method \citep{2017arXiv170202600H,2017ApJ...841...24S}, which assesses the
  response of the shear estimation method to a shear via resimulation of the data itself.  }
\item   \rev{We need improved methods of understanding the impact of \revrev{unrecognized} blends on
  photometric redshifts, given the high \revrev{unrecognized} blend rate in this dataset.  }
\item \rev{Given the
  dominant role of photometric redshift uncertainty in our systematic error budget, it will be
  important to have spectroscopic training samples with better coverage of the faint end of our
  galaxy sample, and use complementary methods such as clustering redshifts
  \citep{2008ApJ...684...88N,2015MNRAS.447.3500R} that benefit from having wider survey areas, as discussed in
  section~\ref{sec:photoz}.}
  \item \rev{The PSF models will need to be improved if we are to meet our requirements in future
  years, and to recover the best-seeing areas that were removed for this analysis.  We anticipate
  that this will involve replacement of the algorithm rather than improvement of the existing
  algorithm\rmrv{; this is actively being worked on now}.  }
\end{enumerate}

\subsection{Outlook for first-year HSC weak lensing science}

\rev{
For weak lensing analyses that will result directly in cosmological parameter constraints, we note that blinding
has been recognized in recent years as a valuable method for reducing confirmation bias.  For
\rmrv{these analyses}, the HSC weak lensing group is adopting a combination of catalog-level and analysis-level
blinding schemes.  However, the \rmrv{unblinded} catalogs are being used directly for non-cosmological
analyses.  A more detailed discussion of the blinding method is deferred to cosmological analysis
papers.}

\rev{To summarize, for the systematics that can be characterized
  with null tests, the 
catalogs presented in this work meet the requirements for first-year weak lensing science with HSC.
This paper has presented requirements on a broader set of weak lensing systematics than can be
  characterized with null tests; additional papers will detail the methods used to assess the
  systematics that were not fully addressed here, referring in all cases to the requirements defined
  in this paper.  Other work has identified the weak lensing signal calibration uncertainty due to
  photometric redshift errors as a systematic that exceeds our nominal (conservative) requirements,
  and thus must be added as a separate term in the systematic error budget for cosmological weak
  lensing analyses.
The tests in this work have helped us identify the dominant sources
of systematic error that will have to be tracked in cosmological weak lensing analyses (as
summarized in Section~\ref{subsec:syserrorbudget}), and also were useful for identifying future
avenues for improvement in subsequent shear catalogs.}

\rev{In this special issue of PASJ, some initial weak lensing science papers will be presented
\rmrv{covering topics such as cluster-galaxy lensing and mass mapping}, with more
to come (\rmrv{including cosmological analyses}) in the following months.  This begins an exciting new era of cosmological weak lensing
analysis with the HSC survey. Also, this catalog will be released publicly when cosmological results
are published.  Details of data access will be made public at that time.
}

\begin{ack}
The Hyper Suprime-Cam (HSC) collaboration includes the astronomical communities of Japan and Taiwan,
and Princeton University. The HSC instrumentation and software were developed by the National
Astronomical Observatory of Japan (NAOJ), the Kavli Institute for the Physics and Mathematics of the
Universe (Kavli IPMU), the University of Tokyo, the High Energy Accelerator Research Organization
(KEK), the Academia Sinica Institute for Astronomy and Astrophysics in Taiwan (ASIAA), and Princeton
University. Funding was contributed by the FIRST program from Japanese Cabinet Office, the Ministry
of Education, Culture, Sports, Science and Technology (MEXT), the Japan Society for the Promotion of
Science (JSPS), Japan Science and Technology Agency (JST), the Toray Science Foundation, NAOJ, Kavli
IPMU, KEK, ASIAA, and Princeton University.

This paper makes use of software developed for the Large Synoptic Survey Telescope. We thank the
LSST Project for making their code available as free software at  \texttt{http://dm.lsst.org}.

The Pan-STARRS1 Surveys (PS1) have been made possible through contributions of the Institute for
Astronomy, the University of Hawaii, the Pan-STARRS Project Office, the Max-Planck Society and its
participating institutes, the Max Planck Institute for Astronomy, Heidelberg and the Max Planck
Institute for Extraterrestrial Physics, Garching, The Johns Hopkins University, Durham University,
the University of Edinburgh, Queen's University Belfast, the Harvard-Smithsonian Center for
Astrophysics, the Las Cumbres Observatory Global Telescope Network Incorporated, the National
Central University of Taiwan, the Space Telescope Science Institute, the National Aeronautics and
Space Administration under Grant No.\ NNX08AR22G issued through the Planetary Science Division of the
NASA Science Mission Directorate, the National Science Foundation under Grant No. AST-1238877, the
University of Maryland, and Eotvos Lorand University (ELTE) and the Los Alamos National Laboratory.

Based on data collected at the Subaru Telescope and retrieved from the HSC data archive system,
which is operated by Subaru Telescope and Astronomy Data Center, National Astronomical Observatory
of Japan.

\rmrv{We thank Josh Meyers, Paul Price\rev{, and the anonymous referee} for helpful  comments that improved the quality of this
  work.}
This work is in part supported by JSPS KAKENHI (Grant Number
26800093, 15H03654, and JP17H01131) as well as
MEXT Grant-in-Aid for Scientific
Research on Innovative Areas (No.~15H05887, 15H05892, 15H05893, 15K21733).
RM is supported by the US Department of Energy Early Career Award Program.
HM is supported by the Jet Propulsion Laboratory, California Institute
of Technology, under a contract with the National Aeronautics and
Space Administration.
MS is supported by the University of California Riverside Office of Research and Economic Development through the FIELDS NASA-MIRO program.
SM is supported by the Japan Society for Promotion of
Science grants JP15K17600 and JP16H01089.
RyM is financially supported by the University of Tokyo-Princeton strategic partnership grant and
Advanced Leading Graduate Course for Photon Science (ALPS).
WC is supported by the WFIRST program.

\end{ack}

\bibliographystyle{apj}
\bibliography{papers}

\appendix

\section{Catalog quantities}\label{app:catalog-def}
The HSC first-year shape catalog is a set of three catalogs which share the same objects in the same
order. The main catalog includes the object positions, photometry and shapes, \rmrv{as} listed
in Table~\ref{tab:shape_catalog}. The photo-z information is stored in two different catalogs. The
\rmrv{first} includes \rmrv{the} point estimates listed in Table~2 in \cite{2017arXiv170405988T}, and the other is the photo-$z$ PDF.

\begin{table*}
  \caption{Quantities in shape catalog.}
  \begin{tabular}{cl} \hline
    Column & Meaning \\ \hline
    \multicolumn{2}{c}{Basic quantities and flags} \\ \hline
    {\tt object\_id} & object ID\\ \hline
    {\tt ira} & right ascension (J2000.0) measured in $i-$band\\ \hline
    {\tt idec} & declination (J2000.0) measured in $i-$band\\ \hline
    {\tt tract} & tract ID\\ \hline
    {\tt patch} & patch ID\\ \hline
    {\tt weak\_lensing\_flag} & weak lensing flag\\ \hline
    {\tt merge\_peak\_[grizy]} & peak detected in $grizy-$band\\ \hline
    {\tt [grizy]countinputs} & number of $grizy-$band visits contributing at center\\ \hline
    {\tt iflags\_pixel\_bright\_object\_center/any} & source center is close to/source footprint includes BRIGHT\_OBJECT pixels\\ \hline
    {\tt iblendedness\_flags} & flag set if {\tt iblendedness\_abs\_flux} could not be measured because a required input was missing\\ \hline
    {\tt iblendedness\_abs\_flux} & measure of how flux is affected by neighbors\\ \hline
    \multicolumn{2}{c}{Photometry} \\ \hline
    {\tt a\_[grizy]} & Galactic extinction in $grizy-$band\\ \hline

    {\tt iflux\_kron}, {\tt iflux\_kron\_err} & kron flux in $i-$band, and its error\\ \hline
    {\tt iflux\_kron\_flags} & kron flux flag in $i-$band\\ \hline

    {\tt imag\_kron}, {\tt imag\_kron\_err} & kron magnitude in $i-$band, and its error\\ \hline
    {\tt imag\_kron\_flags} & kron magnitude flag in $i-$band\\ \hline

    {\tt iflux\_cmodel}, {\tt iflux\_cmodel\_err} & cmodel flux in $i-$band, and its error\\ \hline
    {\tt iflux\_cmodel\_flags} & cmodel flux flag in $i-$band\\ \hline

    {\tt imag\_cmodel}, {\tt imag\_cmodel\_err} & cmodel magnitude in $i-$band, and its error\\ \hline
    {\tt imag\_cmodel\_flags} & cmodel magnitude flag in $i-$band\\ \hline

    {\tt [grizy]flux\_forced\_cmodel} & forced cmodel flux in $grizy-$band\\ \hline
    {\tt [grizy]flux\_forced\_cmodel\_err} & forced cmodel flux error in $grizy-$band\\ \hline
    {\tt [grizy]flux\_forced\_cmodel\_flags} & forced cmodel flag in $grizy-$band (True indicates failure)\\ \hline
    {\tt [grizy]mag\_forced\_cmodel} & forced cmodel magnitude in $grizy-$band\\ \hline
    {\tt [grizy]mag\_forced\_cmodel\_err} & forced cmodel magnitude error in $grizy-$band\\ \hline
    {\tt [grizy]flux\_forced\_kron} & forced kron flux in $grizy-$band\\ \hline
    {\tt [grizy]flux\_forced\_kron\_err} & forced kron flux error in $grizy-$band\\ \hline
    {\tt [grizy]flux\_forced\_kron\_flags} & forced kron flag in $grizy-$band (True indicates failure)\\ \hline
    {\tt [grizy]mag\_forced\_kron} & forced kron magnitude in $grizy-$band\\ \hline
    {\tt [grizy]mag\_forced\_kron\_err} & forced kron magnitude error in $grizy-$band\\ \hline
    \multicolumn{2}{c}{Regaussianization shapes based on data alone} \\ \hline
    {\tt ishape\_hsm\_regauss\_e1/e2} & \rmrv{Distortion} in sky coordinates estimated by regaussianization method defined\\
    & in distortion, i.e., $|e|=(a^2-b^2)/(a^2+b^2)$. \\ \hline
    {\tt ishape\_hsm\_regauss\_sigma} & non-calibrated shape measurement noise\\ \hline
    {\tt ishape\_hsm\_regauss\_resolution} & resolution of galaxy image defined in equation~\eqref{eq:def-r2}\\ \hline
    \multicolumn{2}{c}{Quantities related to regaussianization shapes calibrated based on image simulations} \\ \hline
    {\tt ishape\_hsm\_regauss\_derived\_weight} & weight for galaxy shapes\\ \hline
    {\tt ishape\_hsm\_regauss\_derived\_sigma\_e} & shape measurement noise\\ \hline
    {\tt ishape\_hsm\_regauss\_derived\_rms\_e} & rms galaxy shape of a population of galaxies\\ \hline
    {\tt ishape\_hsm\_regauss\_derived\_bias\_m} & multiplicative bias\\ \hline
    {\tt ishape\_hsm\_regauss\_derived\_bias\_c1} & additive bias for $e_1$\\ \hline
    {\tt ishape\_hsm\_regauss\_derived\_bias\_c2} & additive bias for $e_2$\\ \hline
    \multicolumn{2}{c}{Non PSF-corrected shapes} \\ \hline
    {\tt ishape\_sdss\_ixx/ixy/iyy} & Adaptive moments \hmrv{in arcsec$^2$}\\ \hline
    {\tt ishape\_sdss\_psf\_ixx/ixy/iyy} & Adaptive moments of PSF evaluated at object position
    \hmrv{in arcsec$^2$}\\ \hline
\label{tab:shape_catalog}
\end{tabular}
\end{table*}

\section{Galaxy cuts}\label{app:selection}

Table~\ref{tab:galaxycuts} lists the exact column names used to impose the galaxy selection
criteria discussed in Section~\ref{subsec:selection}.  All cuts are imposed in the $i$-band, as indicated
by the `i' in front of the flag names.
\begin{table*}
  \begin{tabular}{ll} \hline
    Cut & Meaning \\ \hline
\multicolumn{2}{c}{Basic flag cuts} \\ \hline
\texttt{idetect$\_$is$\_$primary} $==$ True & Identify unique detections only \\ \hline
\texttt{ideblend$\_$skipped} $==$ False & Deblender skipped this group of objects \\ \hline
\texttt{iflags$\_$badcentroid} $==$ False & Centroid measurement failed \\ \hline
\hmrv{\texttt{icentroid$\_$sdss$\_$flags} $==$ False} & \hmrv{centroid.sdss measurement failed} \\ \hline
\texttt{iflags$\_$pixel$\_$edge} $==$ False & Object too close to image boundary for reliable measurements \\ \hline
\texttt{iflags$\_$pixel$\_$interpolated$\_$center} $==$ False & A pixel flagged as interpolated is
close to object center \\ \hline
\texttt{iflags$\_$pixel$\_$saturated$\_$center} $==$ False & A pixel flagged as saturated is
close to object center \\ \hline
\texttt{iflags$\_$pixel$\_$cr$\_$center}\footnotemark[*] $==$ False & A pixel flagged as a cosmic ray hit is
close to object center \\ \hline
\texttt{iflags$\_$pixel$\_$bad}\footnotemark[*] $==$ False & A pixel flagged as otherwise bad is
close to object center \\ \hline
\texttt{iflags$\_$pixel$\_$suspect$\_$center} $==$ False & A pixel flagged as near
saturation is
close to object center \\ \hline
\hmrv{\texttt{iflags$\_$pixel$\_$clipped$\_$any} $==$ False} & \hmrv{Flagged as source footprint includes clipped pixels} \\ \hline
\texttt{i\hmrv{shape}$\_$hsm$\_$regauss$\_$flags} $==$ False & Error code returned by shape measurement
code \\ \hline
\texttt{ishape$\_$hsm$\_$regauss$\_$sigma} $!=$ NaN & Shape measurement uncertainty should not be
  NaN \\ \hline
\texttt{iclassification$\_$extendedness} $!= 0$ & Extended object \\ \hline
\multicolumn{2}{c}{Cuts on object properties} \\ \hline
\texttt{iflux$\_$cmodel}/\texttt{iflux$\_$cmodel$\_$err} $\ge 10$ & Galaxy has high enough $S/N$ in
$i$ band\\
\hline
\texttt{ishape$\_$hsm$\_$regauss$\_$resolution} $\ge 0.3$ & Galaxy is sufficiently resolved \\ \hline
(\texttt{ishape$\_$hsm$\_$regauss$\_$e1}$^2 + $\texttt{ishape$\_$hsm$\_$regauss$\_$e2}$^2)^{1/2} < 2$ & Total
\rmrv{distortion} cut \\ \hline
$0\le $\texttt{ishape$\_$hsm$\_$regauss$\_$sigma}  $\le 0.4$ & Estimated shape measurement error is
reasonable \\ \hline
\texttt{imag$\_$cmodel} $-$ \texttt{a$\_$i} $\le 24.5$ & Magnitude cut \\ \hline
\texttt{iblendedness$\_$abs$\_$flux} $< 10^{-0.375}$ & Avoid spurious detections and those
contaminated by blends \\ \hline
\multicolumn{2}{c}{Require that at least two of the following four cuts be passed (not all)} \\ \hline
\texttt{gflux$\_$cmodel}/\texttt{gflux$\_$cmodel$\_$err} $\ge 5$ & Galaxy has high enough $S/N$ in
$g$ band\\
\hline
\texttt{rflux$\_$cmodel}/\texttt{rflux$\_$cmodel$\_$err} $\ge 5$ & Galaxy has high enough $S/N$ in
$r$ band\\
\hline
\texttt{zflux$\_$cmodel}/\texttt{zflux$\_$cmodel$\_$err} $\ge 5$ & Galaxy has high enough $S/N$ in
$z$ band\\
\hline
\texttt{yflux$\_$cmodel}/\texttt{yflux$\_$cmodel$\_$err} $\ge 5$ & Galaxy has high enough $S/N$ in
$y$ band\\
\hline
  \end{tabular}
  \caption{Selection criteria imposed on the shape catalog, as described in
    Section~\ref{subsec:selection}.  \jbrv{The flags marked \textsuperscript{*} come with the
      following caveat: \texttt{hscPipe} does not propagate pixels affected by
  cosmic rays and sensor defects to the coadd, so the actual effect of these
  for our purposes is that the coadded PSF model (which never accounts for
  masked pixels) is subtly incorrect.  Due to a bug \texttt{hscPipe} does not
  set these flags on coadd measurements at all, however, so at present
  including them here is a no-op; in later versions of \texttt{hscPipe} in
  which the bug is fixed, filtering on these flags will reject objects with
  incorrect PSF models.} \label{tab:galaxycuts}}
\end{table*}

\section{Shear estimation}\label{app:shear}


Here we describe the procedure for using the entries in the shear catalog to calculate average shear
signals.  This calculation will use the following shorthand for these catalog entries from
Table~\ref{tab:shape_catalog}:
\begin{itemize}
\item Distortions $e_1$, $e_2$ in sky coordinates: {\tt ishape\_hsm\_regauss\_e1/e2}
\item Shape weights $w$: {\tt ishape\_hsm\_regauss\_derived\_weight}
\item Intrinsic shape dispersion per component $e_\text{rms}$: {\tt ishape\_hsm\_regauss\_derived\_rms\_e} (these
  are defined for the population as a function of galaxy properties like SNR and resolution, and
  hence are not the same for each galaxy)
\item Multiplicative bias $m$: {\tt ishape\_hsm\_regauss\_derived\_bias\_m}
\item Additive biases $c_1$, $c_2$: {\tt ishape\_hsm\_regauss\_derived\_bias\_c1/c2}
\end{itemize}

\subsection{\rmrv{Galaxy-galaxy lensing}}

\subsubsection{Basic calculation}

After using the $e_1$ and $e_2$ values to calculate tangential shear values $e_t$ over which one
wishes to average to get galaxy-galaxy or cluster-galaxy lensing profiles, the average tangential
shear profile can be calculated using the following formulae, where $i$ is used to indicate
lens-source pairs and the calculation is typically done in bins of angle $\theta$ or physical
separation $r_p$.  For simplicity our notation does not explicitly indicate this binning, so the
formulae presented should be read as being used for each bin with lens-source pairs already
identified.

As described in Section~\ref{subsec:shearest}, our distortion estimates do not provide an unbiased
estimator for the shear; we need to calculate a responsivity factor ${\cal R}$.  This can be
calculated \rmrv{based on the inverse variance weights $w_i$ and the per-object estimates of RMS distortion
$e_{\text{RMS},i}$} as
\begin{equation}\label{eq:calr}
{\cal R} = 1 - \frac{\sum_i w_i e_{\text{rms},i}^2}{\sum_i w_i}.
\end{equation}
While this simplified formula can in principle induce some systematic error, the systematic error
will automatically be corrected when we test shear estimation on the simulations.

We can use the catalog estimates of calibration bias $m_i$ to derive an ensemble estimate for the
calibration bias for our sample:
\begin{equation}\label{eq:hatm}
\hat{m} = \frac{\sum_i w_i m_i}{\sum_i w_i}
\end{equation}

Finally, these can be combined with the standard average over tangential shear estimates to get the
stacked shear estimator
\begin{equation}\label{eq:hatgamma}
\hat{g} = \frac{\sum_i w_i e_{t,i}}{2 {\cal R} (1+\hat{m})\sum_i w_i}.
\end{equation}

\subsubsection{Additive bias terms}

In principle, additive systematic errors can induce scale-dependent additive biases in the stacked
shear profiles, particularly if these vary across the sky and/or the signal is calculated on large
scales, where many of the annuli around lenses will be incomplete due to survey boundaries.  While
\rmrv{this systematic can be removed by} subtracting the signal around random lenses that are distributed with the same area coverage as the
real lenses \citep[e.g.,][]{2013MNRAS.432.1544M}, such \rmrv{a}
correction depends on a clear understanding of the factors that determine the lens \rmrv{sample
 angular selection  function} (e.g.,
if the lens selection depends on observing conditions, this must be modeled).  Use of the
\rmrv{additive biases} $c_{1,2}$
values in the catalog can be used to more directly remove the additive systematic due to incomplete
correction for the PSF anisotropy, though some additive systematics due to PSF modeling errors and
selection biases may remain.  To construct the additive term that must be subtracted from
$\hat{g}$, the following calculation must be done.  First, the $c_{1,2}$ values must be rotated
into the tangential frame in the lens-source pair system, just like the $e_{1,2}$ values, to get
$c_t$.  Then \rmrv{define the weighted sum}
\begin{equation}\label{eq:hatc}
\hat{c} = \frac{\sum_i w_i c_{t,i}}{\sum_i w_i}.
\end{equation}
Finally, $\hat{c}/(1+\hat{m})$ should be subtracted from \rmrv{the} $\hat{g}$ \rmrv{from equation~\eqref{eq:hatgamma}}.

The order of operations defining how $\hat{m}$ and $\hat{c}$ should be used comes from defining
shear calibration biases as
\begin{equation}
\frac{\sum_i w_i e_{t,i}}{2 {\cal R} \sum_i w_i} = (1+\hat{m}) g_\text{true} + \hat{c}.
\end{equation}
The first step, dividing by $1+\hat{m}$, is used to define the shear estimator in
equation~\eqref{eq:hatgamma}.  T\rmrv{his definition explains why} the additive term to be removed
is $\hat{c}/(1+\hat{m})$ \rmrv{rather than $\hat{c}$.}

\subsection{\rmrv{Pair-weighted statistics}}

\rmrv{ Some plots in this paper show pair-weighted statistics, such as the star-galaxy shape
  correlation function in figure~\ref{fig:ss-sg}.  Here we describe how to adapt the above calculations
  to such pair-weighted statistics. The star-galaxy correlation
  involves pairs consisting of one star and one galaxy.  Using the galaxy sample, we can use
  equations~\eqref{eq:calr} and~\eqref{eq:hatm} to calculate the responsivity ${\cal R}$ and overall
  calibration bias $\hat{m}$, respectively.  However, the additive bias requires a per-object
  correction.  Hence for this case, we define a per-object galaxy shear estimate as
\begin{equation}
\hat{g}_i = \frac{1}{1+\hat{m}} \left[\frac{e_i}{2{\cal R}} - c_i\right]
\end{equation}
and then the $\hat{g}_i$ values are correlated with the star shapes.  For the star shapes, we
index the stars with a subscript $j$, and define $\hat{g}_j^* = e_j/2$ (using the relation
between distortion and shear for intrinsically round objects).  We equally weight the stars, i.e.,
$w_j=1$, and hence our estimator for the star-galaxy shape correlation function is
\begin{equation}
\xi_\text{sg} = \frac{\sum_i \sum_j w_i w_j \hat{g}_i \hat{g}_j^*}{\sum_i \sum_j w_i w_j}
= \frac{\sum_i \sum_j w_i \hat{g}_i \hat{g}_j^*}{\sum_i \sum_j w_i}
\end{equation}
defined within bins of separation $\theta$.
}

\subsection{Additional complications}

Finally, we note that if additional weight factors are desired, for example, to weight by lens
properties, or to get inverse variance-weighted $\Delta\Sigma$ profiles (which require weighting by
$1/\Sigma_\text{crit}^2$), the weights must be self-consistently updated in all of
equations \rmrv{throughout appendix~\ref{app:shear}}.

\end{document}